\documentclass[a4paper,11pt]{article}
\usepackage{jheppub}
\usepackage{lineno}
\usepackage[english]{babel}

\usepackage{amsmath}
\usepackage{graphicx}
\usepackage{verbatim}

\usepackage{xspace}

\newcommand{\msbar}{\ensuremath{\mathrm{\overline{MS}}}\xspace}
\newcommand{\ttbar}{\ensuremath{\mathrm{t\overline{t}}}\xspace}
\newcommand{\qqbar}{\ensuremath{\mathrm{q\overline{q}}}\xspace}
\newcommand{\bbbar}{\ensuremath{\mathrm{b\overline{b}}}\xspace}

\newcommand{\WbWb}{\ensuremath{\mathrm{WbWb}}\xspace}
\newcommand{\sigmaWbWb}{\ensuremath{\sigma_\WbWb}\xspace}
\newcommand{\sqrts}{\ensuremath{\sqrt{s}}\xspace}

\newcommand{\mt}{\ensuremath{m_\mathrm{t}}\xspace}
\newcommand{\mW}{\ensuremath{m_\mathrm{W}}\xspace}
\newcommand{\mZ}{\ensuremath{m_\mathrm{Z}}\xspace}
\newcommand{\mH}{\ensuremath{m_\mathrm{H}}\xspace}

\newcommand{\mtps}{\ensuremath{m_\mathrm{t}^\mathrm{PS}}\xspace}
\newcommand{\mtones}{\ensuremath{m_\mathrm{t}^\mathrm{1S}}\xspace}

\newcommand{\Gt}{\ensuremath{\Gamma_\mathrm{t}}\xspace}
\newcommand{\GW}{\ensuremath{\Gamma_\mathrm{W}}\xspace}
\newcommand{\GZ}{\ensuremath{\Gamma_\mathrm{Z}}\xspace}
\newcommand{\yt}{\ensuremath{y_\mathrm{t}}\xspace}

\newcommand{\chisq}{\ensuremath{\chi^2}\xspace}
\newcommand{\as}{\ensuremath{\alpha_\mathrm{S}}\xspace}
\newcommand{\aEM}{\ensuremath{\alpha_\mathrm{EM}}\xspace}
\newcommand{\asmz}{\ensuremath{\as(\mZ^2)}\xspace}
\newcommand{\aEMmz}{\ensuremath{\aEM(\mZ^2)}\xspace}
\newcommand{\NNNLO}{\ensuremath{\mathrm{N^3LO}}\xspace}
\newcommand{\qqTh}{\ensuremath{\texttt{QQbar\_Threshold}}\xspace}

\newcommand{\GeV}{\ensuremath{\,\mathrm{GeV}}\xspace}
\newcommand{\MeV}{\ensuremath{\,\mathrm{MeV}}\xspace}
\newcommand{\keV}{\ensuremath{\,\mathrm{keV}}\xspace}

\newcommand{\fbinv}{\ensuremath{\,\mathrm{fb^{-1}}}\xspace}
\newcommand{\abinv}{\ensuremath{\,\mathrm{ab^{-1}}}\xspace}

\newcommand{\epm}{\ensuremath{\mathrm{e^+e^-}}\xspace}
\newcommand{\Nb}{\ensuremath{N_\mathrm{bjets}}\xspace}
\newcommand{\Nj}{\ensuremath{N_\mathrm{jets}}\xspace}

\newcommand{\ie}{i.e.\ }
\newcommand{\eg}{e.g.\ }

\arxivnumber{2503.18713}

\title{A detailed study on the prospects for a \boldmath \texorpdfstring{\ttbar}{tt} threshold scan in \texorpdfstring{\epm}{e+e-} collisions}

\author[a,1]{Matteo M. Defranchis,\note{Corresponding author}}
\author[b]{Jorge de Blas,}
\author[a]{Ankita Mehta,}
\author[a]{Michele Selvaggi,}
\author[c]{and Marcel Vos}

\affiliation[a]{CERN, Geneva, Switzerland}
\affiliation[b]{University of Granada, Granada, Spain}
\affiliation[c]{IFIC (UV/CSIC) Valencia, Spain}

\emailAdd{matteo.defranchis@cern.ch}
\emailAdd{deblasm@ugr.es}
\emailAdd{ankita.mehta@cern.ch}
\emailAdd{michele.selvaggi@cern.ch}
\emailAdd{marcel.vos@ific.uv.es}

\abstract{A scan of the beam energy across the top quark pair (\ttbar) production threshold is part of the program of future  Higgs, top, and electroweak factory projects. In this paper, we provide projections for the achievable precision in the top quark mass (\mt), width (\Gt), and Yukawa coupling (\yt) at the electron-positron (\epm) stage of the Future Circular Collider (FCC-ee). The study includes a detailed assessment of parametric and systematic uncertainties, as well as a rigorous estimate of the effect of point-to-point correlations. We project that \mt and \Gt can be determined with an experimental precision of 6.8 and 11.5\MeV, respectively, when \mt is defined in the potential-subtracted (PS) scheme. The impact of theoretical uncertainties due to missing higher orders is found to be of about 35\,(25)\MeV on \mt (\Gt) at \NNNLO in non-relativistic QCD.
Therefore, improvements in the theoretical accuracy, which is an active area of development, are key to match the achievable experimental precision at a future \epm collider.
We also explore the prospects for a measurement of \yt at FCC-ee via a dedicated run above the \ttbar production threshold.}

\begin{document}

\maketitle

\section{Introduction}
\label{sec:intro}

A scan of the electron-positron (\epm) centre-of-mass energy across the top quark pair (\ttbar) production threshold~\cite{Gusken:1985nf,Fadin:1987wz,Fadin:1988fn, Strassler:1990nw, Guth:1991ab, Jezabek:1992np} is widely recognized as the ``golden'' method for a top quark mass measurement, combining excellent experimental precision with good control over the top quark mass scheme and interpretation.
This method, in fact, relies solely on the measurement of the inclusive production cross sections and does not require any kinematic reconstruction of the top quark pairs. Top quark reconstruction techniques are known to introduce significant dependence on the details of the Monte Carlo (MC) simulation and are often a limiting factor at hadron colliders~\cite{CMS:2024irj}. For these reasons, a dedicated \ttbar threshold scan is part of the program of all proposed Higgs/top/EW factories~\cite{deBlas:2024bmz}. 

At a lepton collider, the shape of the \WbWb production rate (\sigmaWbWb) versus centre-of-mass energy is sensitive to the top quark mass (\mt) and width (\Gt), while the absolute cross section around the threshold is sensitive to the strong coupling and the top quark Yukawa coupling (\yt). Accurate predictions of the total cross section in the threshold region are a key ingredient for this program, as they are used to extract these parameters from experimentally measured cross sections. Calculations of the cross section are now available at \NNNLO~\cite{Beneke:2015kwa} and (partial) NNLL accuracy~\cite{Hoang:2013uda}. A program is publicly available that provides \NNNLO predictions of the cross section~\cite{Beneke:2016kkb} as a function of the centre-of-mass energy, including EW corrections and QED initial-state radiation in the calculation. The code provides predictions for the \WbWb final state, including doubly-resonant (\ie \ttbar), single-resonant (\ie single top production $\epm \to \mathrm{tWb}$), and non-resonant contributions ($\epm \to \mathrm{W^+W^-b\overline{b}}$)~\cite{Beneke:2015lwa}. Other contributions, such as $\epm \rightarrow \mathrm{W^+W^-Z}$ with $\mathrm{Z \to b\overline{b}}$, are not taken into account and are treated as background. Calculations at \NNNLO in QCD based on the principle of maximal conformality (PMC) are also available~\cite{Yan:2023mjj}, but are not considered in this study.

Early experimental studies were performed by Martinez \& Miquel~\cite{Martinez:2002st}. Their experimental strategy used several differential cross section measurements to enhance the sensitivity and disentangle the Standard Model (SM) parameters. While work towards a MC event generator that includes the effect of the Coulomb pseudo-bound-state cross-section enhancement~\cite{Bach:2017ggt} is ongoing, the highest-precision calculations only predict the inclusive cross section. Therefore, more recent studies adopt a simple strategy, first explored by Seidel et al.~\cite{Seidel:2013sqa} and Horiguchi et al.~\cite{Horiguchi:2013wra}. In that study, the parameter extraction is based on a measurement of the inclusive cross section at 10 or 11 different centre-of-mass energies, with points spaced by 1\GeV and a total integrated luminosity of 100--200\fbinv. This approach is adopted as the baseline by ILC and CLIC~\cite{CLICdp:2018esa}. These studies show that a statistical precision of the top quark mass measurement of about 20\MeV can be achieved. The limiting factor is the precision of the predictions, as scale and parametric uncertainties add up to approximately 50\MeV. Theory uncertainties also dominate the uncertainty for the other parameters. The top quark Yukawa coupling, for instance, can be measured to a competitive 4\% statistical precision~\cite{Horiguchi:2013wra}. The potential of this measurement is limited in practice, as the \as and scale uncertainties in the prediction yield a systematic uncertainty of approximately 20\%~\cite{Vos:2016til}. 
Studies by Nowak \& Zarnecki~\cite{Nowak:2021tyn, Nowak:2021xmp} and the CEPC group~\cite{Li:2022iav} optimize the number and spacing of the scan points. A concentration of the integrated luminosity at the steepest section of the threshold can yield a considerable improvement of the statistical uncertainty that can be obtained with a given integrated luminosity, but does not test the agreement between the predicted shape and the data and may be less robust against systematic and theoretical uncertainties. The goal of this work is to outline a robust baseline strategy for the \ttbar threshold scan, derive detailed projections for the FCC-ee operating scenario, and compare our findings with results obtained for other proposed colliders.

\begin{figure}[htbp]
    \centering
    \includegraphics[width=0.495\textwidth]{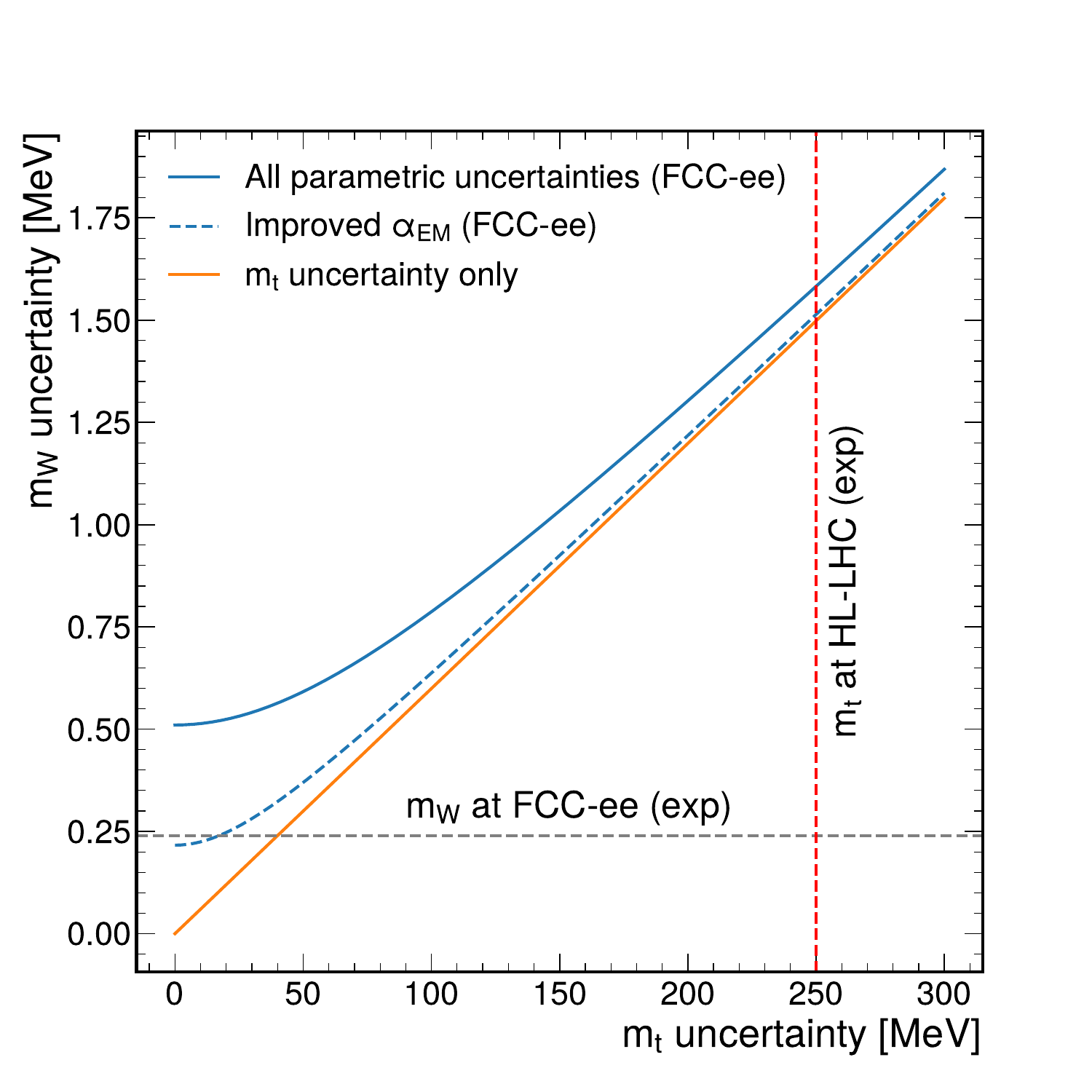} 
    \includegraphics[width=0.46\textwidth]{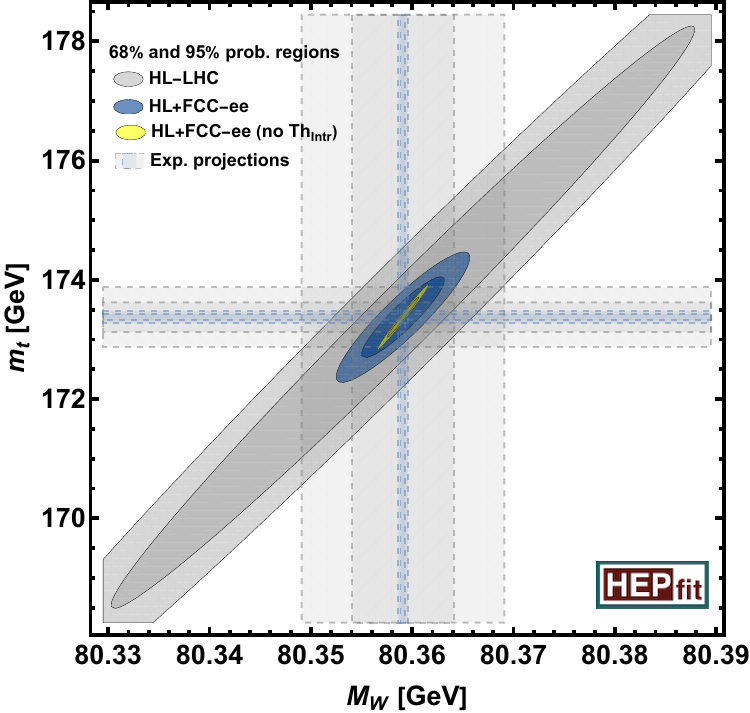} 
    \caption{Left: total parametric uncertainty in the SM prediction of \mW
    as a function of the input uncertainty in \mt, for two different assumptions on \aEM (solid and dashed curves). The solid diagonal line represents the contribution from \mt alone. The horizontal line represents the expected experimental precision in \mW at FCC-ee, while the vertical line indicates the expected precision in \mt after HL-LHC. Uncertainties in \asmz ($10^{-4}$), \aEM ($3\cdot 10^{-5}$ baseline, $10^{-5}$ improved~\cite{Riembau:2025ppc}), \mZ (1\keV), and \mH (3\MeV) according to recent projections for FCC-ee are assumed.
    Right: Comparison of indirect determination of \mt and \mW from the fit to EWPO (elliptical contours) and the projected precision from direct measurements (bands). We show in grey the projections for the HL-LHC, while in blue we show the FCC-ee ones. These results include the projected future intrinsic theory uncertainties in EWPO. The FCC-ee results in a scenario where theory calculations are improved so that these uncertainties become subdominant is shown in the (small) yellow ellipse.
    More details can be found in Appendix~\ref{app:EW_fit}.}
    \label{fig:MW-mt_EWfit}
\end{figure}

To set a target for the desired precision in \mt at future \epm colliders, we investigate the impact of the \mt uncertainty on the SM prediction of electroweak (EW) precision observables (EWPO). This approach exploits the fact that loop diagrams in the predictions for these observables introduce contributions proportional to $\mt^2$. This is particularly relevant in the case of the W boson mass (\mW), which can be measured at FCC-ee with an uncertainty of about 0.24\MeV~\cite{FCC:2025lpp}. We estimate the total uncertainty in the SM prediction of \mW as a function of the uncertainty in \mt, assuming the projected uncertainties for other input parameters after the Z~pole and WW threshold runs at FCC-ee, as detailed in Appendix~\ref{app:EW_fit}. The result, shown in Figure~\ref{fig:MW-mt_EWfit} (left), showcases the need for a measurement of \mt at the level of 50\MeV or better. This target is a factor of five better than the achievable precision at the HL-LHC~\cite{Azzi:2019yne}. With an improved measurement of the electromagnetic constant (\aEM) at the Z~pole from Ref.~\cite{Riembau:2025ppc}, a \mt measurement with an uncertainty of around 20\MeV becomes necessary in order to match the experimental precision in \mW. This is more than a factor of ten better than the projected \mt precision at HL-LHC.
The impact of the intrinsic theoretical uncertainties of the electroweak precision fit is not considered in this result, but it is shown in Figure~\ref{fig:MW-mt_EWfit} (right).

In this paper, we revisit the potential of a \ttbar threshold scan by performing a full analysis including FCC-ee simulated events and a phenomenological study using \NNNLO theory predictions. For this purpose, we consider a \ttbar threshold scan at FCC-ee with a total integrated luminosity of 410\fbinv~\cite{FCC:2025lpp}. In Section~\ref{sec:reco_fit} we present a full demonstration of a cross section measurement using simulated events, including background processes and experimental systematic uncertainties. These results are then used in Section~\ref{sec:threshold_scan} to derive projections for the uncertainties on the top quark mass (\mt) and width (\Gt), including estimates for experimental, machine-related, parametric, and theoretical uncertainties, by performing a phenomenological analysis of the \ttbar threshold scan.
We then investigate the dependence of the results on the assumptions on the systematic uncertainties (Section~\ref{sec:param_scans}) and we explore the possibility of a determination of the top Yukawa coupling above the \ttbar production threshold in Section~\ref{sec:yukawa_fit}. The baseline scenario for this dedicated run assumes an integrated luminosity of 2.65\abinv at $\sqrts = 365\GeV$. Finally, in Section~\ref{sec:summary} we compare the obtained results to previous work and comment on the applicability of our findings to other proposed \epm colliders, and specifically to a linear collider (LC) and CEPC.

\section{Measurement of the WbWb production cross section at FCC-ee}
\label{sec:reco_fit}

In this section we outline an experimental strategy for the measurement of the \WbWb production rate at different centre-of-mass energies at FCC-ee. We focus on three different centre-of-mass energies, corresponding to $\sqrts = 340$, 345, and 365\GeV. These represent \WbWb production below, at, and above the \ttbar threshold, respectively. We restrict our study to the fully hadronic and the semi-hadronic decay channels of the W boson pairs, which together account for a combined branching ratio of about 75\% for this process~\cite{ParticleDataGroup:2024cfk}. Di-leptonic decay channels and the final states involving $\tau$ leptons, which account for the remaining 5\% and 15\% of the branching ratio, respectively, are expected to provide some additional sensitivity and can be included in future work. However, semi-hadronic events in which $\tau$ leptons decay into an electron or a muon are included in this study.
The considered sources of background include \qqbar production, WW and ZZ production, and WWZ production in the $\mathrm{Z} \to \bbbar$ final state. The small interference between the hadronic final states of WW and ZZ production and photon-induced multi-jet production are not expected to have any relevant impact on the result, and is therefore not considered. Similarly, the interference between \WbWb and WWZ ($\mathrm{Z} \to \bbbar$) is expected to be a subleading effect, and is therefore neglected. This is  consistent with the approach followed in the computation of the NNLO EW corrections~\cite{Beneke:2017rdn} included in the \qqTh. Finally, the WWH ($\mathrm{H} \to \bbbar$) background is not considered due to the significantly lower production cross section compared to the other processes.
The \WbWb, WWZ, and \qqbar samples are produced using the WHIZARD generator~\cite{Kilian:2007gr} interfaced with Pythia6~\cite{Sjostrand:2006za}, while ZZ and WW are generated with Pythia8~\cite{Sjostrand:2014zea}. All samples are generated at leading order (LO) in the matrix element, interfaced with a parton shower at leading logarithmic (LL) precision, and are normalised to the cross sections summarised in Table~\ref{tab:MC_samples}. Detector effects are simulated using the \textsc{Delphes}~\cite{deFavereau:2013fsa} parametrisation of the IDEA detector concept~\cite{IDEAStudyGroup:2025gbt,delphes_card_idea} at FCC-ee.

\begin{table}[htbp]
    \centering
    \small
    \begin{tabular}{lllccc}
        \multicolumn{1}{l}{Process} & \multicolumn{1}{l}{Decays}& \multicolumn{1}{l}{Generator}  & \multicolumn{3}{c}{Cross section [pb]} \\
        \multicolumn{3}{c}{} & 340\GeV & 345\GeV & 365\GeV \\

        \hline  
        \WbWb & inclusive & WHIZARD+Pythia6  & 0.1 & 0.5 & 0.5 \\
        WWZ & $\mathrm{Z \to b\bar{b}}$ & WHIZARD+Pythia6  & 2.02 $\times 10^{-3}$ & 1.46 $\times 10^{-3}$ & 1.32 $\times 10^{-3}$ \\
        \qqbar & n.a. & WHIZARD+Pythia6 &  26.3 & 25.6 & 22.8 \\  
        ZZ & inclusive & Pythia8 &  0.932 & 0.916 & 0.643\\  
        WW & inclusive & Pythia8 &  12.1 & 11.9 & 10.7 \\
        \hline
    \end{tabular}
    \caption{Details of the Monte Carlo samples and total cross section for the various processes. All samples are generated at LO in the matrix element interfaced with parton showers at leading-logarithmic accuracy. All processes except \WbWb are normalised to the corresponding LO cross section. The total cross section for \WbWb production, instead, is set to realistic but arbitrary values. The cross section for WWZ also includes the $\mathrm{Z \to \bbbar}$ branching ratio.}
    \label{tab:MC_samples}
\end{table}

\begin{figure}[htbp]
    \centering
    \includegraphics[width=0.495\linewidth]{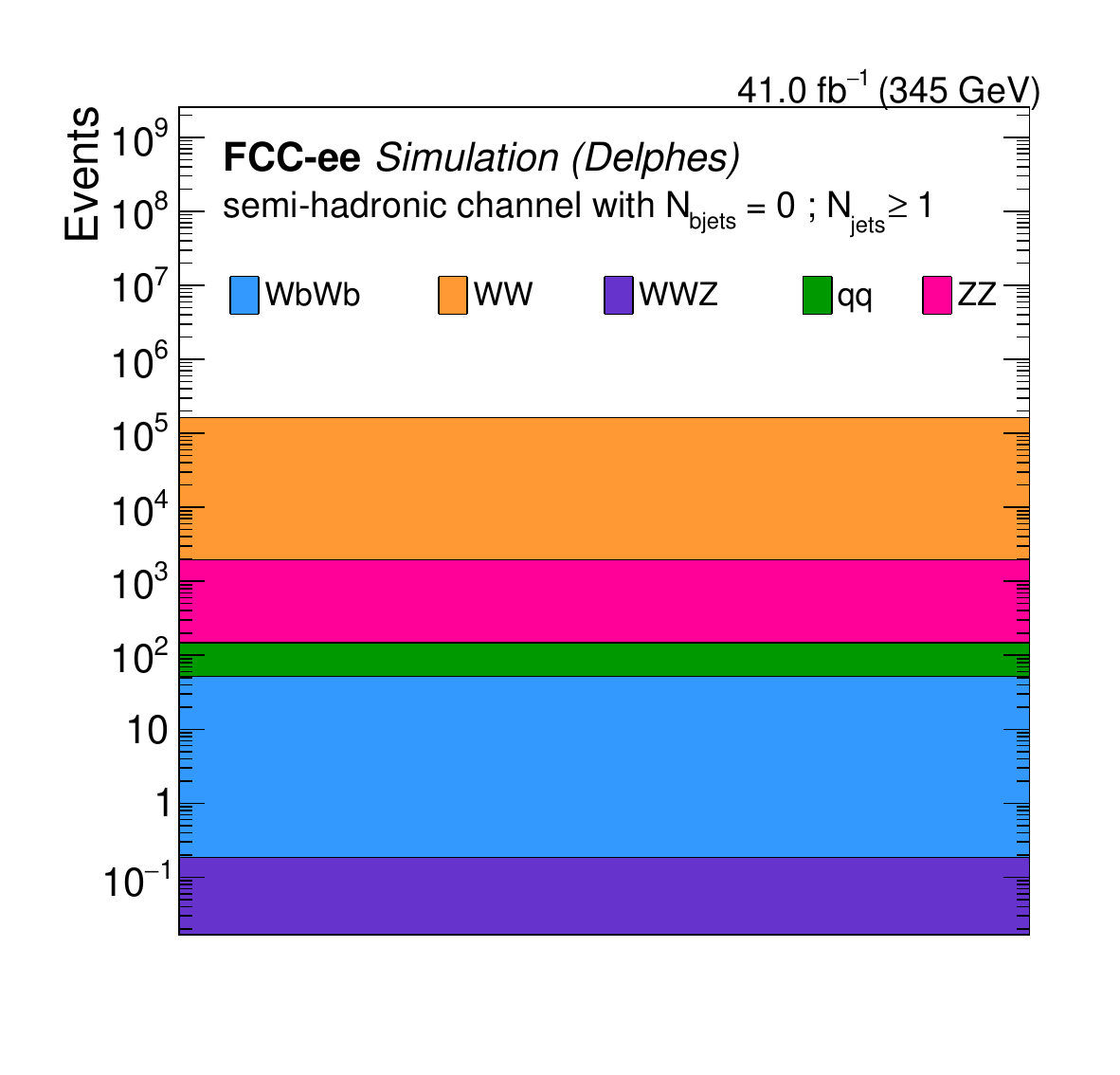} 
    \includegraphics[width=0.495\linewidth]{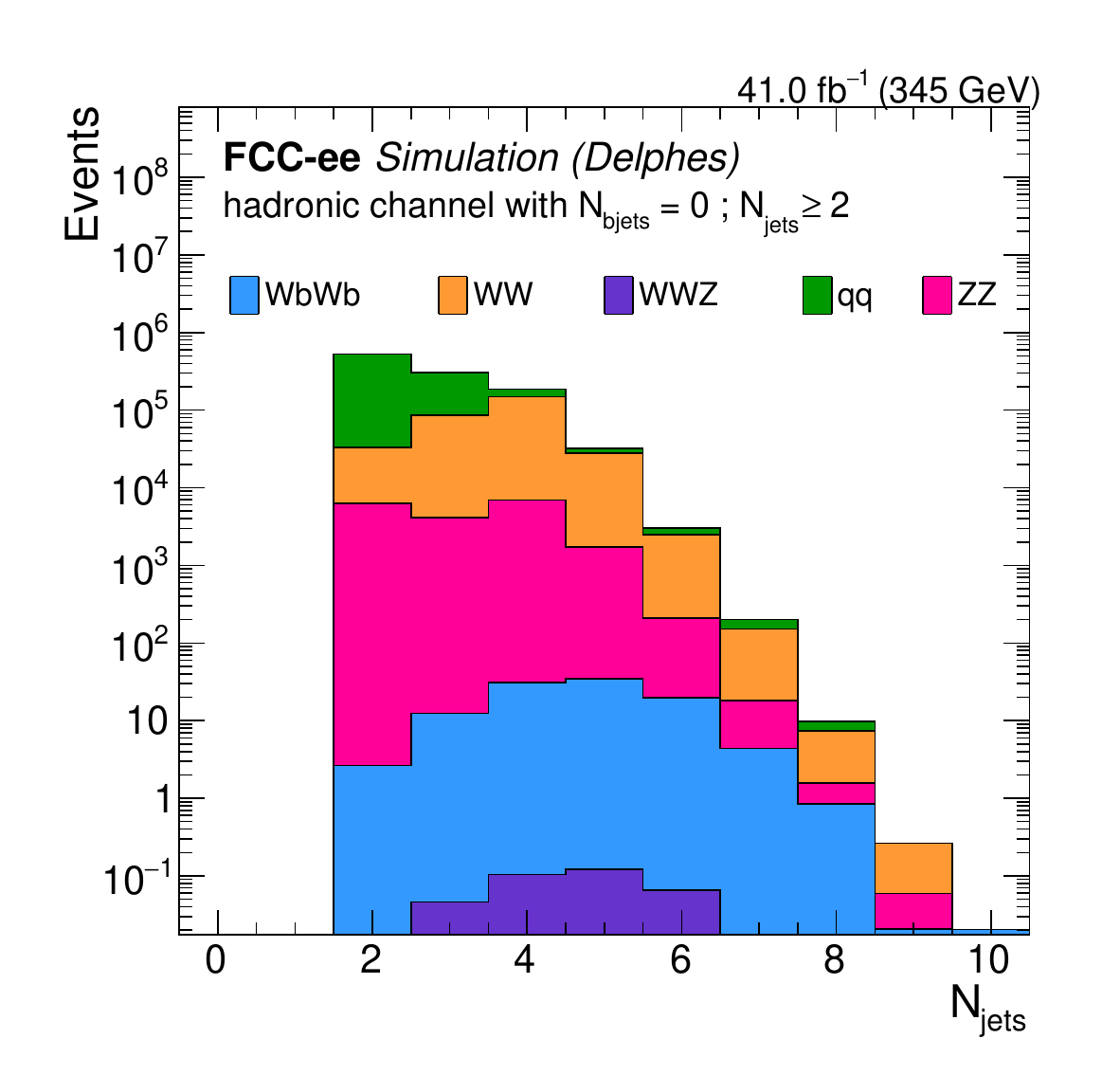}\\
    
    \vspace{-.5em}
    
    \includegraphics[width=0.495\linewidth]{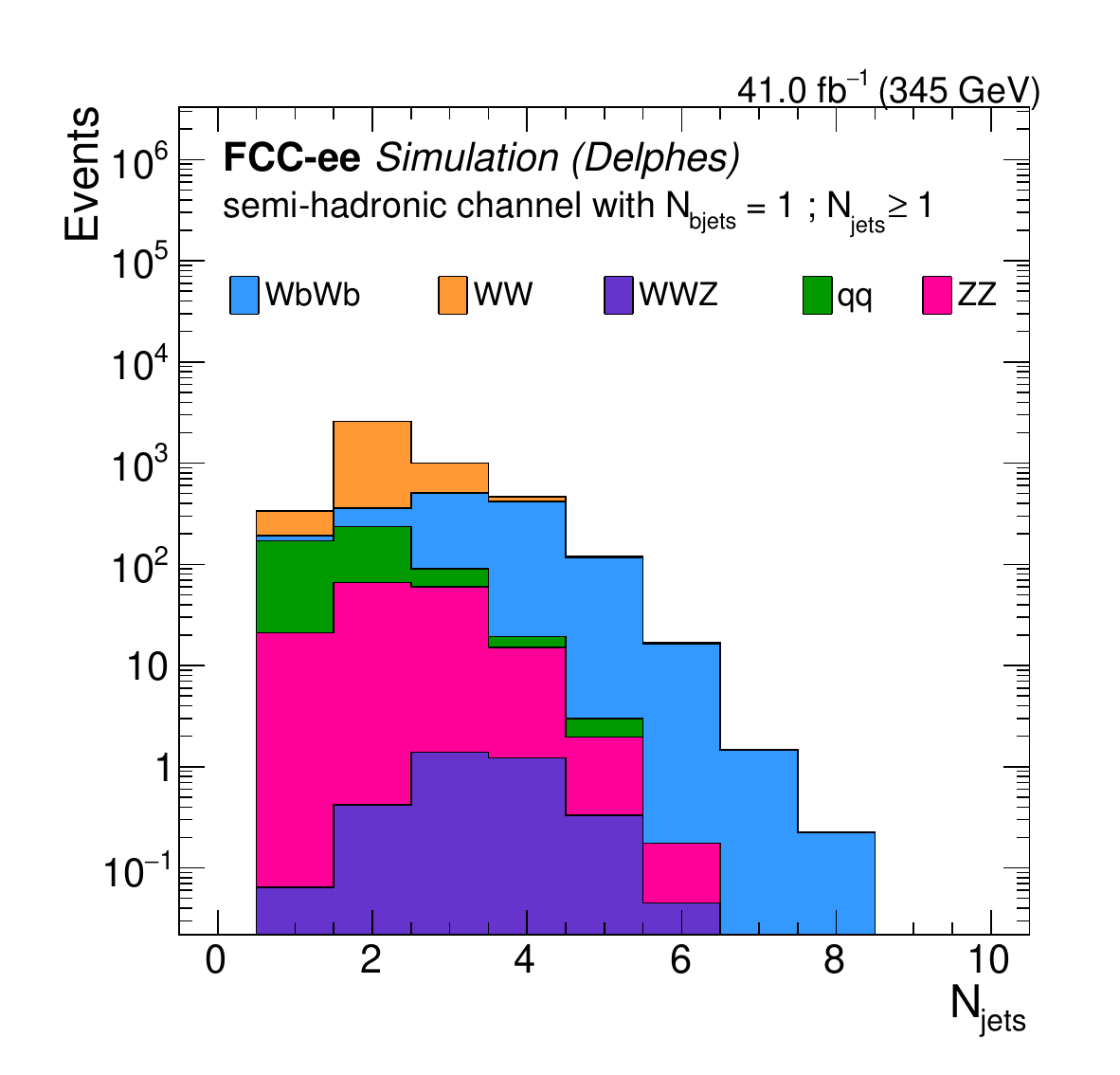}
    \includegraphics[width=0.495\linewidth]{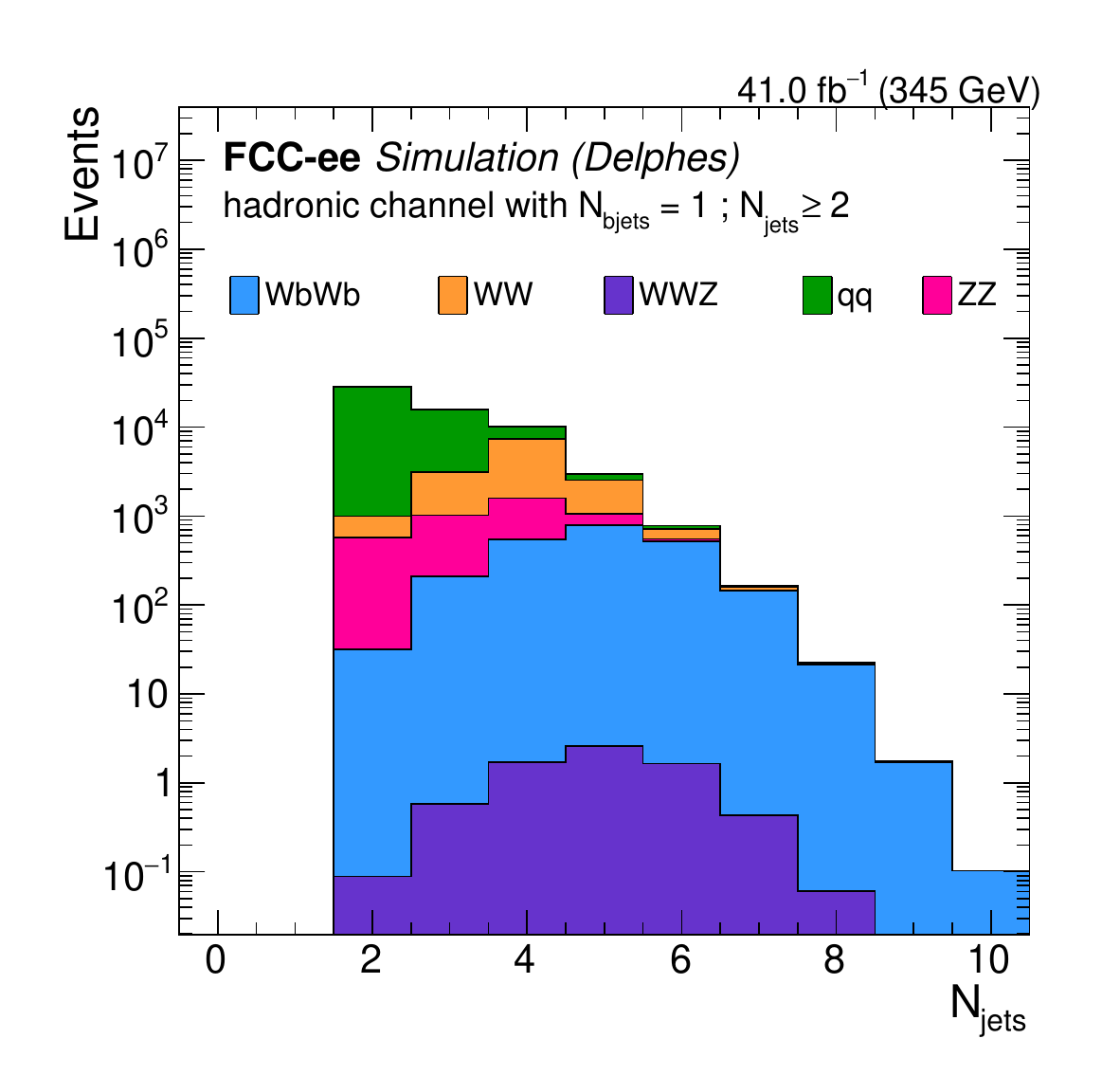}\\

    \vspace{-.5em}

    \includegraphics[width=0.495\linewidth]{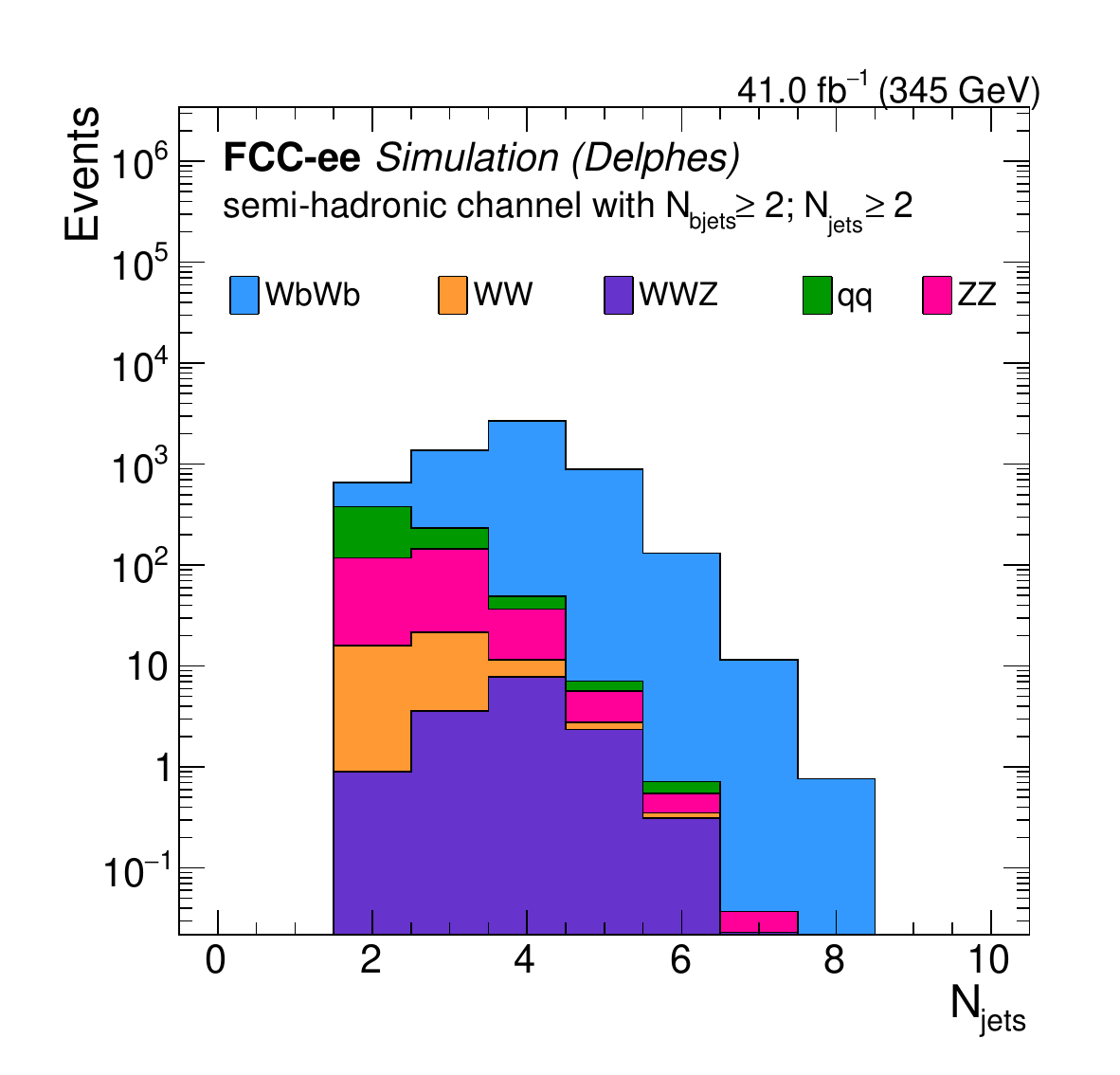}
    \includegraphics[width=0.495\linewidth]{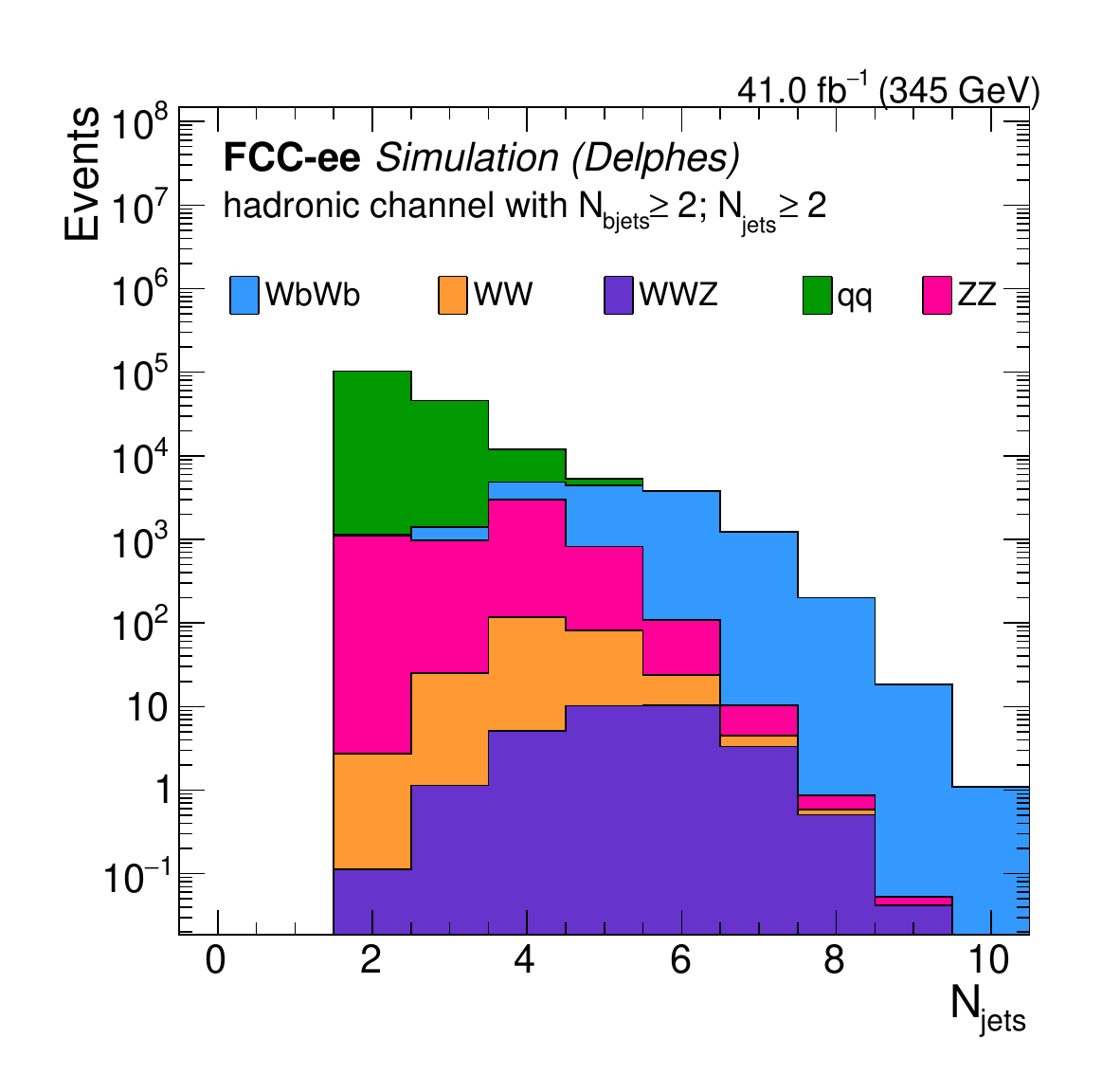} 

    \vspace{-.5em}

    \caption{Final-state distributions used in the fit of the WbWb production cross section at 345\GeV. Events with zero, one, and two or more b-tagged jets are shown in the upper, middle, and lower rows, respectively, for the semi-hadronic (left) and hadronic (right) final states. An integrated luminosity of 41\fbinv is assumed, which corresponds to the total luminosity of 410\fbinv for the \ttbar threshold scan equally split between 10 scan points.}
    \label{fig:fit-distrib-345}
    
\end{figure}

Events are classified in the corresponding decay channel based on the number of reconstructed leptons. Electrons and muons with momenta ($p$) larger than 12\GeV are considered. Taking the geometrical acceptance of the detector into account, this selection was found to have an acceptance of 99.5\% for all centre-of-mass energies.
To suppress the contamination from leptons arising from hadron decays, we also require leptons to be isolated. This is imposed by requiring that the scalar sum of the momenta of all reconstructed particles (other than the lepton itself) falling within a cone of radius 0.5 centred around the lepton does not exceed 25\% of the momentum of the lepton.
In order to exploit the difference in expected jet multiplicity between signal and background processes, an inclusive jet algorithm is applied to all reconstructed particles, excluding the selected electrons and muons. For this purpose, an inclusive generalised $k_\mathrm{T}$ algorithm is used~\cite{Catani:1991hj,Cacciari:2011ma}, which is operated in anti-$k_\mathrm{T}$ mode with a radius of 0.5 and a momentum cut-off of 5\GeV. Events with less than two (one) reconstructed jets are vetoed in the hadronic (semi-hadronic) channels. The jet multiplicity for the considered processes is shown in Figure~\ref{fig:fit-distrib-345}. The b~jet identification (b~tagging) is parametrised according to the efficiency and misidentification rate according to Refs.~\cite{TAGGINGFULLSIM_NOTE,Bedeschi:2022rnj}, which are assumed to be constant with respect to the jet kinematics. As the rest of the analysis does not make use of kinematic information of the b-tagged jets, this approximation does not affect the results. For a b~tagging efficiency of 95\%, we assume misidentification rates of 0.1\%, 3.2\%, and 2.0\% for light, charm, and gluon jets, respectively~\cite{TAGGINGFULLSIM_NOTE}. These estimates were derived using Higgs decays into two jets in ZH production, with Z bosons decaying into a pair of neutrinos. Dedicated studies towards the optimisation of jet clustering and flavour tagging in \ttbar topologies are ongoing, although we do not expect the above assumptions to affect the conclusions of our study.

To measure the total \WbWb production rate, we perform a maximum-likelihood fit to jet and b~jet multiplicities, simultaneously in the hadronic and semi-hadronic channels, using the Combine~\cite{CMS:2024onh} software. This fit strategy, together with the loose event selection requirements described above, allows the effect of the background contamination to be controlled without compromising the signal acceptance. The distributions used in the fit are shown in Figure~\ref{fig:fit-distrib-345}. Events are classified into three categories according to the number of reconstructed b~jets (\Nb). Events with $\Nb = 0$ provide a control region for WW and \qqbar backgrounds, while events with $\Nb \geq 2$ define the signal region. Events with $\Nb = 1$ are enriched in signal events with one unidentified b~jet and provide a powerful handle for an in-situ determination of the b~tagging efficiency. Finally, the jet multiplicity distribution (\Nj) is fitted in each category (with the exception of the $\Nb = 0$ category in the semi-hadronic channel, which is pure in WW events), and provides strong separation between the various processes. The fit distributions for the other centre-of-mass energies can be found in Appendix~\ref{app:plots_distribs}.

The effect of systematic uncertainties is incorporated into the fit via dedicated nuisance parameters. To account for possible differences in detector-related effects, the \qqbar, WW, and ZZ backgrounds are treated as independent processes between the hadronic and semi-hadronic channels. The normalisation of the \qqbar background in the hadronic channel and that of the WW background in both channels are treated as free-floating parameters. Furthermore, the normalisation of the ZZ background in the hadronic channel and that of the WWZ background in both channels are assigned an arbitrary but conservative normalisation uncertainty of 3\%. On the other hand, the \qqbar and ZZ backgrounds in the semi-hadronic channels, which are characterised by misidentified or (in the case of ZZ) unreconstructed leptons, are assigned a conservative normalisation uncertainty of 5\%. A luminosity uncertainty of 0.1\% is assumed for 340 and 345\GeV, which is conservatively scaled down by a factor of five (smaller than the statistical scaling) for 365\GeV. A more detailed discussion of these estimates will be given in Section~\ref{sec:threshold_scan}. Finally, a largely conservative uncertainty of 1\% is assigned to the b~tagging efficiency, with the goal of constraining this effect in situ by exploiting the b~jet multiplicity classification. Other potential sources of uncertainty, such as the dependence of the signal simulation on the value of the top quark mass and variations in the renormalisation scale in the parton shower for all processes, have been considered but were found to have no significant effect on the distributions used in the fit.

The total uncertainty in the \sigmaWbWb and its components are summarised in Table~\ref{tab:impact_WbWb} for the three considered centre-of-mass energies. In all cases, we find that the impact of systematic uncertainties can be controlled at a level well below the statistical uncertainty, allowing for effectively uncorrelated measurements at the different centre-of-mass energies. In particular, all background components, as well as the b~tagging efficiencies, are constrained in the fit at the permille level or below. The impact plots for the fits can be found in Appendix~\ref{app:plots_distribs}. In view of these considerations, we will assume in the following that the \WbWb production cross section can be determined with a total uncertainty equal to the statistical uncertainty on the total rate for any of the considered centre-of-mass energies.

\begin{table}[htbp]
    \centering
    \begin{tabular}{lccc}
        \multicolumn{1}{l}{Uncertainty source} & \multicolumn{3}{c}{Impact on $\sigmaWbWb$ [\%]} \\
        \multicolumn{1}{c}{} & 340\GeV & 345\GeV & 365\GeV \\
        \hline  
        Integrated luminosity  & 0.12 & 0.11 & 0.02 \\ 
        b tagging              & 0.11 & 0.06 & 0.01 \\ 
        ZZ had. norm.          & 0.46 & 0.19 & 0.04 \\ 
        ZZ semihad. norm.      & 0.23 & 0.07 & 0.03 \\ 
        WW had. norm.          & 0.17 & 0.09 & 0.02 \\ 
        WW semihad. norm.      & 0.06 & 0.04 & 0.03 \\ 
        \qqbar had. norm.          & 0.12 & 0.09 & 0.02 \\ 
        \qqbar semihad. norm.      & 0.18 & 0.06 & 0.01 \\ 
        WWZ norm.              & 0.03 & 0.01 & 0.01 \\ 
        \hline
        Total (incl. stat) & 2.31 & 0.89 & 0.12 \\  
    \end{tabular}
    \caption{Total uncertainty and impact of various systematic uncertainty sources on the measured \WbWb\ production cross section (\sigmaWbWb) at different centre-of-mass energies.}
    \label{tab:impact_WbWb}
\end{table}

\section{Phenomenological analysis of the \boldmath \texorpdfstring{\ttbar}{tt} threshold scan}
\label{sec:threshold_scan}

The values of the top quark mass (\mt) and total width (\Gt) can be determined by performing a fit of theoretical predictions to the measured \WbWb cross section as a function of the centre-of-mass energy. For this purpose, we calculate the \WbWb production cross section at \NNNLO in non-relativistic QCD (NR-QCD) using the \qqTh package~\cite{Beneke:2016kkb}. Electroweak corrections to the potential between the \ttbar pair are also included in the calculation, as well as off-shell effects and non-resonant contributions, as explained in Section~\ref{sec:intro}. The code also calculates the effect of initial-state radiation~(ISR) at LL accuracy. In \qqTh, the value of \mt can be set independently to that of \Gt. The values of the strong coupling constant (\as) and the top quark Yukawa coupling~(\yt) can also be varied. In this work, we consider \mt as defined in the potential subtracted~(PS) scheme~(\mtps). The relation between \mtps and other mass schemes is discussed in Section~\ref{sec:summary}. According to the prescription given in Ref.~\cite{Beneke:2024sfa}, we set the renormalisation and factorisation scales to 80\GeV and 350\GeV, respectively, while the scale in the definition of \mtps is set to 20\GeV. The \ttbar threshold lineshape, calculated with \qqTh, is then convoluted with the expected FCC-ee beam energy spread~(BES), which is approximated to be a Gaussian distribution with a standard deviation of 0.184\% per beam. This results in a spread in \sqrts of 0.13\%. Any deviations from the Gaussian approximation of the BES are only expected beyond five standard deviations from the average beam energy. This effect is therefore expected to have a negligible impact on the final result.

\begin{figure}[htbp]
    \centering
    \includegraphics[width=0.495\linewidth]{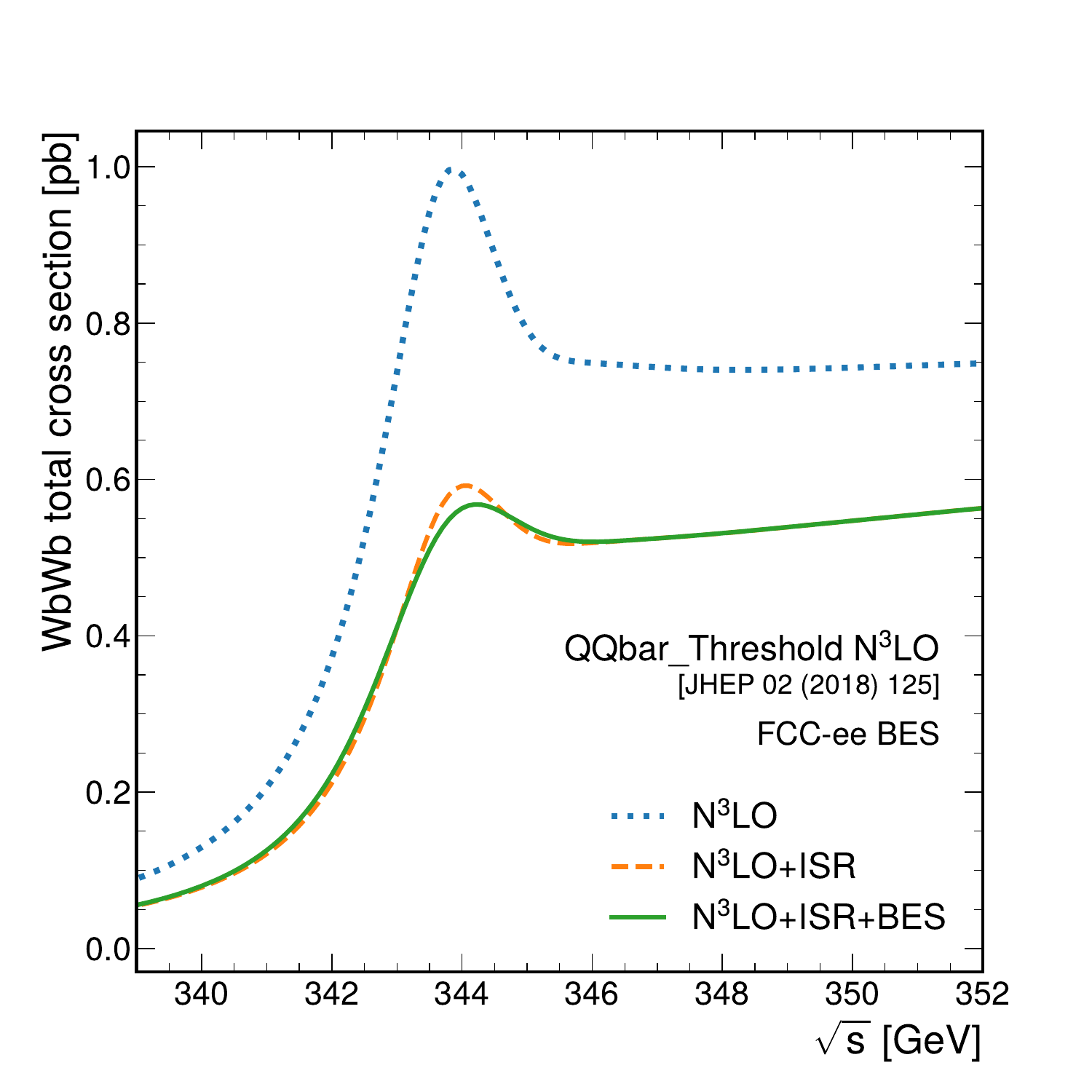}
    \includegraphics[width=0.495\linewidth]{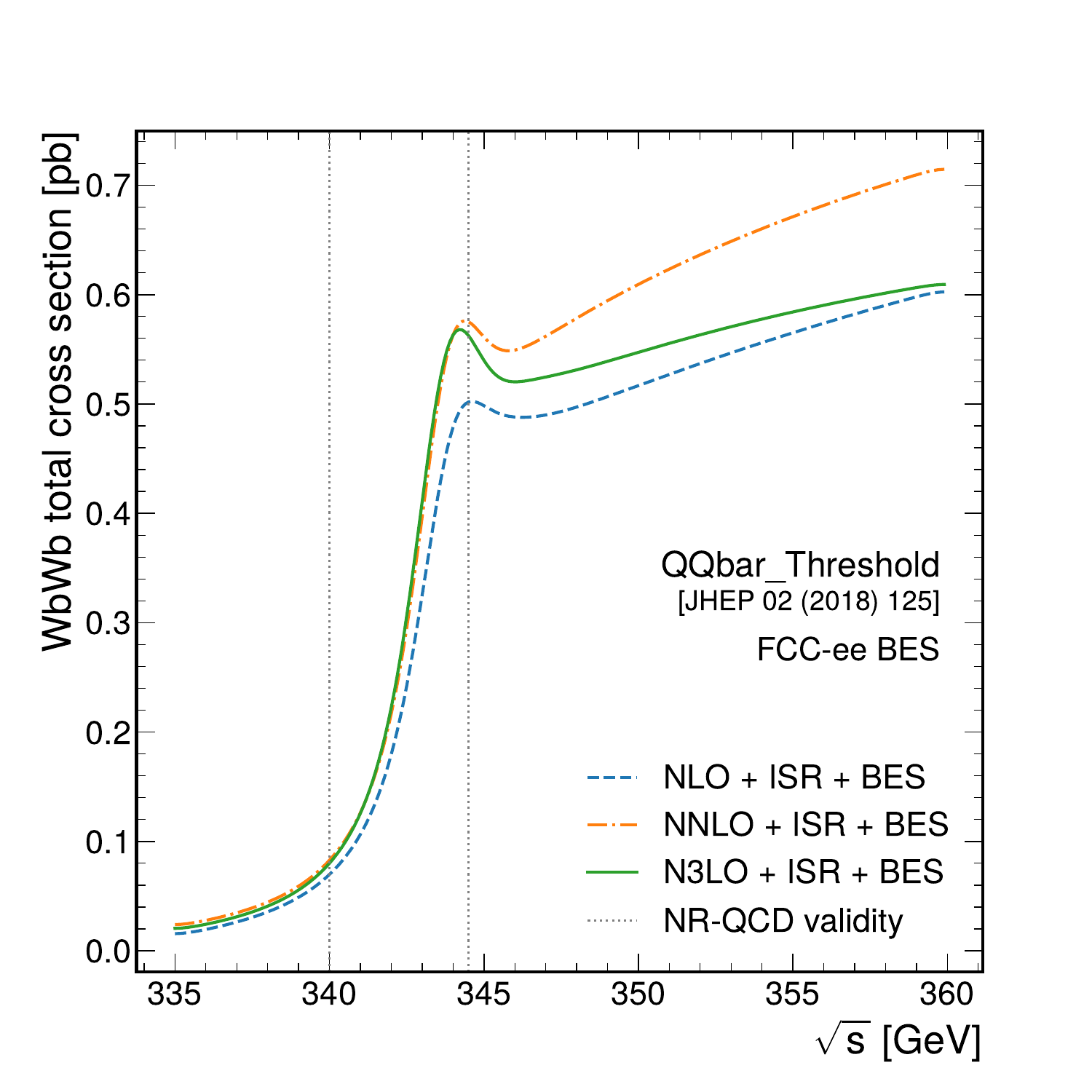}

    \caption{Left: effect of the initial state radiation (ISR) and of the FCC-ee beam energy spectrum (BES) on the \ttbar production threshold prediction at \NNNLO. Right: prediction for the \ttbar threshold scan at different orders in perturbation theory. In both cases, we assume $\mtps = 171.5\GeV$ and $\Gt = 1.33\GeV$.}
    \label{fig:theory}
    \end{figure}

For the sake of illustration, we assume $\mtps = 171.5\GeV$ and $\Gt = 1.33\GeV$ as default values for the predictions. Figure~\ref{fig:theory} (left) shows the effect of the ISR, which is responsible for a decrease in the total \WbWb production cross section, and of the BES, which smears the pseudo-bound-state enhancement at the \ttbar threshold. The perturbative convergence of the calculation is shown in Figure~\ref{fig:theory} (right). This result is used to define a range of validity of the NR-QCD calculation, between $\sqrts = 340.0$ and 344.5\GeV. We therefore focus on this range for the determination of \mt and \Gt.

\begin{figure}[htbp]
    \centering
    \includegraphics[width=0.49\linewidth]{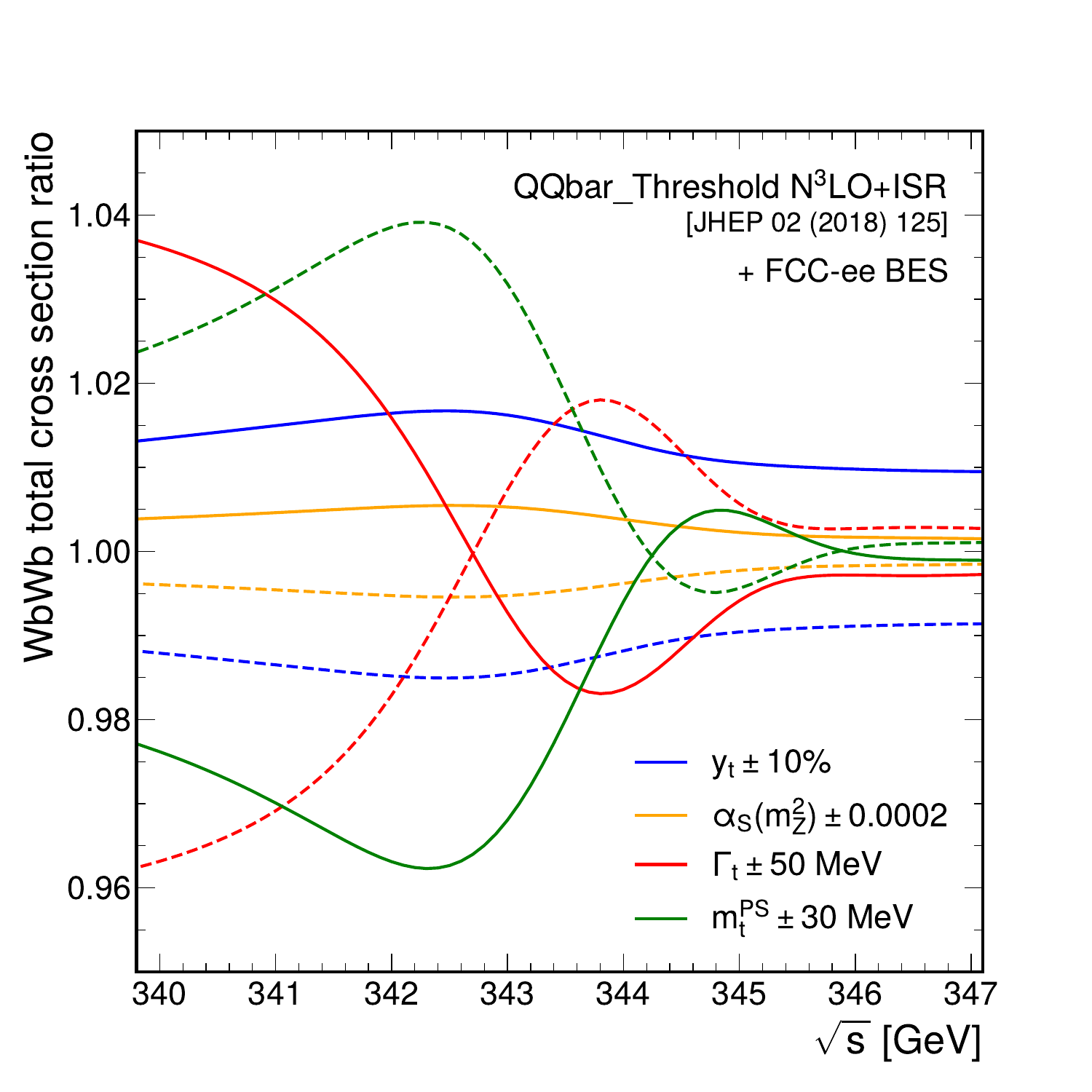}
    \includegraphics[width=0.49\linewidth]{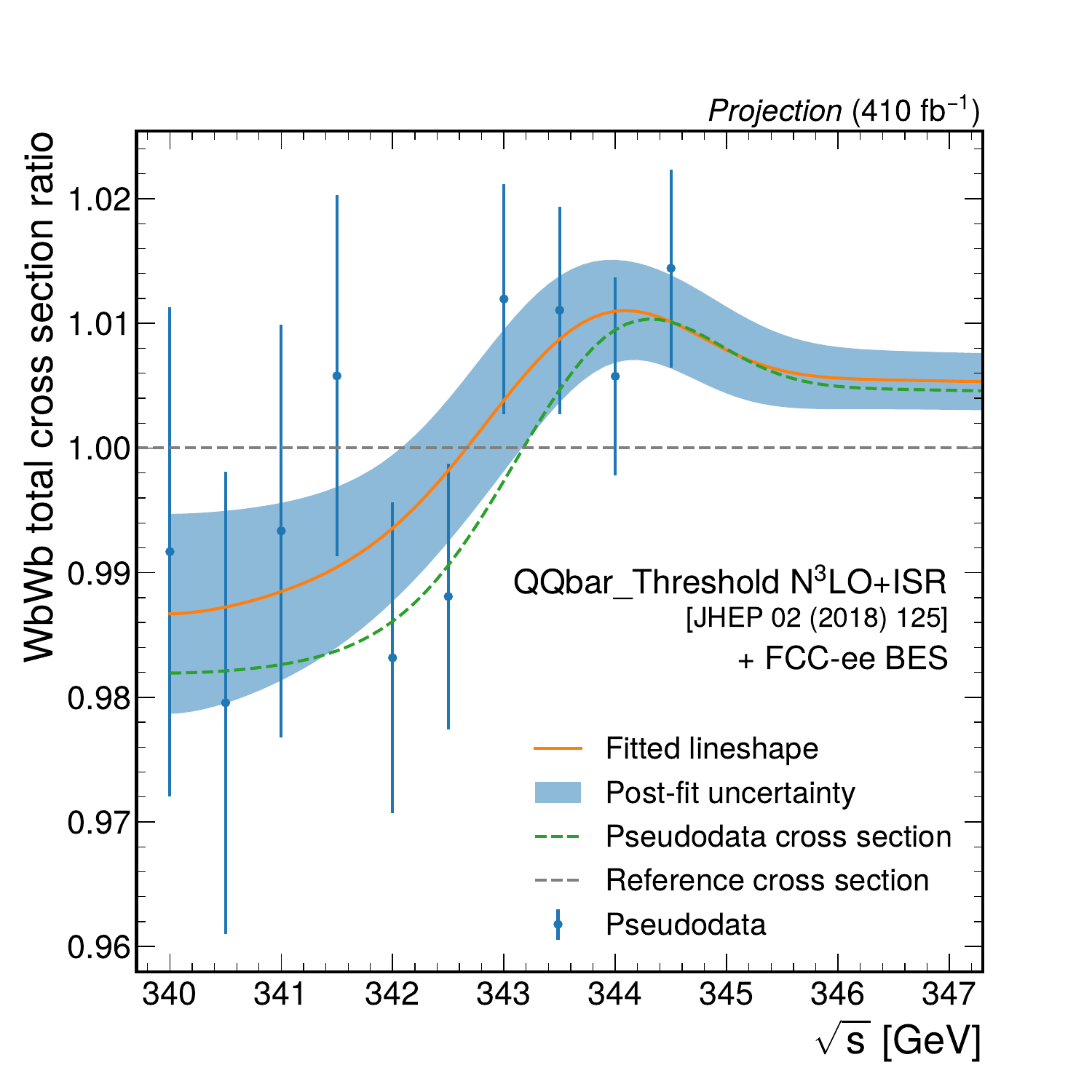}
        \caption{Left: dependence of the \WbWb production cross section on \mt, \Gt, \yt, and \as. The plot shows the ratio of the cross section calculated after varying each parameter to the one obtained using the reference values. The solid lines represent positive variations, while the dashed ones show the corresponding negative variations. Right: Fitted lineshape (solid line) as a function of the centre-of-mass energy, normalised to the reference cross section (horizontal dashed line). The fitted lineshape is derived by applying the parameters obtained in the 10-point fit to the full lineshape calculated with \qqTh, taking correlations into account.
        The fit is performed to pseudo-data (markers) generated according to the pseudo-data cross section (curved dashed line) obtained with different parameters with respect to the reference cross section. The uncertainty in the pseudo-data (vertical bars) represents the experimental uncertainty in the \WbWb cross section. The solid band corresponds to the post-fit uncertainty in the fitted cross section, all profiled uncertainties included.}
    \label{fig:threshold_scan}
\end{figure}

The dependence of the calculated cross section on the relevant parameters (\mt, \Gt, \yt, and \as) is shown in Figure~\ref{fig:threshold_scan} (left). For illustration, the value of \mt (\Gt) is varied by 30 (50)\MeV, \asmz is varied by $2 \times 10^{-4}$, and \yt is varied by 10\%. The \yt and \as variations are found to introduce very similar shape effects on the cross section. This can be explained by the fact that both parameters control the interaction vertex and the NR-QCD potential at the threshold. We therefore expect these two parameters to be largely correlated in the lineshape fit. In contrast, variations in \mt and \Gt introduce significantly different shape distortions to the \ttbar spectrum, as the values of \mt and \Gt control the position and width of the peak, respectively. It is therefore possible to simultaneously determine these two parameters from the \ttbar threshold scan, without assuming any SM relation between the two. This allows for a more model-independent measurement of \Gt, therefore enhancing the sensitivity to any contribution beyond the SM (BSM) in the top quark decay. An alternative approach in which the SM relation between \mt and \Gt is assumed was also implemented and is described later in the text. It should also be noted that the highest sensitivity to \mt and \Gt lies within the range of validity of the NR-QCD calculation (Figure~\ref{fig:theory}). Finally, variations in the weak mixing angle within the expected precision at FCC-ee were also considered but were found to have a negligible effect.

For the fit of the \ttbar threshold lineshape, we choose a baseline scenario with 10 equally-spaced points between $\sqrts = 340.0$ and 344.5\GeV, corresponding to the range of validity of the NR-QCD calculation (Figure~\ref{fig:theory}). This results in a spacing of 0.5\GeV between the different centre-of-mass energies, corresponding to about one standard deviation of the BES. We further assume that the total integrated luminosity planned for the \ttbar threshold run is split equally between the different scan points, resulting in 41\fbinv per scan point. While it was demonstrated that the overall precision can be maximised by choosing a smaller number of scan points~\cite{Li:2022iav}, this approach would not allow testing the validity of the theoretical prediction and may undermine the robustness of the result. We therefore choose a scenario that allows for stringent tests of the theoretical calculation, without significantly compromising on the achievable precision in the parameters of interest. Alternative scenarios with a different number of scan points, spacing between the points, or distribution of the integrated luminosity between scan points were tested. However, none of these scenarios showed a significant improvement over the baseline scenario.

The fit of the \ttbar lineshape is performed by maximising a \chisq function.
The dependence of the predicted cross sections on \mt, \Gt, \yt, and \as, as well as on the BES, and beam energy calibration (BEC) is incorporated in the \chisq in the form of nuisance parameters. The effect of these parameters is evaluated using discrete two-point variations, which are then interpolated linearly.
A covariance matrix is built using the uncertainty in the integrated luminosity and the statistical uncertainty of the measured cross sections. As shown in Section~\ref{sec:reco_fit}, the measured values of \sigmaWbWb at the different scan points are limited by statistical uncertainties and can therefore be considered statistically independent. For each scan point $i$, the statistical uncertainty in the measured cross section ($\sigma_i$) is estimated as $\sqrt{\sigma_i / \mathcal{L}_i}$, where $\mathcal{L}_i$ is the assumed integrated luminosity for that scan point. In order to approximately match the total uncertainties of Table~\ref{tab:impact_WbWb}, we inflate this uncertainty by a factor of 1.2 for all centre-of-mass energies. This is necessary in order to take into account the effect of the background contamination. The values of \mt and \Gt are treated as free-floating parameters in the fit, while the other parameters are constrained using Gaussian priors representing external constraints. Further details on the statistical model can be found in Appendix~\ref{app:fit_model}. 

In the fit, the value of \asmz is constrained within an uncertainty of $10^{-4}$ according to FCC-ee projections~\cite{dEnterria:2020cpv}, while the value of \yt is constrained within 3\% according to the expected precision at HL-LHC~\cite{Azzi:2019yne} combined with the absolute measurement of other Higgs couplings at FCC-ee. 
At the energy of the \ttbar production threshold, significantly larger energy deposits in the luminosity calorimeters are expected compared to lower energy runs, , and the luminosity calibration via small-angle Bahaba scattering remains to be demonstrated in this regime. We therefore assume that we can measure the integrated luminosity using central di-photon events. A conservative back-of-the-envelope estimate results in a luminosity uncertainty of 0.1\% per scan point. We also assume that a beam energy calibration uncertainty of 5\MeV can be achieved using WW events using the \mW constraint from the WW threshold scan~\cite{Blondel:2019jmp,Beguin:2710098}. Finally, we estimate a BES uncertainty of 1\% per scan point from di-muon events based on the results of Ref.~\cite{Blondel:2019jmp}. The longitudinal boost of the centre-of-mass frame and the crossing angle between the beams can also be measured precisely using di-muon events~\cite{Blondel:2019jmp}. This effects, which mainly affects the detector acceptance, is however subleading and is not considered in this study.

We expect the above estimates of beam-related uncertainties to be limited by statistical uncertainties, and therefore to be uncorrelated between the different centre-of-mass energies. However, we also evaluate the effect of any correlated component among the different scan points, \eg arising from common systematic effects. For the sake of illustration, we assume the size of the correlated component to be half of the corresponding uncorrelated component. While this choice is entirely arbitrary, the effect of correlated components was neglected in previous studies. In addition, in Section~\ref{sec:param_scans} we present a detailed evaluation of the impact of the various systematic components as a function of the assumptions on the input uncertainties. This would allow the results presented in this work to be updated following any refinement in the estimate of the systematic uncertainties.

Figure~\ref{fig:threshold_scan} (right) shows the fitted lineshape of the \ttbar production threshold. For illustration, pseudo-data are generated according to the statistical uncertainty of the measured cross sections. In the pseudo-data cross section, the central values of \mt and \Gt are varied by $+10$ and $-20\MeV$, respectively, compared to the reference values. The fitted lineshape is then derived by applying the parameters obtained in the 10-point fit to the full lineshape calculated with \qqTh, taking correlations into account.
Good closure is observed between the fitted and the predicted cross section, within the considered uncertainties. This plot validates the closure of the fit with respect to arbitrary choices of \mt and \Gt. The rest of the results presented in this section, instead, are obtained using an Asimov fit~\cite{Cowan:2010js}.

\begin{table}[htpb]
    \centering
\begin{tabular}{l|c|c|l}
Uncertainty source  & \mtps [\MeV{}] & \Gt [\MeV{}] & Input values \\ \hline
Experimental (stat. $\times 1.2$)         & 4.3 & 10.4 & $L = 410\fbinv$ (FCC-ee) \\
Parametric \yt &  4.2   & 3.6    &  $\delta y_t = 3\%$ \\
Parametric \as &  2.2   & 1.7    &  $\delta \asmz = 10^{-4}$ \\ \hline
Luminosity calibration (uncorr.) &  0.5 & 1.0  & $\delta L/L = 0.1\%$ \\
Luminosity calibration (corr.) &  0.4 & 0.4  & $\delta L/L = 0.05\%$ \\
Beam energy calibration (uncorr.) & 1.2 &  1.8  & $ \delta \sqrt{s} = 5\MeV$ \cite{Blondel:2019jmp,Beguin:2710098} \\
Beam energy calibration (corr.) & 1.2 &  0.1  & $ \delta \sqrt{s} = 2.5\MeV$ \\
Beam energy spread (uncorr.) &  0.3 & 0.8  & $\delta \Delta E =  1\%$ \cite{Blondel:2019jmp} \\
Beam energy spread (corr.) &  0.1 & 1.1  & $\delta \Delta E = 0.5\%$ \\
\hline
Total profiled & 6.8 & 11.5 & \\
Theory, unprofiled (scale)       & 35 & 25 & $\mathrm{N^3LO}$ NR-QCD~\cite{Beneke:2016kkb} \\ 
\end{tabular}
     \caption{Impact of the various sources of systematic uncertainties to the total uncertainty in \mt and \Gt. The impacts are estimated as the difference in quadrature between the total uncertainty and an alternative fit in which the corresponding nuisance parameters are removed. The uncertainty from the theory prediction, instead, is estimated separately.}
     \label{tab:breakdown_2D}
\end{table}

\begin{figure}[htpb]
    \centering
     \includegraphics[width=0.49\linewidth]{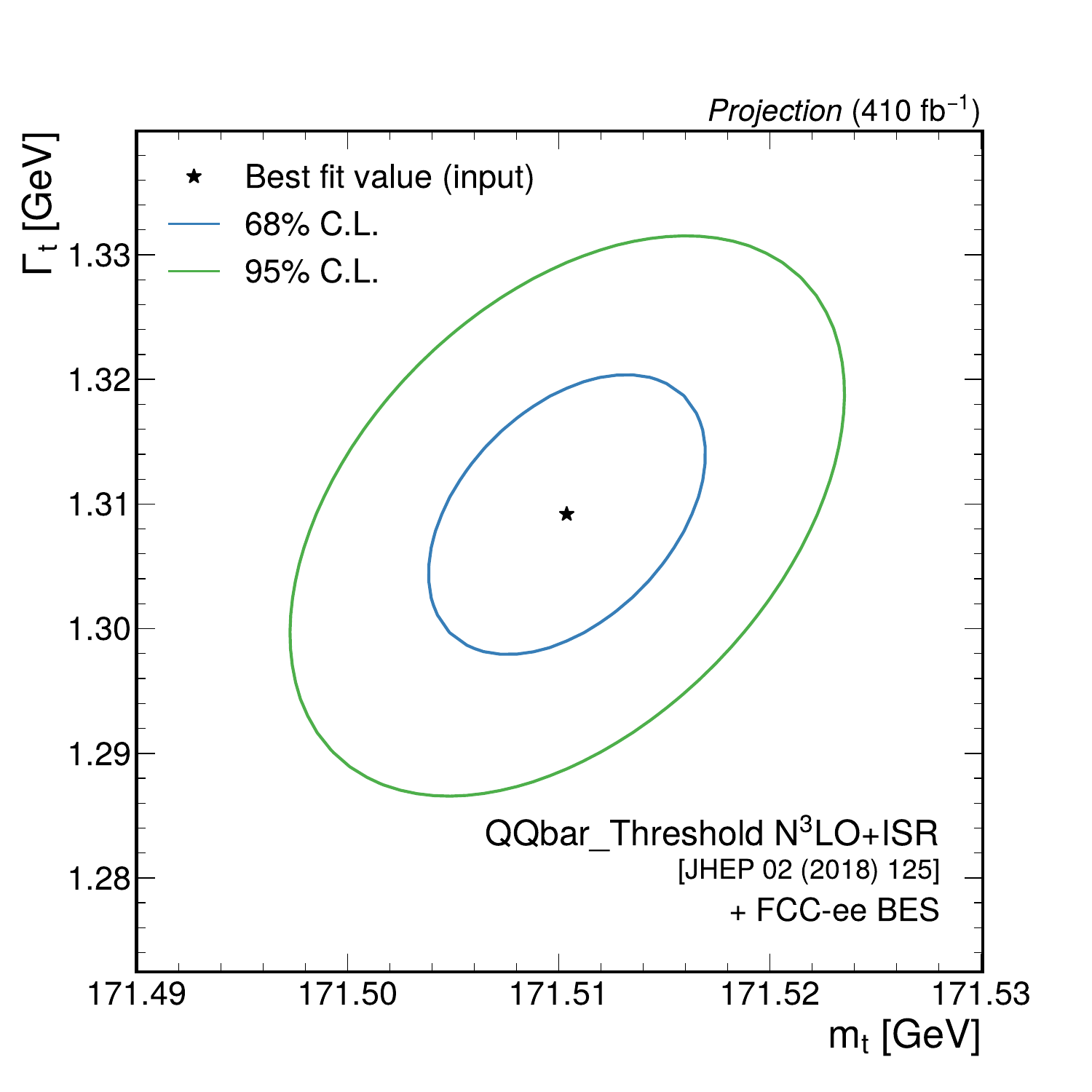}
    \includegraphics[width=0.49\linewidth]{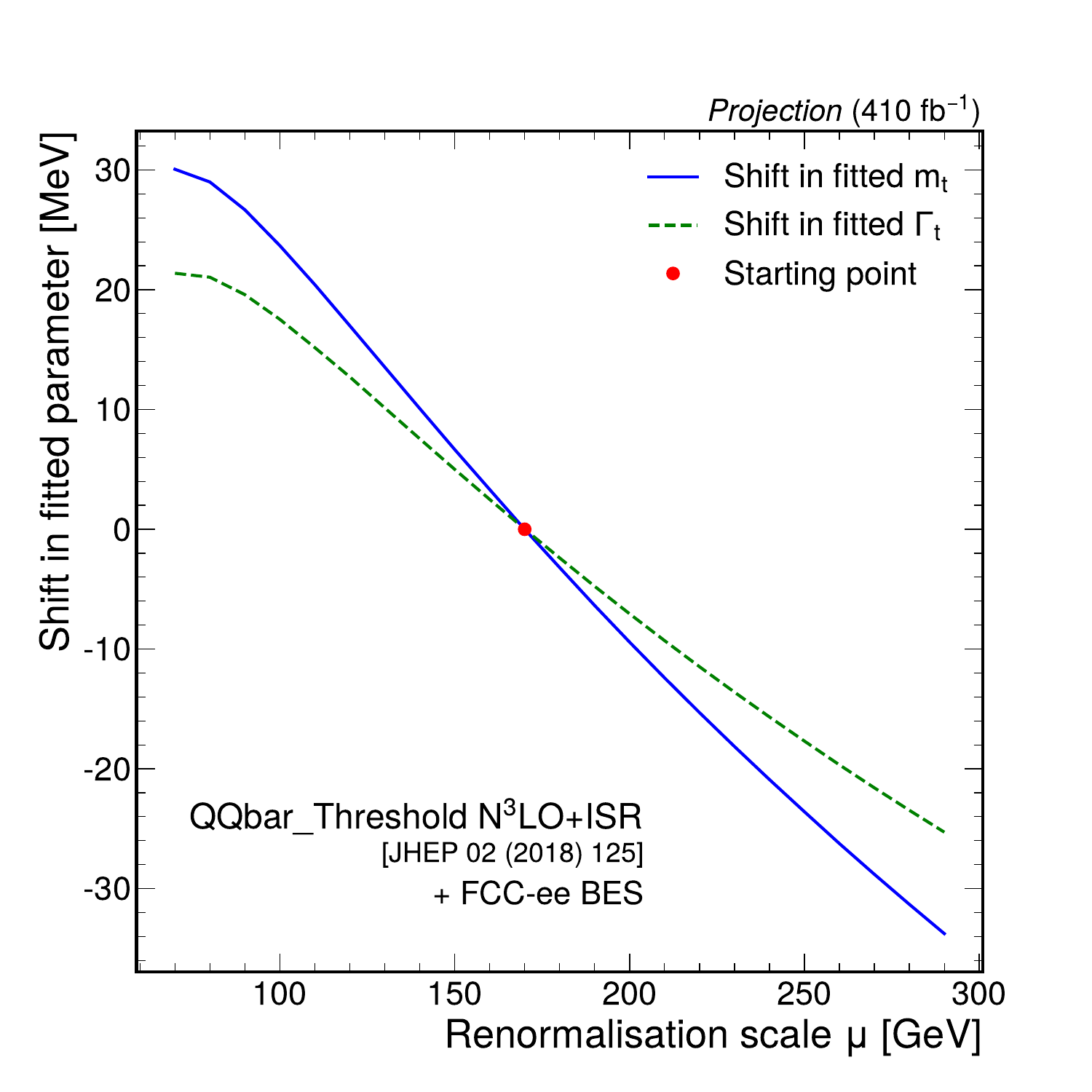}
        \caption{Left: two-dimensional confidence intervals in \mt and \Gt, corresponding to 68\% (inner ellipse) and 95\% (outer ellipse) confidence level (C.L.). Right: shift in fitted \mt and \Gt as a function of the choice of the renormalisation scale in the \NNNLO calculation, with respect to a reference value (starting point) of 170\GeV.}
    \label{fig:scan_theory}
\end{figure}

The results of the Asimov fit are shown in Table~\ref{tab:breakdown_2D}. We find that \mt and \Gt can be determined with a total uncertainty of 6.8 and 11.5\MeV, respectively, taking into account all experimental, parametric, and machine-related uncertainties. The experimental uncertainty in the \WbWb cross section measurements is the leading component for both \mt and \Gt, and it largely dominates the uncertainty in \Gt. The uncertainty in \yt is found to have a large impact on the measurement of both parameters, and in particular of \mt. A correlation of 42\% between \mt and \Gt is obtained after the fit, as shown in Figure~\ref{fig:scan_theory} (left). If all systematic uncertainties are neglected, the statistical correlation decreases to 25\%. We then investigate the impact of theoretical uncertainties by varying the renormalisation scale in the calculation by a factor of two in both directions. Here, we set the central value of the renormalisation scale to 170\GeV. The result is shown in Figure~\ref{fig:scan_theory} (right). A shift in \mt and \Gt of about 35 and 25\MeV, respectively, is observed. We therefore take these numbers as a proxy for the theoretical uncertainty in the measurement, as reported in Table~\ref{tab:breakdown_2D}. We also considered variations in the factorisation scale between resonant and non-resonant effects in the calculation~\cite{Beneke:2016kkb}, and the impact was found to be negligible. While improvements in the theoretical calculations are an active area of development~\cite{Beneke:2024sfa}, the impact from missing higher-order corrections is found to be the limiting factor with the current state-of-the-art theoretical tools. We also find our estimate of the theoretical uncertainty is similar to the one obtained in Ref.~\cite{Simon:2016pwp}, despite a different procedure being followed.

Finally, we consider an alternative version of the fit where the SM relation between \mt and \Gt is assumed within a theoretical uncertainty of 5\MeV, according to Ref.~\cite{Chen:2022wit}. Here, the \yt modifier is set to unity according to the SM expectation. With this approach, a total profiled uncertainty of 5.1\MeV on \mt is obtained. If only the SM relation between \mt and \Gt is assumed, while \yt is left unconstrained, a total profiled uncertainty of 6.2\MeV is obtained. While these numbers marginally improve upon the results shown in Table~\ref{tab:breakdown_2D}, we consider this approach less robust due to the significant loss of flexibility with respect to BSM effects in \Gt and \yt. We finally note that calculations of \Gt at \NNNLO in QCD obtained with the PMC method are also available~\cite{Yan:2024hbz}, but are not used in this study.

The \mt results in Tab.~\ref{tab:breakdown_2D} correspond to the potential-subtracted mass, a scheme designed for calculations in the threshold region. To obtain the mass value in more general schemes, such as the \msbar mass, the result must be converted. The relations between mass schemes for heavy quarks are known to four loops~\cite{Marquard:2015qpa}. The conversion of the threshold mass to the \msbar scheme introduces an uncertainty due to missing higher orders of 7--23\MeV, depending on the threshold mass scheme (where the smallest number corresponds to the so-called ``1S'' scheme and the largest number to the PS scheme). The parametric uncertainty in the conversion from \as, instead, is estimated to be about 7--8\MeV assuming a projected uncertainty in \asmz of $10^{-4}$, depending on the chosen mass scheme. As a consistency check we also performed a simplified version of the fit in the 1S scheme, excluding the (subleading) BES and BEC uncertainties. The fit yields an uncertainty on \mtones of about 5.9\MeV, compared to 6.6\MeV for the equivalent fit of \mtps, while the uncertianty in the other parameters remains unchanged. The effect of the BES and BEC uncertainties is expected to be identical in the two cases, as they are independent of the details of the theory calculation. In this fit we assumed $\mtones = \mtps + 0.4\GeV$ in order to match the position of the cross section peak in the two schemes. The smaller uncertainty in \mtones compared to \mtps was identified in the reduced post-fit correlation between \mtones and the other input parameters (\Gt, \as, and \yt) compared to \mtps. No difference was instead observed between the two schemes in terms of  impact of the theoretical uncertainties on \mt and \Gt. While the detailed study of these aspects is beyond the scope of this work, this test demonstrates the robustness of the presented results against the choice of the \mt renormalisation scheme.

\section{Dependence of the results on the assumptions on systematic uncertainties}
\label{sec:param_scans}

The results presented in Section~\ref{sec:threshold_scan}, and in particular in Table~\ref{tab:breakdown_2D}, rely on assumptions on the input values for the various sources of systematic uncertainties. While these are based on reasonable estimates reflecting our current expectations, it is important to investigate the dependence of the results on these assumptions. For this reason, in this section we present scans for all the profiled uncertainties in Table~\ref{tab:breakdown_2D}, within reasonable but conservative ranges. Additionally, we estimate the impact of the beam-related uncertainties separately for the uncorrelated and the correlated components. With the results presented in this section, the numbers in Table~\ref{tab:breakdown_2D} can be rescaled to reflect any refinement in the estimate of the input uncertainties. 

\begin{figure}[htpb]
    \centering
     \includegraphics[width=0.49\linewidth]{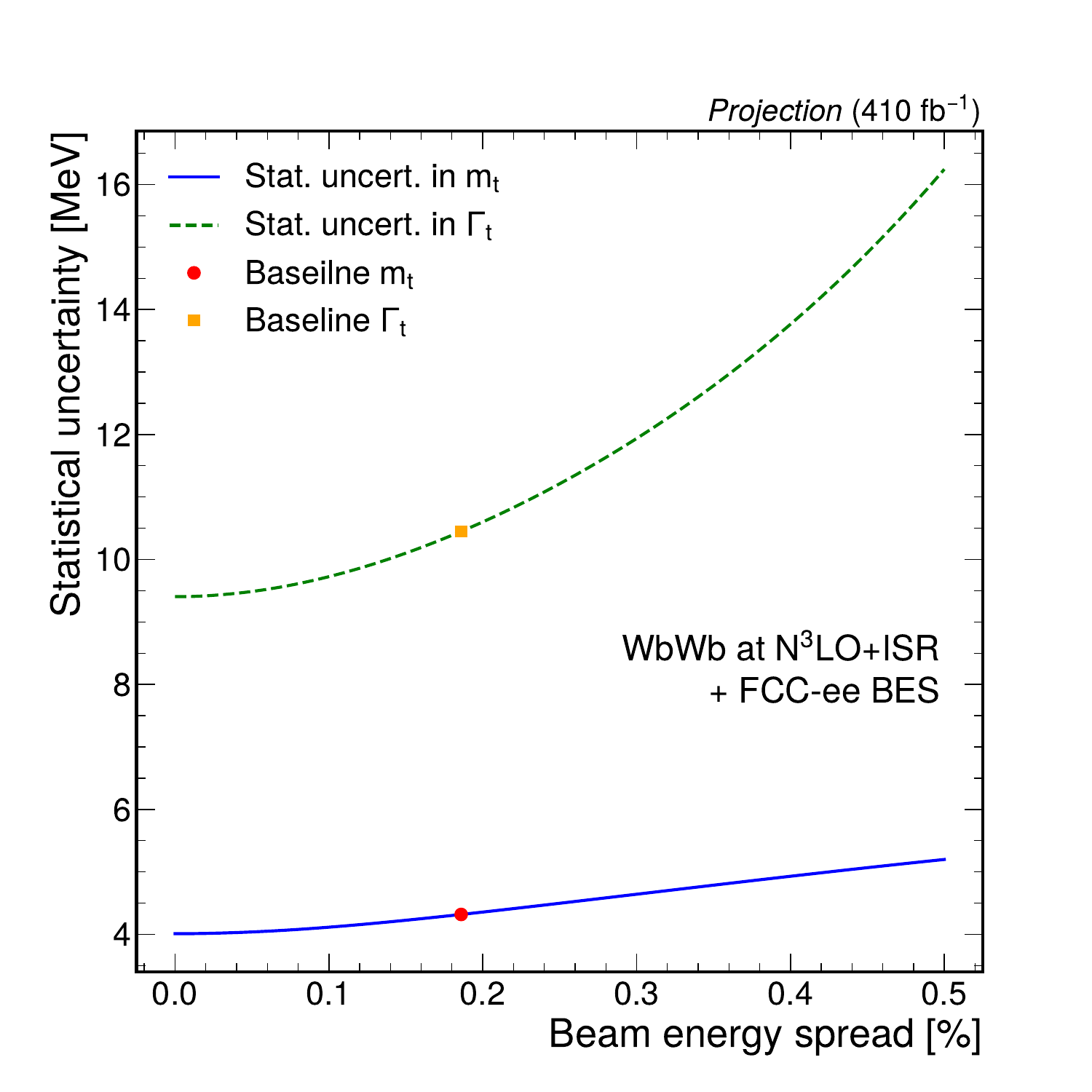}
    \includegraphics[width=0.49\linewidth]{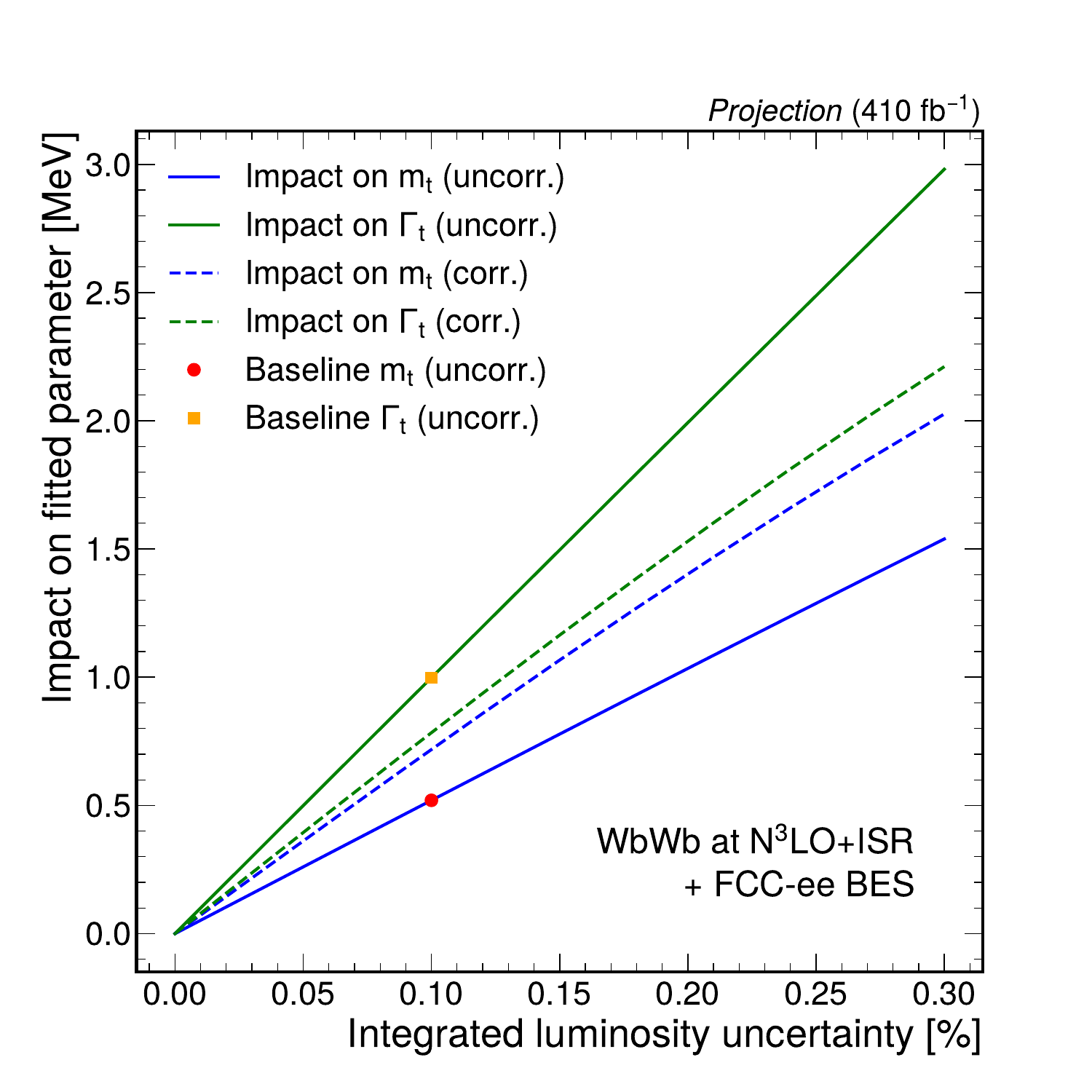}
        \caption{Left: statistical uncertainty in \mt (solid line) and \Gt (dashed line) as a function of the spread in the energy of the FCC-ee beams. The markers represent the baseline value of 0.23\%. Right: Impact on \mt and \Gt of the uncorrelated (solid lines) and correlated (dashed lines) components of the uncertainty in the integrated luminosity. The markers represent the baseline assumption for the uncorrelated component of 0.1\%.}
    \label{fig:statVsBES_lumi}
\end{figure}

Furthermore, we provide an estimate of the statistical uncertainty in \mt and \Gt as a function of the assumed central value for the FCC-ee BES. The result, shown in Figure~\ref{fig:statVsBES_lumi} (left), demonstrates that it is crucial to control the spread in the beam energy to maximise the precision on \Gt. The effect on \mt, instead, is found to be smaller. This can be explained by the fact that \Gt is directly related to the width of the \ttbar threshold lineshape, which can be smeared by the beam spread, while the sensitivity to \mt is driven by the position of the \ttbar threshold in \sqrts, which mostly depends on the precise calibration of the beam energy.

The impact of the uncertainty in the integrated luminosity on \mt and \Gt is shown in Figure~\ref{fig:statVsBES_lumi} (right). This is estimated by repeating the fit with different values for the input luminosity uncertainty, independently for the point-to-point correlated and uncorrelated components. The impacts are then estimated as explained in Section~\ref{sec:threshold_scan}, and the markers labelled as ``baseline'' correspond to the estimates for the uncorrelated components in Table~\ref{tab:breakdown_2D}. The impact of the integrated luminosity is found to be larger in absolute terms for \Gt compared to \mt, while the relative contribution to the total uncertainty is comparable, as expected. The effect of the correlated component, instead, is of similar size in absolute terms for the two parameters.

\begin{figure}[htpb]
    \centering
     \includegraphics[width=0.49\linewidth]{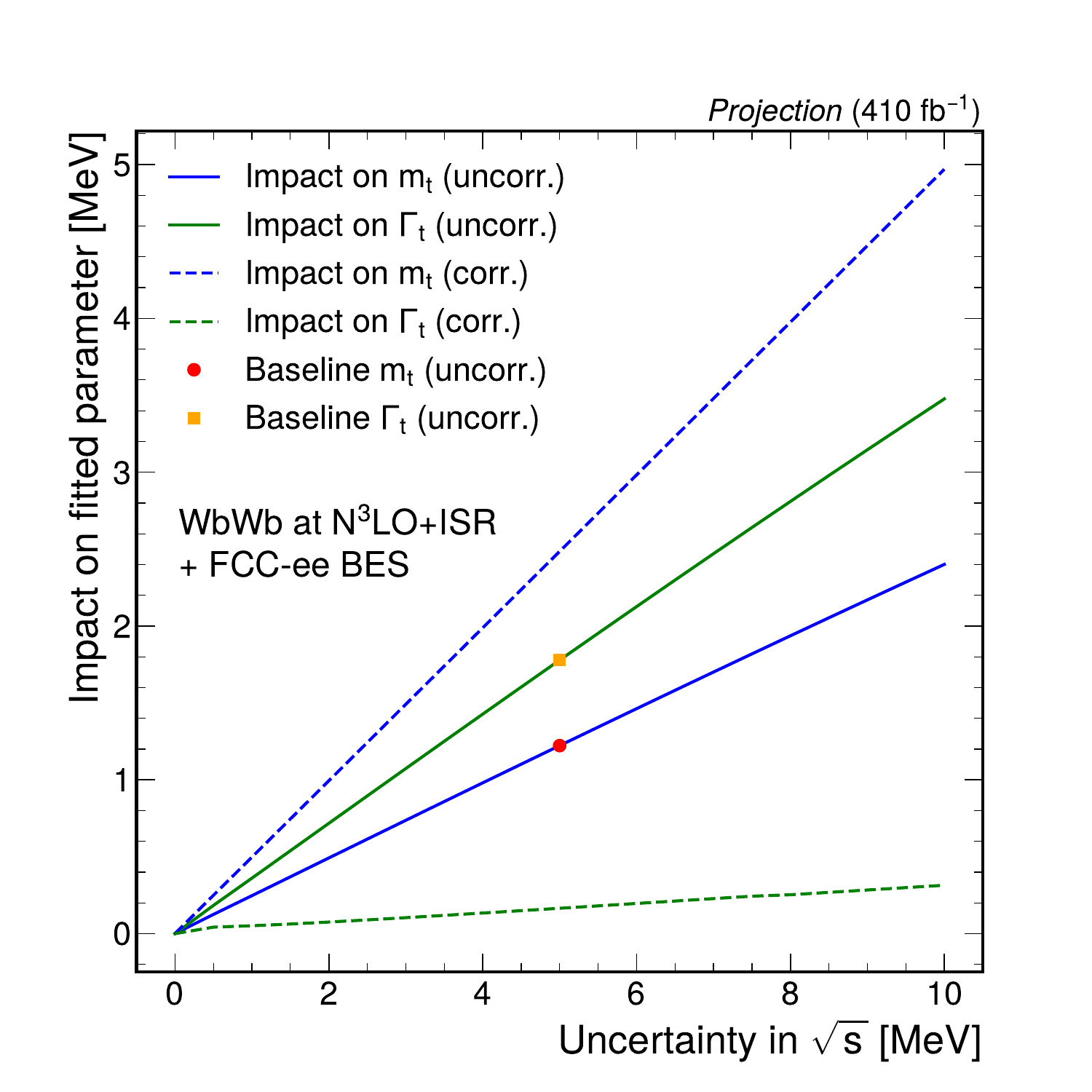}
    \includegraphics[width=0.49\linewidth]{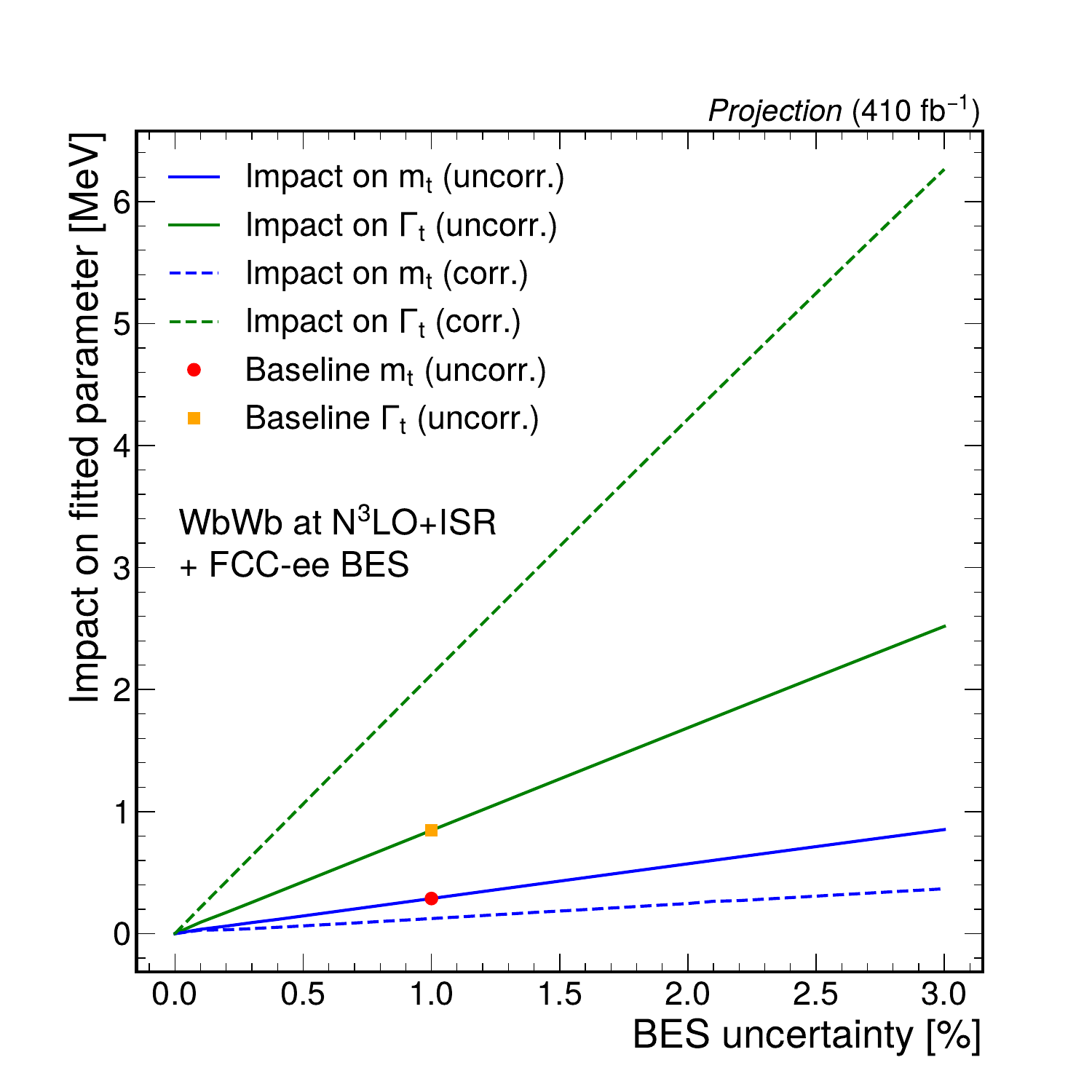}
        \caption{Impact of the uncertainties in the BEC (left) and BES (right) calibrations on \mt and \Gt. The solid and dashed lines correspond to the uncorrelated and correlated components, respectively. The markers represent the baseline values for the BEC and BES uncertainties of 5\MeV and 1\%, respectively.}
    \label{fig:uncertainties_machine}
\end{figure}

The impact of the BEC and BES uncertainties is shown in Figure~\ref{fig:uncertainties_machine}. From the left plot it can be seen that any correlated BEC uncertainty, which would cause an overall shift in the spectrum, translates into a corresponding shift in the measured value of \mt by $\delta \sqrts / 2$, as expected by the fact that the peak of the \ttbar threshold scan lies at around $2 \mt$. On the contrary, a similar shift would have nearly no effect on \Gt, as it can be fully reabsorbed by the \mt parameter in the fit. The uncorrelated component, instead, is found to have a more (less) sizeable effect on \Gt (\mt). The scan of the BES uncertainty, instead, is shown in  Figure~\ref{fig:uncertainties_machine} (right). The result shows that the impact on \Gt is significantly larger than that on \mt, for both the correlated and uncorrelated components, as expected. We also find that any residual correlated component would have a strong impact on \Gt, as it would directly impact the width of the \ttbar threshold lineshape.

\begin{figure}[htpb]
    \centering
        \includegraphics[width=0.49\linewidth]{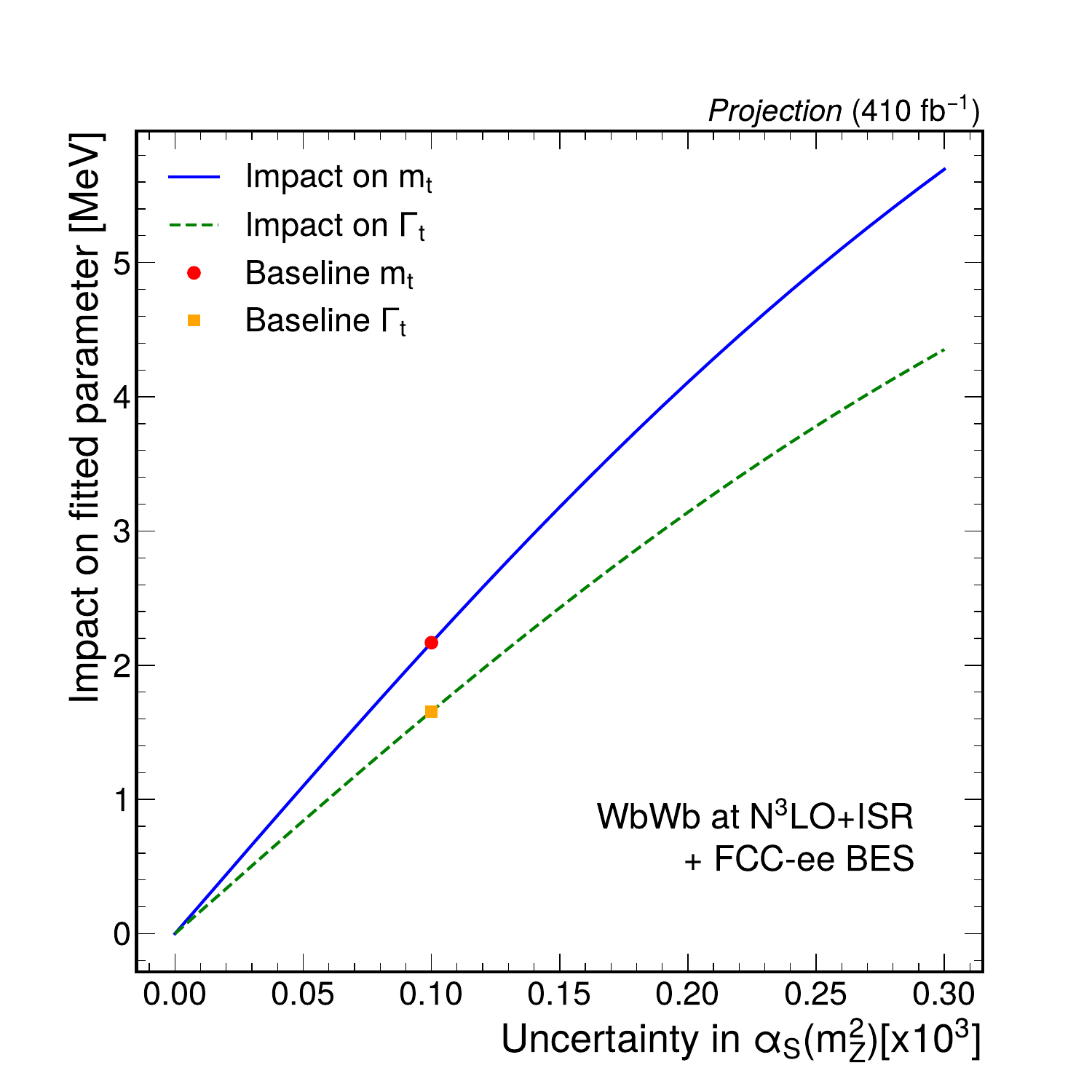}
         \includegraphics[width=0.49\linewidth]{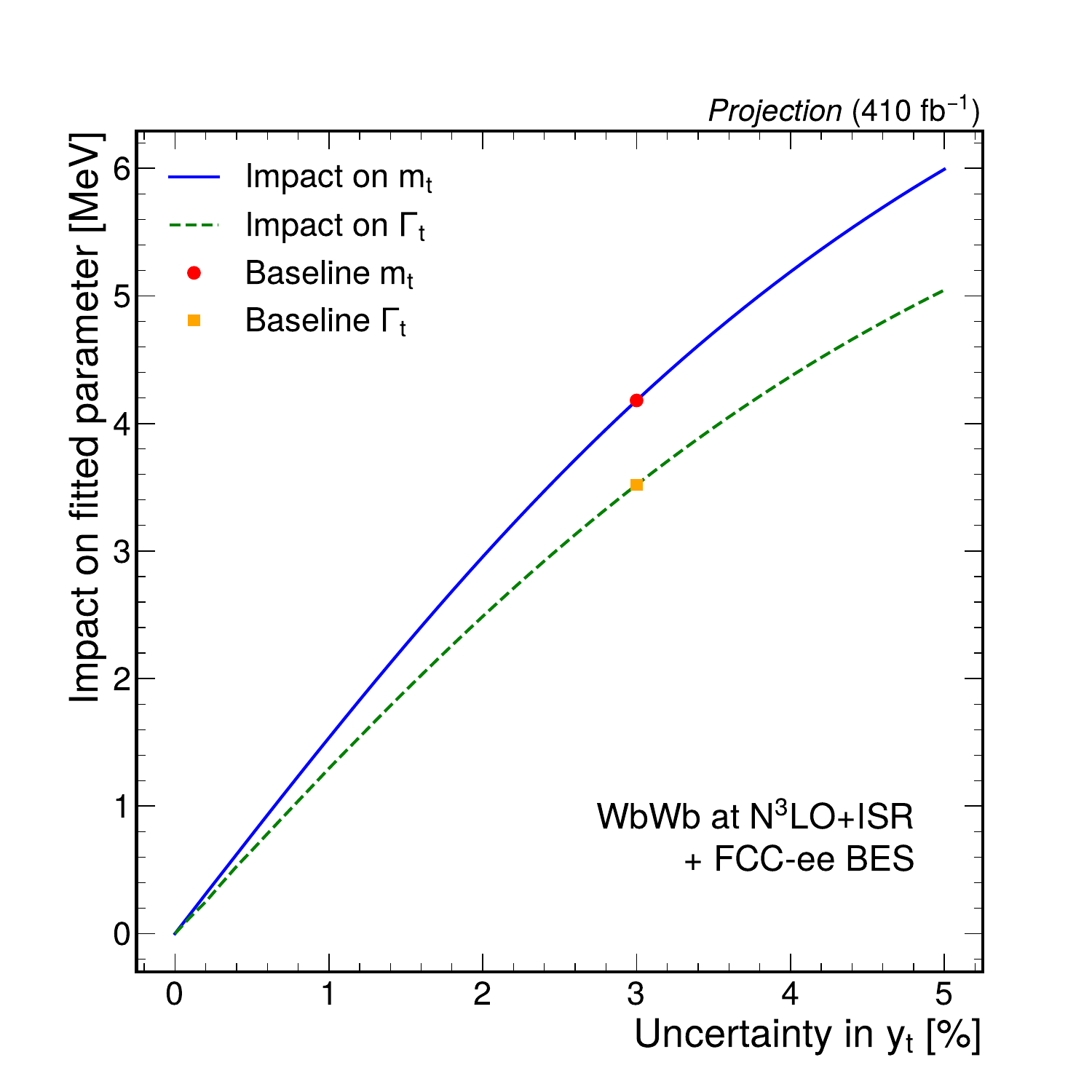}
    \caption{Impact of the parametric uncertainty in \asmz (left) and \yt (right) on \mt (solid curves) and \Gt (dashed curves). The markers correspond to the baseline uncertainties of $10^{-4}$ and 3\% on \asmz and \Gt, respectively.}
    \label{fig:uncertainties_parameters}
\end{figure}

Finally, we investigate the dependence of the results on the parametric uncertainty in \as and \yt. The results are shown in Figure~\ref{fig:uncertainties_parameters}. In both cases, we find a similar impact in absolute terms on \mt and \Gt, corresponding to a larger impact on \mt in relative terms. We also find that the dependence is very similar for the two parameters of interest. It can be noted that deviations from linearity arise at large values of the uncertainty in \yt. This is because this parameter can be further constrained in the fit of the \ttbar threshold scan, provided that the value of \as is known precisely. This is also confirmed by the fact that in the baseline fit we obtain a post-fit uncertainty for \yt of 2.6\% for an input uncertainty of 3\%. For this reason, we further investigated the sensitivity to \yt by performing a three-dimensional fit of the \ttbar threshold scan, promoting \yt to an unconstrained parameter. This yields a total uncertainty in \yt of 5.8\%, limited by the experimental uncertainties in the measured cross section (2.3\%), and the parametric uncertainties in \mt (5.1\%) and \Gt (3.5\%). In this fit, \mt and \Gt are measured with a total uncertainty of 10.9 and 13.6\MeV, respectively. The increase in total uncertainty is due to the strong correlation between the \yt and \mt (87\%) and between \yt and \Gt (60\%). As a result, the correlation between \mt and \Gt also increases to 64\%. This demonstrates that a precise measurement of \yt is a crucial input to the measurement of \mt and \Gt in a \ttbar threshold scan. In Section~\ref{sec:yukawa_fit}, this approach is extended to the above-threshold run at $\sqrts = 365\GeV$.

\section{Sensitivity to the top quark Yukawa coupling at 365 GeV}
\label{sec:yukawa_fit}

The baseline plan for FCC-ee also foresees a four-year-long run at the highest achievable energy of 365\GeV, targeting a total integrated luminosity of 2.65\abinv~\cite{FCC:2025lpp}. As shown in Section~\ref{sec:reco_fit}, this will allow a precise determination of the \WbWb production cross section, at the level of 0.12\%. Figure~\ref{fig:threshold_scan} (left) shows that the sensitivity to \mt, \Gt, and \as significantly decreases above the \ttbar production threshold, while the sensitivity to \yt remains significant. In particular, we find that a 10\% variation of \yt results in a 0.5\% shift in the calculated cross section at $\sqrts = 365\GeV$. Considering the experimental precision on \sigmaWbWb, this translates into a sensitivity to \yt at the level of about 2\%.
Also in this case we consider any possible impact of the expected uncertainty in the electroweak mixing angle, and we find the effect to be more than two orders of magnitude smaller than that of \yt. BSM contributions to the electroweak couplings of the top quark, which may become relevant in this context, can also be constrained at FCC-ee by means of the 365\GeV run using differential distributions in the final-state leptons~\cite{Janot:2015yza}. A simultaneous measurement of these couplings with \mt, \Gt, and \yt would provide the most robust result, but is beyond the scope of this work. We therefore assume the SM coupling between the top quark and the Z boson in the following.

We then perform a three-dimensional fit of \mt, \Gt, and \yt, including both the \ttbar threshold scan and the 365\GeV run, for a total integrated luminosity of about 3\abinv. All sources of uncertainty considered in Section~\ref{sec:threshold_scan} are propagated to the 365\GeV run. Also in this case we assume the parametric uncertainties to be correlated between centre-of-mass energies, and we split the input systematic uncertainties into correlated and uncorrelated components. The statistical uncertainty in \sigmaWbWb at $\sqrts = 365\GeV$ is also increased by 20\% according to the results of Section~\ref{sec:reco_fit}. Finally, we conservatively scale down the uncorrelated uncertainty in the integrated luminosity by a factor of five for the 365\GeV point, which is below the statistical scaling with the integrated luminosity. As the effect of the BES and BEC uncertainties on \yt is found to be negligible (as expected) we do not further scale down these uncertainties according to the integrated luminosity.

\begin{table}[htpb]
    \centering
\begin{tabular}{l|c|c|c}
    Uncertainty source  & \mtps [\MeV{}] & \Gt [\MeV{}] & \yt [\%] \\ 
    \hline
    Experimental (stat. $\times 1.2$)  & 4.2  & 10.0 & 1.5  \\  
    Parametric \mt    & --   & 5.3  & 1.2  \\  
    Parametric \Gt   & 3.0  & --   & 0.8  \\  
    Parametric \yt  & 3.8  & 4.8  & --   \\  
    Parametric \as & 2.2  & 1.6  & 0.2  \\  \hline
    Luminosity calibration (uncorr.)  & 0.6  & 1.1  & 0.2  \\  
    Luminosity calibration (corr.)    & 1.0  & 0.7  & 0.9  \\  
    Beam energy calibration (uncorr.) & 1.3  & 1.9  & 0.1  \\  
    Beam energy calibration (corr.)   & 1.3  & $<0.1$  & $<0.1$  \\  
    Beam energy spread (uncorr.)  & 0.3  & 0.9  & $<0.1$  \\  
    Beam energy spread (corr.)    & $<0.1$  & 1.1  & $<0.1$  \\  
    \hline
    Total profiled & 6.5 & 11.7 & 2.1  \\ 
\end{tabular}
    \caption{Same as Table~\ref{tab:breakdown_2D}, but for the three-dimensional fit of \mt, \Gt, and \yt, including the 365\GeV run. Here the parametric uncertainties due to the post-fit uncertainties in \mt and \Gt are explicitly shown in order to highlight their effect on the determination of \yt. However, the parametric impact of \yt on other parameters is not directly comparable to the corresponding impact in Table~\ref{tab:breakdown_2D}, as here \yt is an unconstrained parameter. The theoretical uncertainties in \mt and \Gt are the same as in Table~\ref{tab:breakdown_2D}, while the one in \yt is a back-of-the-envelope estimate based on the impact of the uncorrelated component of the integrated luminosity.}
    \label{tab:breakdown_3D}
\end{table}

The results of the fit are shown in Table~\ref{tab:breakdown_3D}. Unlike in Table~\ref{tab:breakdown_2D}, we explicitly factor out the parametric impacts of \mt and \Gt from the experimental uncertainty in order to emphasise their contribution to the total uncertainty in \yt. We find that, in this scenario, \yt can be determined with a total uncertainty of 2.1\%, in line with our expectations. The precision in this measurement is found to be limited by the experimental uncertainties in the measured \WbWb cross sections, the correlated component of the integrated luminosity, and the post-fit uncertainties in \mt and \Gt. Given the poor convergence of the NR-QCD calculation above the \ttbar production threshold (Figure~\ref{fig:theory}, right), variations of the renormalisation scale in \qqTh would not provide a reliable assessment of the theoretical uncertainty. We however estimate that for this measurement to become competitive, a calculation with an accuracy of 0.1\% or better would be needed. This would result in a theoretical uncertainty on \yt of 1.9\%, at the same level as the experimental uncertainty. Therefore, the precision of this measurement also relies crucially on progress in the theoretical predictions, which is an active area of research~\cite{Chen:2022vzo}.

\section{Comparison with other proposed \boldmath \texorpdfstring{\epm}{e+e-} colliders}

The qualitative conclusions of this study apply more generally and can be transferred to other proposed \epm colliders with some adaptations. The CEPC environment is very similar to that of FCC-ee. The project envisages collecting an integrated luminosity of 100\fbinv at the \ttbar production threshold, increasing by a factor of two the uncertainties of statistical nature with respect to Table~\ref{tab:breakdown_2D}. The authors of Ref.~\cite{Li:2022iav} project machine calibration uncertainties of 2--5\MeV on \mt, in agreement with our findings. 
While previous studies at linear colliders~\cite{Seidel:2013sqa,Horiguchi:2013wra,CLICdp:2018esa} typically envisaged an integrated luminosity of 100--200\fbinv, the instantaneous luminosity at circular and linear colliders is expected to be very similar at the energy of the \ttbar threshold~\cite{LinearColliderVision:2025hlt}. An important difference at LCs is the shape of the luminosity spectrum, which develops a tail towards lower energy due to beamstrahlung. This effect is responsible for an effective reduction in the \WbWb production cross section by 50--60\% at the \ttbar threshold~\cite{Simon:2019axh,CLICdp:2018esa} compared to a circular collider. However, this is partially compensated for by the left-right beam polarization, which can lead to an increase in the total cross section of about 43\% compared to unpolarised beams when the running scenario of Ref.~\cite{LinearColliderVision:2025hlt} is considered. The shape of the luminosity spectrum must be precisely calibrated with Bhabha events for the most precise determination of the top quark mass and width~\cite{Poss:2013oea}. 
The interplay between the various parametric uncertainties also changes, as \as is less constrained at the Z~pole, while \yt can be measured in $\epm \to \ttbar\mathrm{H}$ production at a similar level of precision as at HL-LHC~\cite{CLICdp:2018esa}. We however note that measurements at both circular and linear collider will profit from advances in \as measurements from lattice QCD.

\begin{table}[htbp]
\centering
\begin{tabular}{r|ccc}
 & FCC-ee & ILC~\cite{Simon:2019axh} & CLIC~\cite{CLICdp:2018esa}  \\
\hline
\mt stat.\ uncert.\ [\MeV{}] & 4.3 & 13 & 21  \\
\Gt stat.\ uncert.\ [\MeV{}] & 10.4 & 30 & 51  \\
\hline
Integrated luminosity [\fbinv{}] & 410 & 200 & 100  \\
Signal acceptance & 0.75 & 0.7 & 0.7  \\
Beam polarisation: gain on \sigmaWbWb~\cite{LinearColliderVision:2025hlt} & -- & 1.43 & 1.43  \\
FCC scan range: gain on \mt & -- & 1.39 & 1.39  \\
FCC scan range: gain on \Gt & -- & 1.34 & 1.34 \\
Stat.\ uncert.\ inflation & 1.2 & -- & --  \\
\hline
\mt stat.\ uncert.\ rescaled [MeV] & 4.3 & 6.3 & 7.2 \\
\Gt stat.\ uncert.\ rescaled [MeV] & 10.4 & 15.1 & 18.1  \\
\end{tabular}
\caption{Comparison of achievable precision on \mt and \Gt at FCC-ee, ILC, and CLIC for equal assumptions on integrated luminosity, signal acceptance, and scan points, including the effect of beam polarisation at linear colliders according to the scenario of Ref.~\cite{LinearColliderVision:2025hlt}.}
\label{tab:comparison}
\end{table}

In order to compare on the same footing results obtained for different colliders, we consider previous studies for ILC~\cite{Simon:2019axh} and CLIC~\cite{CLICdp:2018esa} and attempt to factor out any assumptions that are not specific to the considered detector technology. These include assumptions on the integrated luminosity, the signal acceptance, as well as the choice of the \ttbar threshold scan points. The nominal projections for FCC-ee, ILC, and CLIC are summarized in the first two rows of Table~\ref{tab:comparison}. We then rescale the marginalised statistical uncertainties in \mt and \Gt for ILC and CLIC to match our assumptions for FCC-ee. In particular, we re-scale to an integrated luminosity of 410\fbinv, to a signal acceptance of 75\%, and to the expected increase in \sigmaWbWb from beam polarisation, which was neglected in Refs.~\cite{Simon:2019axh,CLICdp:2018esa}. In addition, we estimate that our choice of scan points leads to an improvement of 39 (34)\% on \mt (\Gt) with respect to the 10-point scans with 1\GeV spacing of Refs.~\cite{Simon:2019axh,CLICdp:2018esa}, and we down-scale the LC uncertainties accordingly. This was estimated by repeating the FCC fit of Section~\ref{sec:threshold_scan} with that particular scenario. Finally, we inflate the LC statistical uncertainties by 20\% as done for FCC-ee in Section~\ref{sec:threshold_scan}. The rescaling factors described above are summarised in Table~\ref{tab:comparison}. The last two rows of the same table show the statistical uncertainties in \mt and \Gt after rescaling, for the three colliders. The observed residual differences are to be attributed to the details of the BES at the different colliders. In particular, we find a 45\% larger statistical uncertainty at ILC, and a 65 (75)\% larger uncertainty at CLIC in the case of \mt (\Gt), compared to FCC-ee. From this we can conclude that, in order to reach the same statistical precision as FCC-ee, a two (three)-times larger dataset would be needed at ILC (CLIC). The obtained results are consistent with the expectation that the ILC luminosity spectrum has intermediate properties between FCC-ee and CLIC, as explicitly shown in Ref.~\cite{Simon:2016pwp} for the \ttbar threshold scan.

\section{Summary and prospects}
\label{sec:summary}

We have presented a projection of the achievable precision in the measurement of the top quark mass (\mt) and width (\Gt) at FCC-ee, including a detailed assessment of experimental, parametric, and machine-related uncertainties. A detector-level measurement of the \WbWb production cross section at different centre-of-mass energies was demonstrated using fast simulation of the IDEA detector at FCC-ee, showing that all background components and systematic uncertainties can be controlled to a level well below the expected statistical uncertainty. 
We then used these results as input to a phenomenological analysis of the \ttbar threshold scan where the value of \mt is fitted simultaneously with \Gt, assuming the FCC-ee scenario for the integrated luminosity and machine-related parameters. We show that a precision of 6.8 (11.5)\MeV can be achieved for \mt (\Gt), considering all experimental, parametric, and machine-related uncertainties, when \mt is defined in the potential-subtracted (PS) scheme.
The statistical uncertainty is estimated to be 4.3 and 10.4\MeV for \mt and \Gt, respectively. We note however that the theoretical uncertainty remains the limiting factor with state-of-the-art theoretical predictions, with an impact of about 35 (25)\MeV on \mt (\Gt) at \NNNLO in non-relativistic QCD. Reducing the impact of theoretical uncertainties to match the expected experimental precision is an active area of development~\cite{Beneke:2024sfa}.
We also considered calculations in the so-called ``1S'' renormalisation scheme, and similar results were obtained. However, the detailed investigation of this aspect is beyond the scope of this work, and can be object of future studies.
We also conducted a detailed investigation of the dependence of these results on the assumption on the various systematic uncertainties. This allows our results to be updated in a straightforward (although approximate) way following any refinement in the estimates of the input uncertainties. Finally, we show that a high-luminosity run at the centre-of-mass energy of 365\GeV could achieve a determination of the top quark Yukawa coupling (\yt) at the level of 2.1\%, excluding theoretical uncertainties. However, we estimate that a cross section calculation with an uncertainty of 0.1\% or better is needed to match the experimental precision. Finally, we presented a detailed comparison between the statistical uncertainties in \mt and \Gt at FCC-ee, ILC, and CLIC under equal assumptions on the integrated luminosity, signal acceptance, and scan points, including the effect of beam polarisation at linear colliders. This results in a 45\% (65--75\%) larger uncertainty at ILC (CLIC) compared to FCC-ee. The residual differences can be attributed to the details of the BES.

In conclusion, this work demonstrates the remarkable physics reach of a \ttbar threshold run at future \epm colliders, and in particular at FCC-ee, which would allow \mt and \Gt to be measured in a theoretically and experimentally robust way, and with unprecendented precision. We demonstrate that this level of precision in \mt is necessary in order to match the SM prediction for \mW to the corresponding expected experimental precision at FCC-ee, allowing for a stringent test of the SM. The achievable precision in \Gt would allow for setting exclusion limits on BSM decays of top quarks with branching ratios larger than 1\%. This can also be directly probed using the large statistics of the 365\GeV run. A precise \Gt measurement can also be used as an input to global interpretations in the context of SM effective field theories (SMEFT)~\cite{Durieux:2018tev}.
Finally, we show that significant theoretical progress is crucial in order to fully exploit the physics potential of a future \epm machine.

\section*{Acknowledgements}
This study was initiated as a ``focus topic'' of the ECFA Higgs/top/EW factory studies~\cite{deBlas:2024bmz}. The authors would like to thank Martin Beneke, David d'Enterria, Patrick Janot, Andr\'e Hoang, and Frank Simon for discussions during the development of this study. 
We also thank the FCC-ee simulation team, and in particular Louis Portales, for help in producing the Monte Carlo samples used in this study, and Andrew Mehta for the valuable suggestions.
The work of J.B. has been partially funded by MICIU/AEI/10.13039/501100011033 and FEDER/UE
(grant PID2022-139466NB-C21). 
M.V. acknowledges financial support from the Spanish Ministry of Science and Universities under grant number PID2021-122134NB-C21, from the Generalitat Valenciana under grant number CIPROM/2021/073, and from the ``Severo Ochoa'' excellence programme.

\bibliographystyle{JHEP}
\bibliography{biblio}

@article{Fadin:1987wz,
    author = "Fadin, Victor S. and Khoze, Valery A.",
    title = "{Threshold Behavior of Heavy Top Production in \epm Collisions}",
    reportNumber = "LENINGRAD-87-1333",
    journal = "JETP Lett.",
    volume = "46",
    pages = "525--529",
    year = "1987"
}

@article{Fadin:1988fn,
    author = "Fadin, Victor S. and Khoze, Valery A.",
    title = "{Production of a pair of heavy quarks in $e^+ e^-$ annihilation in the threshold region}",
    journal = "Sov. J. Nucl. Phys.",
    volume = "48",
    pages = "309--313",
    year = "1988"
}

@article{Gusken:1985nf,
    author = "Gusken, S. and Kuhn, Johann H. and Zerwas, P. M.",
    title = "{Threshold Behavior of Top Production in $e^+ e^-$ Annihilation}",
    reportNumber = "CERN-TH-4106/84",
    doi = "10.1016/0370-2693(85)90983-9",
    journal = "Phys. Lett. B",
    volume = "155",
    pages = "185",
    year = "1985"
}

@article{Strassler:1990nw,
    author = "Strassler, Matthew J. and Peskin, Michael E.",
    title = "{The Heavy top quark threshold: QCD and the Higgs}",
    reportNumber = "SLAC-PUB-5308",
    doi = "10.1103/PhysRevD.43.1500",
    journal = "Phys. Rev. D",
    volume = "43",
    pages = "1500--1514",
    year = "1991"
}

@article{Guth:1991ab,
    author = "Guth, R. J. and Kuhn, Johann H.",
    title = "{Top quark threshold and radiative corrections}",
    reportNumber = "TKP-91-3",
    doi = "10.1016/0550-3213(92)90196-I",
    journal = "Nucl. Phys. B",
    volume = "368",
    pages = "38--56",
    year = "1992"
}

@article{Jezabek:1992np,
    author = "Jezabek, M. and Kuhn, Johann H. and Teubner, T.",
    title = "{Momentum distributions in $t\bar{t}$ production and decay near threshold}",
    reportNumber = "TTP-92-16",
    doi = "10.1007/BF01474740",
    journal = "Z. Phys. C",
    volume = "56",
    pages = "653--660",
    year = "1992"
}

@article{Beneke:2016kkb,
    author = "Beneke, M. and Kiyo, Y. and Maier, A. and Piclum, J.",
    title = "{Near-threshold production of heavy quarks with $\tt{QQbar\_threshold}$}",
    eprint = "1605.03010",
    archivePrefix = "arXiv",
    primaryClass = "hep-ph",
    doi = "10.1016/j.cpc.2016.07.026",
    journal = "Comput. Phys. Commun.",
    volume = "209",
    pages = "96--115",
    year = "2016"
}

@article{Hoang:2013uda,
    author = "Hoang, Andr\'e H. and Stahlhofen, Maximilian",
    title = "{The Top-Antitop Threshold at the ILC: NNLL QCD Uncertainties}",
    eprint = "1309.6323",
    archivePrefix = "arXiv",
    primaryClass = "hep-ph",
    reportNumber = "UWTHPH-2013-23, DESY-13-168",
    doi = "10.1007/JHEP05(2014)121",
    journal = "JHEP",
    volume = "05",
    pages = "121",
    year = "2014"
}

@article{Beneke:2015kwa,
    author = "Beneke, Martin and Kiyo, Yuichiro and Marquard, Peter and Penin, Alexander and Piclum, Jan and Steinhauser, Matthias",
    title = "{Next-to-Next-to-Next-to-Leading Order QCD Prediction for the Top Antitop $S$-Wave Pair Production Cross Section Near Threshold in $e^+e^-$ Annihilation}",
    eprint = "1506.06864",
    archivePrefix = "arXiv",
    primaryClass = "hep-ph",
    reportNumber = "ALBERTA-THY-12-15, DESY-15-100, SFB-CPP-14-127, TTP15-021, TUM-HEP-1001-15",
    doi = "10.1103/PhysRevLett.115.192001",
    journal = "Phys. Rev. Lett.",
    volume = "115",
    number = "19",
    pages = "192001",
    year = "2015"
}

@article{Vos:2016til,
    author = "Vos, M. and others",
    title = "{Top physics at high-energy lepton colliders}",
    eprint = "1604.08122",
    archivePrefix = "arXiv",
    primaryClass = "hep-ex",
    reportNumber = "DESY-16-038, IFIC-16-12",
    month = "4",
    year = "2016"
}

@article{Bach:2017ggt,
    author = {Bach, Fabian and Nejad, Bijan Chokouf\'e and Hoang, Andre and Kilian, Wolfgang and Reuter, J\"urgen and Stahlhofen, Maximilian and Teubner, Thomas and Weiss, Christian},
    title = "{Fully-differential Top-Pair Production at a Lepton Collider: From Threshold to Continuum}",
    eprint = "1712.02220",
    archivePrefix = "arXiv",
    primaryClass = "hep-ph",
    reportNumber = "DESY-17-158, LTH-1143, MITP-17-077, SI-HEP-2017-20, UWTHPH2017-35",
    doi = "10.1007/JHEP03(2018)184",
    journal = "JHEP",
    volume = "03",
    pages = "184",
    year = "2018"
}

@article{Martinez:2002st,
    author = "Martinez, Manel and Miquel, Ramon",
    title = "{Multiparameter fits to the $t\bar{t}$ threshold observables at a future $e^+e^-$ linear collider}",
    eprint = "hep-ph/0207315",
    archivePrefix = "arXiv",
    doi = "10.1140/epjc/s2002-01094-1",
    journal = "Eur. Phys. J. C",
    volume = "27",
    pages = "49--55",
    year = "2003"
}

@article{Seidel:2013sqa,
    author = "Seidel, Katja and Simon, Frank and Tesar, Michal and Poss, Stephane",
    title = "{Top quark mass measurements at and above threshold at CLIC}",
    eprint = "1303.3758",
    archivePrefix = "arXiv",
    primaryClass = "hep-ex",
    reportNumber = "MPP-2013-85",
    doi = "10.1140/epjc/s10052-013-2530-7",
    journal = "Eur. Phys. J. C",
    volume = "73",
    number = "8",
    pages = "2530",
    year = "2013"
}

@article{CLICdp:2018esa,
    author = "Abramowicz, H. and others",
    collaboration = "CLICdp",
    title = "{Top-Quark Physics at the CLIC Electron-Positron Linear Collider}",
    eprint = "1807.02441",
    archivePrefix = "arXiv",
    primaryClass = "hep-ex",
    reportNumber = "CLICdp-Pub-2018-003, CLICDP-PUB-2018-003",
    doi = "10.1007/JHEP11(2019)003",
    journal = "JHEP",
    volume = "11",
    pages = "003",
    year = "2019"
}

@article{Durieux:2018tev,
    author = "Durieux, Gauthier and Perell\'o, Mart\'\i{}n and Vos, Marcel and Zhang, Cen",
    title = "{Global and optimal probes for the top-quark effective field theory at future lepton colliders}",
    eprint = "1807.02121",
    archivePrefix = "arXiv",
    primaryClass = "hep-ph",
    reportNumber = "DESY-18-096, IFIC-18-27, DESY 18-096, IFIC 18-27",
    doi = "10.1007/JHEP10(2018)168",
    journal = "JHEP",
    volume = "10",
    pages = "168",
    year = "2018"
}

@article{Nowak:2021xmp,
    author = "Nowak, Kacper and Zarnecki, Aleksander Filip",
    title = "{Optimising top-quark threshold scan at CLIC using genetic algorithm}",
    eprint = "2103.00522",
    archivePrefix = "arXiv",
    primaryClass = "hep-ex",
    reportNumber = "CLICdp-Pub-2021-002",
    doi = "10.1007/JHEP07(2021)070",
    journal = "JHEP",
    volume = "07",
    pages = "070",
    year = "2021"
}

@unpublished{Blondel:2019jmp,
    author = "Blondel, Alain and others",
    title = "{Polarization and Centre-of-mass Energy Calibration at FCC-ee}",
    eprint = "1909.12245",
    archivePrefix = "arXiv",
    primaryClass = "physics.acc-ph",
    reportNumber = "FERMILAB-PUB-19-495-APC",
    month = "9",
    year = "2019",
}

@article{Chen:2022wit,
    author = "Chen, Long-Bin and Li, Hai Tao and Wang, Jian and Wang, Yefan",
    title = "{Analytic result for the top-quark width at next-to-next-to-leading order in QCD}",
    eprint = "2212.06341",
    archivePrefix = "arXiv",
    primaryClass = "hep-ph",
    doi = "10.1103/PhysRevD.108.054003",
    journal = "Phys. Rev. D",
    volume = "108",
    number = "5",
    pages = "054003",
    year = "2023"
}

@article{Poss:2013oea,
    author = "Poss, St\'ephane and Sailer, Andr\'e",
    title = "{Luminosity Spectrum Reconstruction at Linear Colliders}",
    eprint = "1309.0372",
    archivePrefix = "arXiv",
    primaryClass = "physics.ins-det",
    reportNumber = "LCD-NOTE-2013-008",
    doi = "10.1140/epjc/s10052-014-2833-3",
    journal = "Eur. Phys. J. C",
    volume = "74",
    number = "4",
    pages = "2833",
    year = "2014"
}

@article{Chen:2022vzo,
    author = "Chen, Xiang and Guan, Xin and He, Chuan-Qi and Liu, Xiao and Ma, Yan-Qing",
    title = "{Heavy-Quark Pair Production at Lepton Colliders at NNNLO in QCD}",
    eprint = "2209.14259",
    archivePrefix = "arXiv",
    primaryClass = "hep-ph",
    doi = "10.1103/PhysRevLett.132.101901",
    journal = "Phys. Rev. Lett.",
    volume = "132",
    number = "10",
    pages = "101901",
    year = "2024"
}

@article{deBlas:2024bmz,
    author = "de Blas, Jorge and others",
    title = "{Focus topics for the ECFA study on Higgs / Top / EW factories}",
    eprint = "2401.07564",
    archivePrefix = "arXiv",
    primaryClass = "hep-ph",
    month = "1",
    year = "2024"
}

@article{Li:2022iav,
    author = "Li, Zhan and Sun, Xiaohu and Fang, Yaquan and Li, Gang and Xin, Shuiting and Wang, Shudong and Wang, Yiwei and Zhang, Yuan and Zhang, Hao and Liang, Zhijun",
    title = "{Top quark mass measurements at the $t\bar{t}$ threshold with CEPC}",
    eprint = "2207.12177",
    archivePrefix = "arXiv",
    primaryClass = "hep-ex",
    doi = "10.1140/epjc/s10052-023-11421-1",
    journal = "Eur. Phys. J. C",
    volume = "83",
    number = "4",
    pages = "269",
    year = "2023",
    note = "[Erratum: Eur.Phys.J.C 83, 501 (2023)]"
}

@phdthesis{Beguin:2710098,
      author        = "Beguin, Marina",
      title         = "{Calorimetry and W mass measurement for future
                       experiments}",
      school        = "Université Paris-Saclay",
      year          = "2019",
      url           = "https://cds.cern.ch/record/2710098",
      note          = "Presented 10 Dec 2019",
}

@article{Horiguchi:2013wra,
    author = "Horiguchi, Tomohiro and Ishikawa, Akimasa and Suehara, Taikan and Fujii, Keisuke and Sumino, Yukinari and Kiyo, Yuichiro and Yamamoto, Hitoshi",
    title = "{Study of top quark pair production near threshold at the ILC}",
    eprint = "1310.0563",
    archivePrefix = "arXiv",
    primaryClass = "hep-ex",
    month = "10",
    year = "2013"
}

@article{Marquard:2015qpa,
    author = "Marquard, Peter and Smirnov, Alexander V. and Smirnov, Vladimir A. and Steinhauser, Matthias",
    title = "{Quark Mass Relations to Four-Loop Order in Perturbative QCD}",
    eprint = "1502.01030",
    archivePrefix = "arXiv",
    primaryClass = "hep-ph",
    reportNumber = "DESY-15-013, TTP15-01, SFB-CPP-14-120",
    doi = "10.1103/PhysRevLett.114.142002",
    journal = "Phys. Rev. Lett.",
    volume = "114",
    number = "14",
    pages = "142002",
    year = "2015"
}

@article{Nowak:2021tyn,
    author = "Nowak, Kacper and Zarnecki, Aleksander Filip",
    title = "{Top-quark mass determination in the optimised threshold scan}",
    eprint = "2107.12647",
    archivePrefix = "arXiv",
    primaryClass = "hep-ex",
    reportNumber = "CLICdp-Conf-2021-005",
    doi = "10.21468/SciPostPhysProc.8.173",
    journal = "SciPost Phys. Proc.",
    volume = "8",
    pages = "173",
    year = "2022"
}

@article{deFavereau:2013fsa,
    author = "de Favereau, J. and Delaere, C. and Demin, P. and Giammanco, A. and Lemaitre, V. and Mertens, A. and Selvaggi, M.",
    title = "{DELPHES} 3, A modular framework for fast simulation of a generic collider experiment",
    eprint = "1307.6346",
    archivePrefix = "arXiv",
    primaryClass = "hep-ex",
    doi = "10.1007/JHEP02(2014)057",
    journal = "JHEP",
    volume = "02",
    pages = "057",
    year = "2014"
}

@article{Janot:2015gjr,
    author = "Janot, Patrick",
    title = "{Direct measurement of $\alpha_{QED}(m_{Z}^{2})$ at the FCC-ee}",
    eprint = "1512.05544",
    archivePrefix = "arXiv",
    primaryClass = "hep-ph",
    doi = "10.1007/JHEP02(2016)053",
    journal = "JHEP",
    volume = "02",
    pages = "053",
    year = "2016",
    note = "[Erratum: JHEP 11, 164 (2017)]"
}

@article{Bernardi:2022hny,
    author = "Bernardi, G. and others",
    title = "{The Future Circular Collider: a Summary for the US 2021 Snowmass Process}",
    eprint = "2203.06520",
    archivePrefix = "arXiv",
    primaryClass = "hep-ex",
    reportNumber = "FERMILAB-PUB-22-494-SCD",
    month = "3",
    year = "2022"
}

@article{deBlas:2021wap,
    author = "de Blas, J. and Ciuchini, M. and Franco, E. and Goncalves, A. and Mishima, S. and Pierini, M. and Reina, L. and Silvestrini, L.",
    title = "{Global analysis of electroweak data in the Standard Model}",
    eprint = "2112.07274",
    archivePrefix = "arXiv",
    primaryClass = "hep-ph",
    reportNumber = "KEK-TH-2378",
    doi = "10.1103/PhysRevD.106.033003",
    journal = "Phys. Rev. D",
    volume = "106",
    number = "3",
    pages = "033003",
    year = "2022"
}

@article{deBlas:2016ojx,
    author = "de Blas, Jorge and Ciuchini, Marco and Franco, Enrico and Mishima, Satoshi and Pierini, Maurizio and Reina, Laura and Silvestrini, Luca",
    title = "{Electroweak precision observables and Higgs-boson signal strengths in the Standard Model and beyond: present and future}",
    eprint = "1608.01509",
    archivePrefix = "arXiv",
    primaryClass = "hep-ph",
    reportNumber = "KEK-TH-1919",
    doi = "10.1007/JHEP12(2016)135",
    journal = "JHEP",
    volume = "12",
    pages = "135",
    year = "2016"
}

@article{Belloni:2022due,
    author = "Belloni, Alberto and others",
    title = "{Report of the Topical Group on Electroweak Precision Physics and Constraining New Physics for Snowmass 2021}",
    eprint = "2209.08078",
    archivePrefix = "arXiv",
    primaryClass = "hep-ph",
    month = "9",
    year = "2022"
}

@article{Blondel:2019qlh,
    author = "Blondel, Alain and Freitas, Ayres and Gluza, Janusz and Riemann, Tord and Heinemeyer, Sven and Jadach, Stanislaw and Janot, Patrick",
    title = "{Theory Requirements and Possibilities for the FCC-ee and other Future High Energy and Precision Frontier Lepton Colliders}",
    eprint = "1901.02648",
    archivePrefix = "arXiv",
    primaryClass = "hep-ph",
    month = "1",
    year = "2019"
}

@article{Freitas:2019bre,
    author = "Freitas, A. and others",
    title = "{Theoretical uncertainties for electroweak and Higgs-boson precision measurements at FCC-ee}",
    eprint = "1906.05379",
    archivePrefix = "arXiv",
    primaryClass = "hep-ph",
    reportNumber = "IFT-UAM/CSIC-18-021, TUM-HEP-1185/19, KW 19-001, TTK-19-20, UWThPh
  2019-16, UWThPh 2019-16, IFJPAN-IV-2019-8, FR-PHENO-2019-010, DESY 19-105, DESY-19-105",
    month = "6",
    year = "2019"
}

@article{Riembau:2025ppc,
    author = "Riembau, Marc",
    title = "{On the extraction of $\alpha_\textit{em}(m_Z^2)$ at Tera-$Z$}",
    eprint = "2501.05508",
    archivePrefix = "arXiv",
    primaryClass = "hep-ph",
    reportNumber = "CERN-TH-2025-005",
    month = "1",
    year = "2025"
}

@article{dEnterria:2020cpv,
    author = "d'Enterria, David and Jacobsen, Villads",
    title = "{Improved strong coupling determinations from hadronic decays of electroweak bosons at N$^3$LO accuracy}",
    eprint = "2005.04545",
    archivePrefix = "arXiv",
    primaryClass = "hep-ph",
    month = "5",
    year = "2020"
}

@article{Beneke:2024sfa,
    author = "Beneke, M. and Kiyo, Y.",
    title = "{Third-order correction to top-quark pair production near threshold II. Potential contributions}",
    eprint = "2409.05960",
    archivePrefix = "arXiv",
    primaryClass = "hep-ph",
    reportNumber = "TUM-HEP-1525/24",
    month = "9",
    year = "2024"
}

@article{Simon:2016pwp,
    author = "Simon, Frank",
    title = "{Impact of Theory Uncertainties on the Precision of the Top Quark Mass in a Threshold Scan at Future $e^+ e^-$ Colliders}",
    eprint = "1611.03399",
    archivePrefix = "arXiv",
    primaryClass = "hep-ex",
    reportNumber = "MPP-2016-325",
    doi = "10.22323/1.282.0872",
    journal = "PoS",
    volume = "ICHEP2016",
    pages = "872",
    year = "2017"
}

@article{IDEAStudyGroup:2025gbt,
    collaboration = "IDEA Study Group",
    title = "{The IDEA detector concept for FCC-ee}",
    eprint = "2502.21223",
    archivePrefix = "arXiv",
    primaryClass = "physics.ins-det",
    month = "2",
    year = "2025"
}

@misc{delphes_card_idea,
  author={{Various authors}},
  title = {{FCC-ee IDEA detector \textsc{DELPHES} card}},
  howpublished = {\url{https://github.com/delphes/delphes/blob/master/cards/delphes_card_IDEA.tcl}},
}

@misc{TAGGINGFULLSIM_NOTE,
   title="{Jet Flavor Tagging Performance at FCC-ee}",
   author={Aumiller, Sara and Garcia, Dolores and Selvaggi, Michele},
   doi = "10.17181/8g834-jv464",
   year = "2024"  
}

@article{Catani:1991hj,
    author = "Catani, S. and Dokshitzer, Yuri L. and Olsson, M. and Turnock, G. and Webber, B. R.",
    title = "{New clustering algorithm for multi-jet cross-sections in $e^+ e^-$ annihilation}",
    reportNumber = "CAVENDISH-HEP-91-5",
    doi = "10.1016/0370-2693(91)90196-W",
    journal = "Phys. Lett. B",
    volume = "269",
    pages = "432",
    year = "1991"
}

@article{Bedeschi:2022rnj,
    author = "Bedeschi, Franco and Gouskos, Loukas and Selvaggi, Michele",
    title = "{Jet flavour tagging for future colliders with fast simulation}",
    eprint = "2202.03285",
    archivePrefix = "arXiv",
    primaryClass = "hep-ex",
    doi = "10.1140/epjc/s10052-022-10609-1",
    journal = "Eur. Phys. J. C",
    volume = "82",
    number = "7",
    pages = "646",
    year = "2022"
}

@article{ParticleDataGroup:2024cfk,
    author = "Navas, S. and others",
    collaboration = "Particle Data Group",
    title = "{Review of particle physics}",
    doi = "10.1103/PhysRevD.110.030001",
    journal = "Phys. Rev. D",
    volume = "110",
    number = "3",
    pages = "030001",
    year = "2024"
}

@article{Sjostrand:2006za,
    author = "Sjostrand, Torbjorn and Mrenna, Stephen and Skands, Peter Z.",
    title = "{PYTHIA 6.4 Physics and Manual}",
    eprint = "hep-ph/0603175",
    archivePrefix = "arXiv",
    reportNumber = "FERMILAB-PUB-06-052-CD-T, LU-TP-06-13",
    doi = "10.1088/1126-6708/2006/05/026",
    journal = "JHEP",
    volume = "05",
    pages = "026",
    year = "2006"
}

@article{Sjostrand:2014zea,
    author = {Sj\"ostrand, Torbj\"orn and Ask, Stefan and Christiansen, Jesper R. and Corke, Richard and Desai, Nishita and Ilten, Philip and Mrenna, Stephen and Prestel, Stefan and Rasmussen, Christine O. and Skands, Peter Z.},
    title = "{An introduction to PYTHIA 8.2}",
    eprint = "1410.3012",
    archivePrefix = "arXiv",
    primaryClass = "hep-ph",
    reportNumber = "LU-TP-14-36, MCNET-14-22, CERN-PH-TH-2014-190, FERMILAB-PUB-14-316-CD, DESY-14-178, SLAC-PUB-16122",
    doi = "10.1016/j.cpc.2015.01.024",
    journal = "Comput. Phys. Commun.",
    volume = "191",
    pages = "159--177",
    year = "2015"
}

@article{Azzi:2019yne,
    author = "Azzi, P. and others",
    editor = "Dainese, Andrea and Mangano, Michelangelo and Meyer, Andreas B. and Nisati, Aleandro and Salam, Gavin and Vesterinen, Mika Anton",
    title = "{Report from Working Group 1}: {Standard Model Physics at the HL-LHC and HE-LHC}",
    eprint = "1902.04070",
    archivePrefix = "arXiv",
    primaryClass = "hep-ph",
    reportNumber = "CERN-LPCC-2018-03",
    doi = "10.23731/CYRM-2019-007.1",
    journal = "CERN Yellow Rep. Monogr.",
    volume = "7",
    pages = "1--220",
    year = "2019"
}

@article{Kilian:2007gr,
    author = "Kilian, Wolfgang and Ohl, Thorsten and Reuter, Jurgen",
    title = "{WHIZARD: Simulating Multi-Particle Processes at LHC and ILC}",
    eprint = "0708.4233",
    archivePrefix = "arXiv",
    primaryClass = "hep-ph",
    reportNumber = "DESY-11-126, EDINBURGH-2010-36, FR-PHENO-2010-037, SI-HEP-2010-18",
    doi = "10.1140/epjc/s10052-011-1742-y",
    journal = "Eur. Phys. J. C",
    volume = "71",
    pages = "1742",
    year = "2011"
}

@article{CMS:2024onh,
    author = "Hayrapetyan, Aram and others",
    collaboration = "CMS",
    title = "{The CMS Statistical Analysis and Combination Tool: Combine}",
    eprint = "2404.06614",
    archivePrefix = "arXiv",
    primaryClass = "physics.data-an",
    reportNumber = "CMS-CAT-23-001, CERN-EP-2024-078",
    doi = "10.1007/s41781-024-00121-4",
    journal = "Comput. Softw. Big Sci.",
    volume = "8",
    number = "1",
    pages = "19",
    year = "2024"
}

@article{FCC:2025lpp,
    author = "Benedikt, M. and others",
    collaboration = "FCC",
    title = "{Future Circular Collider Feasibility Study Report: Volume 1, Physics, Experiments, Detectors}",
    eprint = "2505.00272",
    archivePrefix = "arXiv",
    primaryClass = "hep-ex",
    reportNumber = "CERN-FCC-PHYS-2025-0002",
    doi = "10.17181/CERN.9DKX.TDH9",
    month = "4",
    year = "2025"
}

@article{Janot:2015yza,
    author = "Janot, Patrick",
    title = "{Top-quark electroweak couplings at the FCC-ee}",
    eprint = "1503.01325",
    archivePrefix = "arXiv",
    primaryClass = "hep-ph",
    doi = "10.1007/JHEP04(2015)182",
    journal = "JHEP",
    volume = "04",
    pages = "182",
    year = "2015"
}

@article{Cowan:2010js,
    author = "Cowan, Glen and Cranmer, Kyle and Gross, Eilam and Vitells, Ofer",
    title = "{Asymptotic formulae for likelihood-based tests of new physics}",
    eprint = "1007.1727",
    archivePrefix = "arXiv",
    primaryClass = "physics.data-an",
    doi = "10.1140/epjc/s10052-011-1554-0",
    journal = "Eur. Phys. J. C",
    volume = "71",
    pages = "1554",
    year = "2011",
    note = "[Erratum: Eur.Phys.J.C 73, 2501 (2013)]"
}

@article{CMS:2024irj,
    author = "Hayrapetyan, Aram and others",
    collaboration = "CMS",
    title = "{Review of top quark mass measurements in CMS}",
    eprint = "2403.01313",
    archivePrefix = "arXiv",
    primaryClass = "hep-ex",
    reportNumber = "CMS-TOP-23-003, CERN-EP-2024-005",
    month = "3",
    year = "2024"
}

@misc{iminuit,
  author={Hans Dembinski and Piti Ongmongkolkul et al.},
  title={scikit-hep/iminuit},
  DOI={10.5281/zenodo.3949207},
  url={https://doi.org/10.5281/zenodo.3949207}
}

@article{Beneke:2015lwa,
    author = "Beneke, Martin and Maier, Andreas and Piclum, Jan and Rauh, Thomas",
    title = "{Higgs effects in top anti-top production near threshold in $e^+e^-$ annihilation}",
    eprint = "1506.06865",
    archivePrefix = "arXiv",
    primaryClass = "hep-ph",
    reportNumber = "TUM-HEP-1002-15",
    doi = "10.1016/j.nuclphysb.2015.07.034",
    journal = "Nucl. Phys. B",
    volume = "899",
    pages = "180--193",
    year = "2015",
}

@article{Yan:2023mjj,
    author = "Yan, Jiang and Wu, Xing-Gang and Wu, Zhi-Fei and Shan, Jing-Hao and Zhou, Hua",
    title = "{Reanalysis of the top-quark pair production via the e+e{\ensuremath{-}} annihilation near the threshold region up to N3LO QCD corrections}",
    eprint = "2312.15442",
    archivePrefix = "arXiv",
    primaryClass = "hep-ph",
    doi = "10.1016/j.physletb.2024.138664",
    journal = "Phys. Lett. B",
    volume = "853",
    pages = "138664",
    year = "2024"
}

@article{Yan:2024hbz,
    author = "Yan, Jiang and Wu, Xing-Gang and Zhou, Hua and Li, Hong-Tai and Shan, Jing-Hao",
    title = "{Improved analysis of the decay width of t{\textrightarrow}Wb up to N3LO QCD corrections}",
    eprint = "2404.11133",
    archivePrefix = "arXiv",
    primaryClass = "hep-ph",
    doi = "10.1103/PhysRevD.109.114026",
    journal = "Phys. Rev. D",
    volume = "109",
    number = "11",
    pages = "114026",
    year = "2024"
}

@article{Cacciari:2011ma,
    author = "Cacciari, Matteo and Salam, Gavin P. and Soyez, Gregory",
    title = "{FastJet} User Manual",
    eprint = "1111.6097",
    archivePrefix = "arXiv",
    primaryClass = "hep-ph",
    reportNumber = "CERN-PH-TH-2011-297",
    doi = "10.1140/epjc/s10052-012-1896-2",
    journal = "Eur. Phys. J. C",
    volume = "72",
    pages = "1896",
    year = "2012"
}

@article{Beneke:2017rdn,
    author = "Beneke, Martin and Maier, Andreas and Rauh, Thomas and Ruiz-Femenia, Pedro",
    title = "Non-resonant and electroweak {NNLO} correction to the $e^+ e^-$ top anti-top threshold",
    eprint = "1711.10429",
    archivePrefix = "arXiv",
    primaryClass = "hep-ph",
    reportNumber = "TUM-HEP-1113-17, IPPP-17-87, FTUAM-17-27, IFT-UAM-CSIC-17-111",
    doi = "10.1007/JHEP02(2018)125",
    journal = "JHEP",
    volume = "02",
    pages = "125",
    year = "2018",  
}

@article{LinearColliderVision:2025hlt,
    collaboration = "Linear Collider Vision",
    title = "{A Linear Collider Vision for the Future of Particle Physics}",
    eprint = "2503.19983",
    archivePrefix = "arXiv",
    primaryClass = "hep-ex",
    reportNumber = "FERMILAB-PUB-25-0216-CSAID-TD",
    month = "3",
    year = "2025"
}

@inproceedings{Simon:2019axh,
    author = "Simon, Frank",
    title = "Scanning Strategies at the Top Threshold at {ILC}",
    booktitle = "{International Workshop on Future Linear Colliders}",
    eprint = "1902.07246",
    archivePrefix = "arXiv",
    primaryClass = "hep-ex",
    reportNumber = "MPP-2019-35",
    year = "2019"
}

\clearpage

\appendix

\section{Impact of the top quark mass precision in the Standard Model electroweak fit at FCC-ee}
\label{app:EW_fit}

The measurements of the so-called electroweak precision observables (EWPO) provide one of the most stringent tests of the validity of the SM description of the EW interactions. Through radiative corrections, the SM predictions of EWPO depend on all the inputs of the theory. 
Of particular importance are the top-quark loop effects, which introduce
a quadratic dependence on \mt. This parameter can be 
tested by very precise measurements, \eg of the W boson mass, 
and indirectly determined from a global fit to EWPO. 
The resulting value can then be compared with direct measurements of \mW, providing a test of the consistency of the SM. 
This, however, requires a precise direct determination of the top-quark mass, with similar or better precision than the indirect determination.
A precise direct determination of \mt (and the other SM parameters) is also needed for the interpretation of the EWPO in terms of BSM physics, so the SM parametric uncertainties associated with the finite precision of the input parameters are not to be confounded with possible new effects that could be small compared with current experimental precision, but could be tested in the electroweak physics programme of future \epm colliders.

To evaluate the importance of reducing the top-quark mass uncertainty to the level discussed in this paper, we present in Table \ref{tab:SMparUnc} the expected parametric uncertainties on different EWPO that could be measured at future colliders, taking as a baseline the FCC-ee electroweak precision programme. We use as inputs the projections reported in \cite{Bernardi:2022hny,Belloni:2022due}, extrapolated to 4 interaction points~\cite{FCC:2025lpp}. For each of the observables in that table, we show the parametric error associated with an experimental precision of: 0.0001 for \asmz~\cite{dEnterria:2020cpv},
0.003\% in \aEMmz~\cite{Janot:2015gjr}, 0.1\MeV for \mZ~\cite{Bernardi:2022hny,FCC:2025lpp}, 3\MeV for \mH~\cite{FCC:2025lpp} and, for \mt, considering 250\MeV as an estimate of the expected HL-LHC precision, and 50\MeV as a proxy for the precision that could be obtained at \epm colliders in a scan of the \ttbar threshold. These last two estimates are purposely chosen for the sake of illustration. The different contributions can be compared to the expected experimental precision, in the last column. 
Compared to the case of current electroweak measurements, where the parametric uncertainties of all SM parameters are subdominant~\cite{deBlas:2016ojx,deBlas:2021wap}, it is clear that this may not be the case with the future precision of EWPO. 
In particular, from the point of view of the top-quark mass, while the HL-LHC determination would be enough to keep the \mt parametric uncertainty under control for several of the quantities in the table, it is insufficient for some key observables, such as \mW, \GZ, or the leptonic asymmetries. In particular, a precision of $\sim 175\MeV$  is required to match the projected \mW experimental precision of about 1\MeV, which is expected at most future \epm colliders. This is, therefore, expected to be beyond the HL-LHC reach. For high-luminosity circular colliders, like the FCC-ee, where \mW is expected to be measured with even higher precision, a determination of \mt with an uncertainty of less than 40\MeV would be needed to match the experimental precision, see left panel in Figure~\ref{fig:MW-mt_EWfit}. At this point, however, the total parametric uncertainty would be dominated by the uncertainty in the electromagnetic constant. With an improved measurement of \aEMmz, as suggested in~\cite{Riembau:2025ppc}, a \mt precision of around 30\MeV would bring the overall parametric uncertainty down to the expected 0.24\MeV experimental precision of \mW.

\begin{table}[htbp]
    \centering
    \small
    \begin{tabular}{cccccc|c} 
    \hline
Observable  & \asmz & \aEMmz & \mZ & \mt  & Total  & Exp.  \\  
            &                     &               &         & ($\Delta_{\mt}=50/250\MeV$)   &  &   \\  
\hline
\mW [\MeV{}]  & 0.07 & 0.48 & 0.13 & 0.3 / 1.5 & 0.59 & 0.24\\  
\GW [\MeV{}]  & 0.04 & 0.04 & 0.01 & 0.02 / 0.12 & 0.06 & 1\\  
\GZ [\MeV{}]  & 0.05 & 0.03 & 0.01 & 0.01 / 0.06 & 0.06 & 0.02\\  
$\sigma_{\textrm{had}}$ [pb]  & 0.49 & 0.04 & 0.09 & 0.03 / 0.17 & 0.5 & 4\\  
$A_{\ell}~[\times 10^{-5}] $  & 0.26 & 7.29 & 0.57 & 1.17 / 5.84 & 7.41 & 2\\  
$A_{c}~[\times 10^{-5}]$  & 0.12 & 3.22 & 0.25 & 0.54 / 2.71 & 3.27 & 24\\  
$A_{b}~[\times 10^{-5}]$  & 0.01 & 0.6 & 0.05 & 0.04 / 0.18 & 0.6 & 21\\  
$R_{\ell}~[\times 10^{-3}]$ & 0.62 & 0.16 & -- & -- / 0.11 & 0.64 & 1\\  
$R_{c}~[\times 10^{-5}]$  & 0.2 & 0.05 & -- & 0.06 / 0.29 & 0.21 & 26\\  
$R_{b}~[\times 10^{-5}]$ & 0.11 & 0.03 & -- & 0.17 / 0.84 & 0.2 & 6.5\\  
        \hline
    \end{tabular}
    \caption{Summary of SM parametric uncertainties for the main EWPO. For the top quark mass, we show the results assuming an uncertainty of 50\MeV, as a proxy of the \epm precision, and also assuming 250\MeV, as the expected uncertainty in \mt to be obtained at the HL-LHC. Entries marked with ``--'' denote a small contribution to the total uncertainty, which we omit in the table. }
    \label{tab:SMparUnc}
\end{table}

In the right panel of Figure~\ref{fig:MW-mt_EWfit} we also present a projection for the consistency test of the SM in the \mW--\mt plane, where we compare the expected precision of both quantities, as predicted from the EW fit (\ie removing the corresponding inputs from the likelihood) with that of the projected experimental measurements. We compare the HL-LHC and FCC-ee precision, considering two scenarios for the future SM intrinsic theory uncertainties (\ie those associated with the effects of missing higher-order corrections) for FCC-ee. The first scenario, shown by the blue ellipses, assumes the projected future intrinsic theory errors after including the general 3-loop (and leading 4-loop) corrections in the SM predictions for EWPO~\cite{Blondel:2019qlh,Freitas:2019bre}. The second one, shown in yellow, presents the results assuming the theory calculations are further improved to the point where the intrinsic uncertainties become subdominant. As can be seen, this has an important impact on the results, but the direct top-quark mass precision is, in any case, better than the indirect determination. 
Alternatively, one can compare the expected experimental precision of \mt with the 1D indirect determination from the future EW fit, which turns out to be approximately 60\MeV, if one assumes the SM intrinsic theory uncertainties are subdominant, indicating that a direct measurement of similar or better precision would be needed to test the SM predictions with future electroweak precision data. On the other hand, using the projected theory uncertainties for EWPO, these become the main limiting factor, and the \mt indirect precision would degrade to approximately 185\MeV. This fact highlights once more the need for significant theoretical developments in order to match the achievable experimental precision at future \epm colliders.

\section{Statistical model for the fit of the  \boldmath \texorpdfstring{\ttbar}{tt} threshold scan}
\label{app:fit_model}

The \chisq function used for the analysis of the \ttbar threshold scan in Section~\ref{sec:threshold_scan} depends on the parameters of interest (\mt and \Gt), on the profiled SM parameters (\yt and \as), and on nuisance parameters that model the impact of the systematic uncertainties ($\Vec{\lambda}$). All systematic uncertainties (except the integrated luminosity) are convoluted with the calculated cross section and are therefore implemented as modifiers of the predicted cross section. The \chisq function is defined as
\begin{align}
    \chisq (\mt, \Gt, \yt, \as, \Vec{\lambda}) =& ~ \Vec{r}\, (\mt, \Gt, \yt, \as, \Vec{\lambda})^\mathrm{T} ~ C_\mathrm{tot}^{-1} ~ \Vec{r}\, (\mt, \Gt, \yt, \as, \Vec{\lambda}) + \\ &  + \left ( \frac{\as -\as^0}{\delta \as } \right )^2 + \left ( \frac{\yt -\yt^0}{\delta \yt } \right )^2 + \sum_i \lambda_i^2, 
    \label{eq:chisq}
\end{align}
where
\begin{equation}
    \Vec{r}\, (\mt, \Gt, \yt, \as, \Vec{\lambda}) =  \Vec{\sigma}_\mathrm{exp} - \Vec{\sigma}_\mathrm{th} (\mt, \Gt, \yt, \as, \Vec{\lambda}) 
\end{equation}
represent the residuals between the calculated cross section and the experimentally measured ones, which here correspond to an Asimov data set or a pseudo-dataset. In the case where a pseudo-dataset is used, the $\Vec{\sigma}_\mathrm{exp}$ are generated according to a Gaussian distribution around the central value of the cross section and according to the expected experimental uncertainty. In the case of the Asimov data set, the $\Vec{\sigma}_\mathrm{exp}$ correspond to the central value of the cross section. The last three terms represent Gaussian penalty terms for the nuisance parameters and for the parametric uncertainty in \as and \yt. All nuisance parameters are defined in a way that all penalty terms follow a standard normal distribution. Finally, $\as^0$ ($\yt^0)$ and $\delta\as$ ($\delta\yt$) represent the assumed central value and input uncertainty for \as (\yt), respectively.  The covariance matrix in Eq.~\ref{eq:chisq} is defined as
\begin{equation}
    C_\mathrm{tot} = C_\mathrm{exp} + C_\mathrm{lumi}^\mathrm{uncorr} + C_\mathrm{lumi}^\mathrm{corr},
\end{equation}
where $C_\mathrm{exp}$ is the covariance matrix of the experimentally measured cross sections, while $C_\mathrm{lumi}^\mathrm{uncorr}$ and $C_\mathrm{lumi}^\mathrm{corr}$ represent the uncorrelated and correlated component of the luminosity uncertainty, respectively. The matrix $C_\mathrm{exp}$ is assumed to be diagonal and is calculated as
\begin{equation}
        \left[C_\mathrm{exp}\right]_{i,j} = \delta_{i,j} \sqrt{\frac{\sigma_\mathrm{exp}^i}{\mathcal{L}_i}} \sqrt{\frac{\sigma_\mathrm{exp}^j}{\mathcal{L}_j}},
\end{equation}
where $\mathcal{L}_i$ and $\sigma_\mathrm{exp}^i$ are the integrated luminosity and measured cross sections for scan point~$i$, respectively. As explained in Section~\ref{sec:threshold_scan}, these uncertainties are inflated by a factor of 1.2 in order to take into account the effect of background contamination, according to the results of Section~\ref{sec:reco_fit}. Similarly, $C_\mathrm{lumi}^\mathrm{uncorr}$ and $C_\mathrm{lumi}^\mathrm{corr}$ are built according to the uncorrelated (correlated) luminosity uncertainty of 0.1 (0.05)\% for all centre-of-mass energies (see Table~\ref{tab:breakdown_2D}). Therefore:
\begin{align}
    \left[C_\mathrm{lumi}^\mathrm{uncorr}\right]_{i,j} &= \delta_{i,j} \left[\sigma_\mathrm{exp}^i \cdot 0.1\% \right] \left[\sigma_\mathrm{exp}^j \cdot 0.1\% \right] \\
    \left[C_\mathrm{lumi}^\mathrm{corr}\right]_{i,j} &=  \left[\sigma_\mathrm{exp}^i \cdot 0.05\% \right] \left[\sigma_\mathrm{exp}^j \cdot 0.05\% \right]
    \label{eq:covariance}
\end{align}

The dependence of the calculated cross section on all parameters is modelled via a linear interpolation of two-point variations. We define nuisance parameters so that $\lambda_i = 0$ corresponds to the baseline value of the systematic sources, and $\lambda_i = 1$ corresponds to the systematic sources varied by one standard deviation. In this way, the cross section for any value of $\lambda_i$ can be approximated using the following expression:
\begin{equation}
    \Vec{\sigma}_\mathrm{th} (\lambda_i) = \sigma _\mathrm{th} (\lambda_i = 0) \left\{ 1 + \lambda_i \left[\frac{\Vec{\sigma}_\mathrm{th} (\lambda_i = 1)}{\Vec{\sigma}_\mathrm{th} (\lambda_i = 0)} -1 \right ] \right\}
\end{equation}
When estimating the correlated components, a single parameter $\lambda_i$ induces a simultaneous variation on all the scan points. In contrast, the uncorrelated components are estimated by performing the variation independently for each scan point. The cross section for any set of arbitrary values of the nuisance parameters can then be estimated as:
\begin{equation}
    \Vec{\sigma}_\mathrm{th} (\Vec{\lambda}) = \sigma _\mathrm{th} (\Vec{\lambda} = \Vec{0}) \prod_i  \left\{ 1 + \lambda_i \left[\frac{\Vec{\sigma}_\mathrm{th} (\lambda_i = 1)}{\Vec{\sigma}_\mathrm{th} (\lambda_i = 0)} -1 \right ] \right\},
\end{equation}
assuming the full factorisation of all variations. The same procedure is followed for the other parameters (\mt, \Gt, \yt, and \as), enforcing the effect to be correlated between the different centre-of-mass energies. For example, in the case of \yt, we use the following approximation:
\begin{equation}
    \Vec{\sigma}_\mathrm{th} (\yt) = \sigma _\mathrm{th} (\yt = \yt^0) \left\{ 1 + \left(\frac{\yt -\yt^0}{\delta \yt} \right)\left[\frac{\Vec{\sigma}_\mathrm{th} (\yt = \yt^0 + \delta \yt)}{\Vec{\sigma}_\mathrm{th} (\yt = \yt^0)} -1 \right ] \right\}.
\end{equation}
In the case of \yt and \as we use values of $\delta\yt$ and $\delta \as$ corresponding to the values in Table~\ref{tab:breakdown_2D}. The same values are also used in Eq.~\ref{eq:chisq}. In the case of \mt and \Gt, we use arbitrary variations of 30 and 50\MeV, respectively. This choice does not impact the result of the fit, as these variations do not appear in Eq.~\ref{eq:chisq}. The factorisation of the variations and the validity of the linear interpolation have been explicitly checked for \mt, \Gt, \yt, and \as, and closure is achieved well below the permille level for arbitrary simultaneous variations of these parameters. The \chisq function is then minimised using the iminuit package~\cite{iminuit} to obtain the best-fit values for the parameters and the corresponding covariance matrix.
The parameter scans presented in Section~\ref{sec:param_scans} are obtained by rescaling the penalty terms in Eq.~\ref{eq:chisq} and, in the case of the integrated luminosity, the covariances in Eq.~\ref{eq:covariance}. Finally, the \chisq function used in Section~\ref{sec:yukawa_fit} for the fit of \yt is the same as the one in Eq.~\ref{eq:chisq}, but without the penalty term for \yt.

\section{Fit distributions and impact plots}
\label{app:plots_distribs}

\begin{figure}[htbp]
    \centering
    \includegraphics[width=0.495\linewidth]{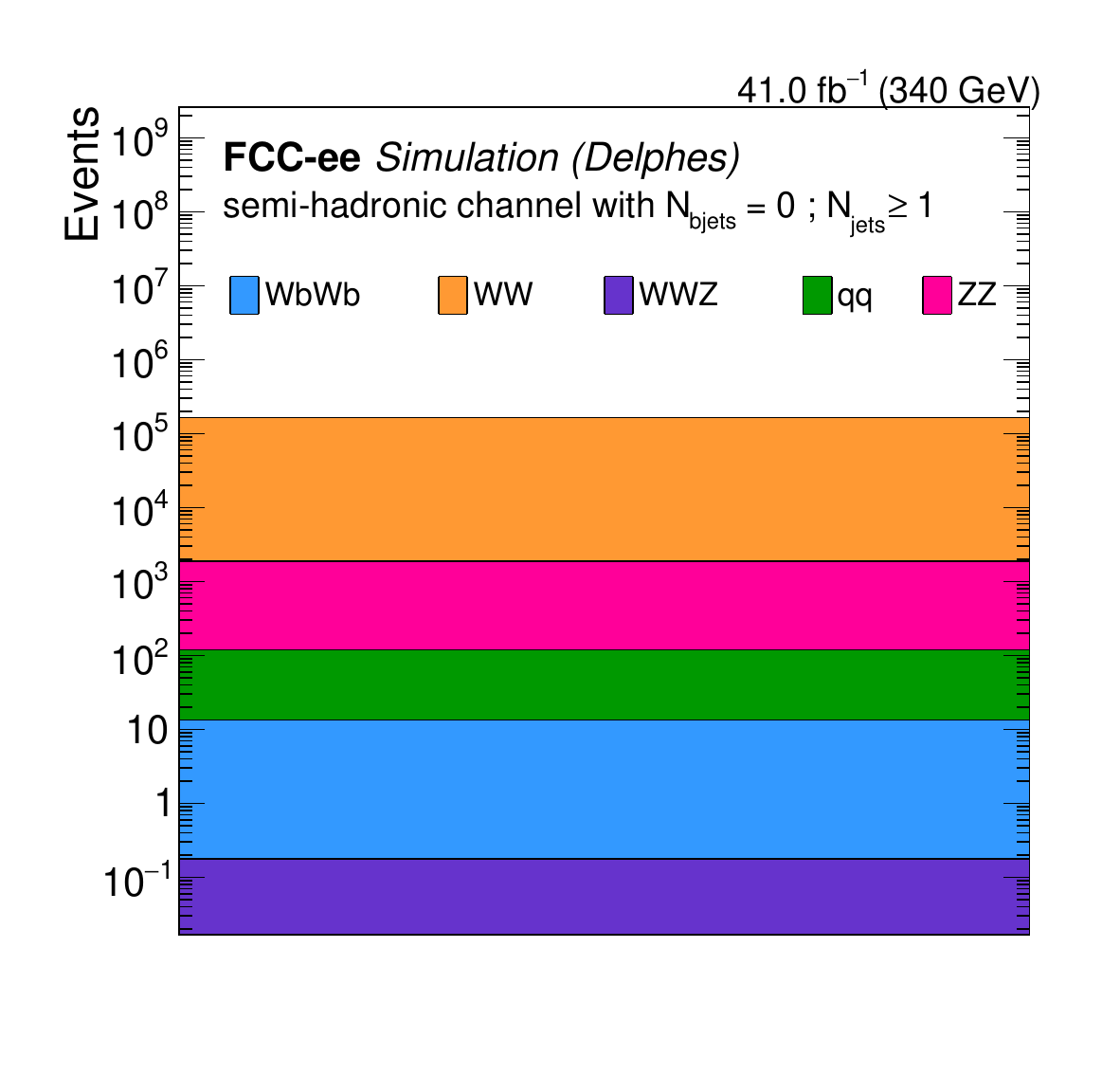}
    \includegraphics[width=0.495\linewidth]{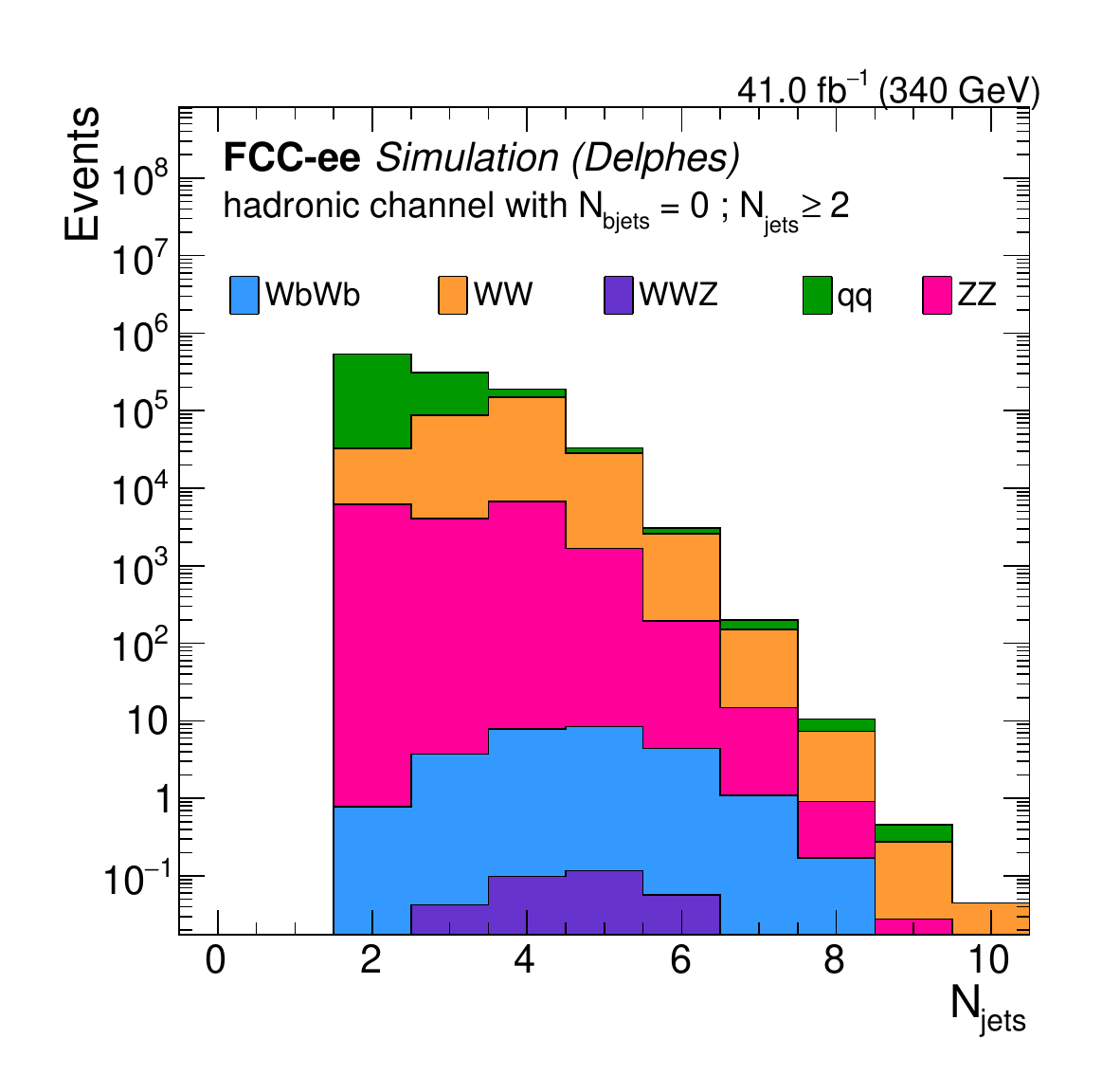}\\

    \includegraphics[width=0.495\linewidth]{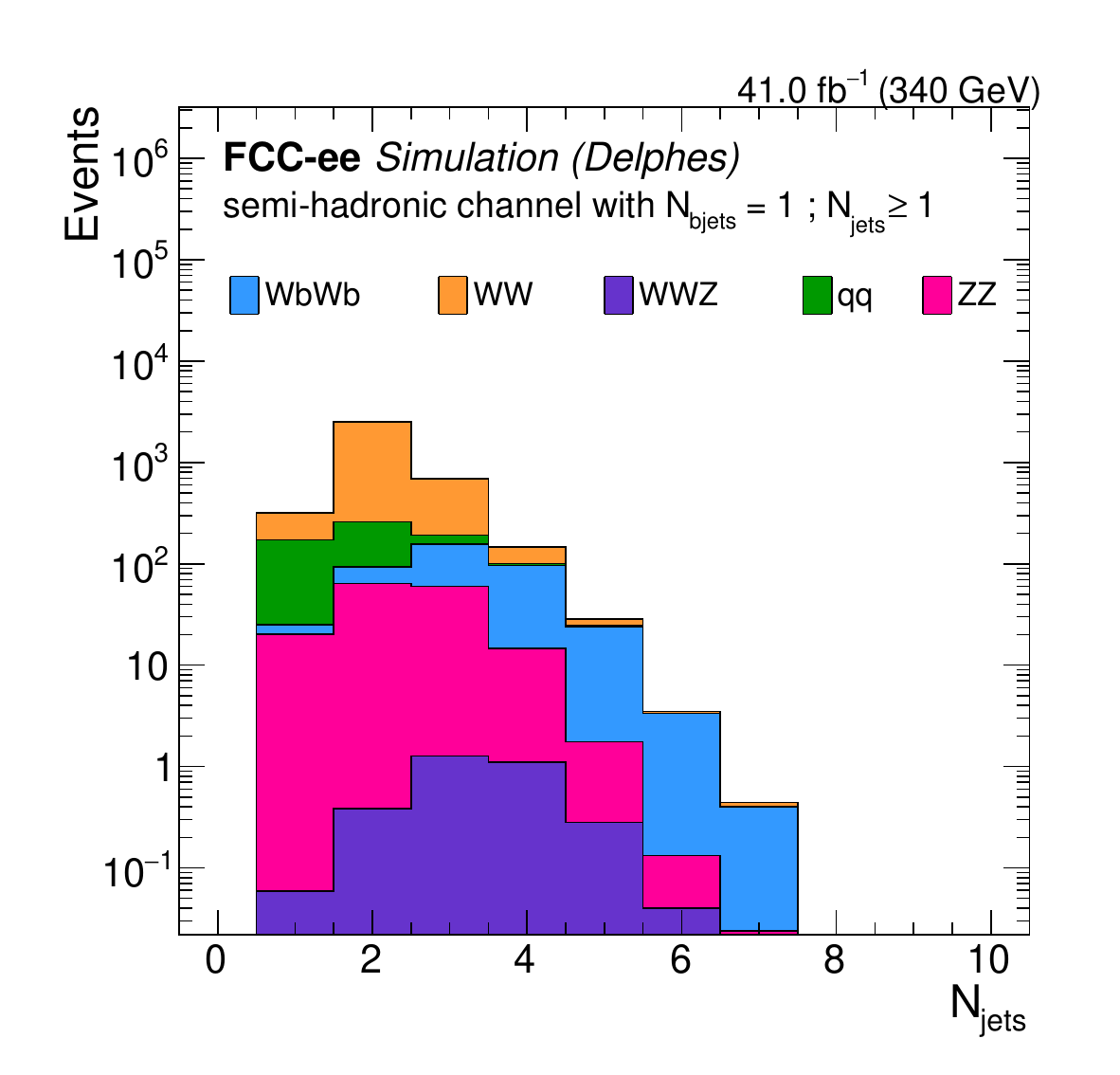}
    \includegraphics[width=0.495\linewidth]{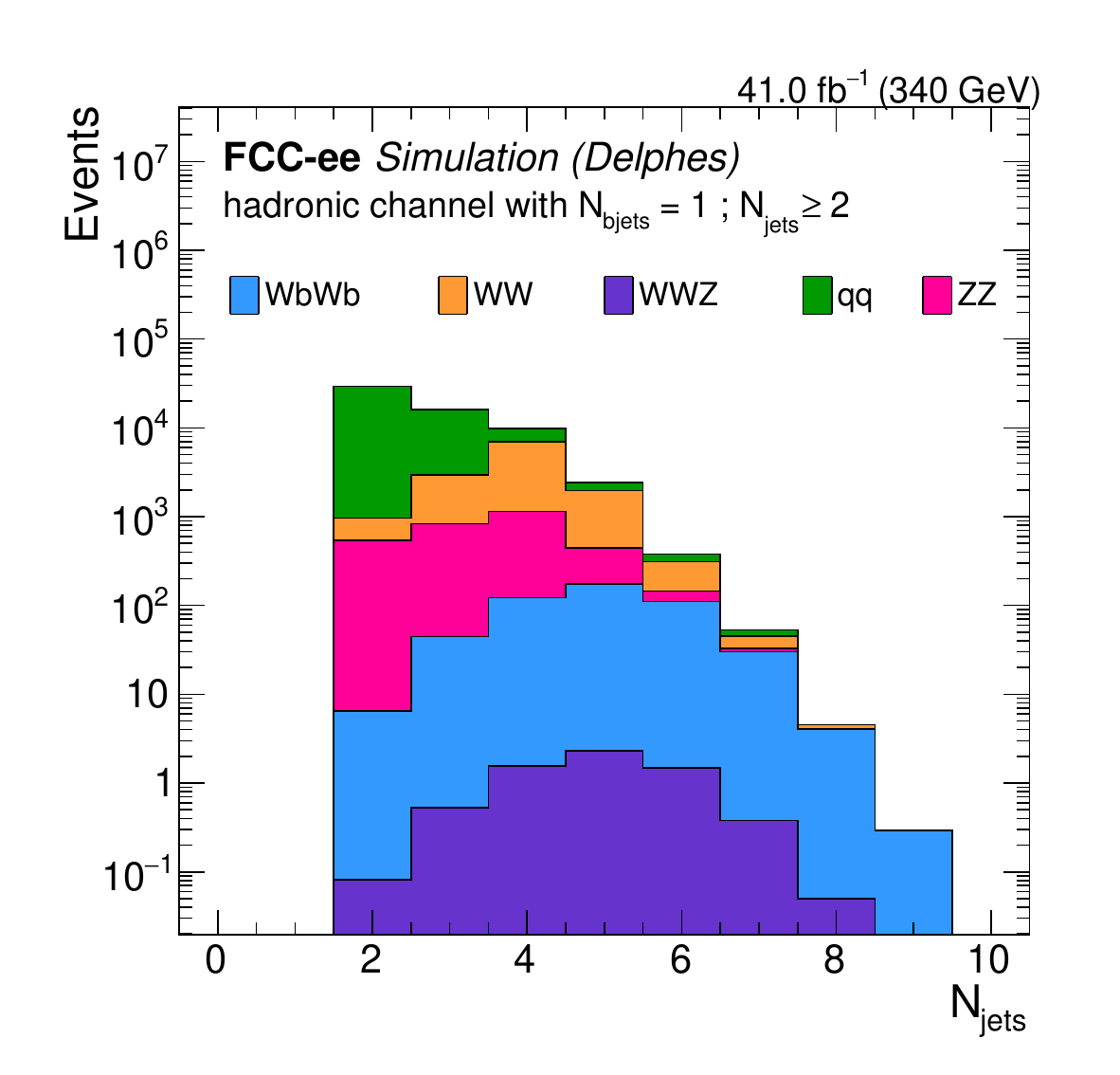}\\

    \includegraphics[width=0.495\linewidth]{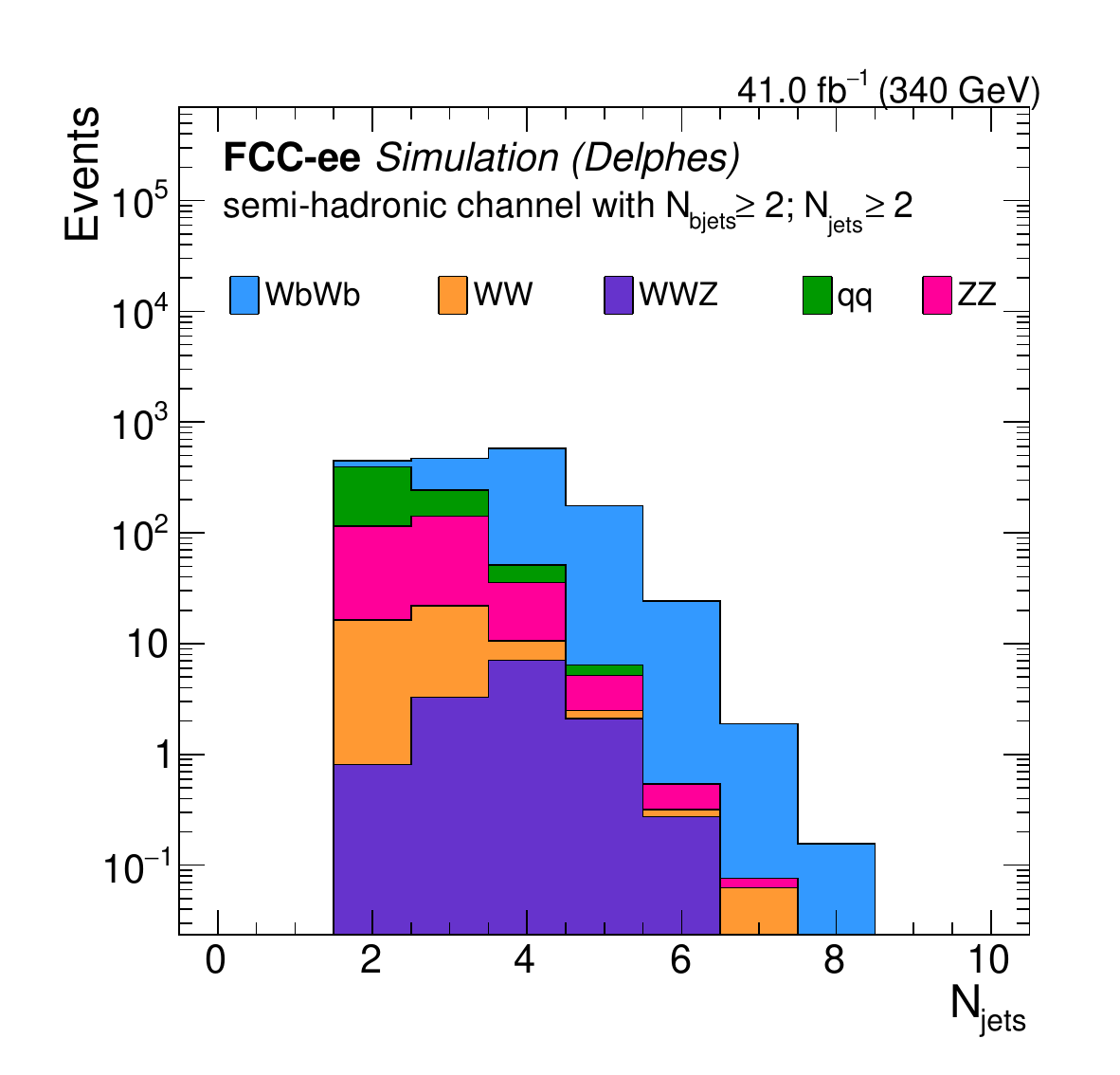}
    \includegraphics[width=0.495\linewidth]{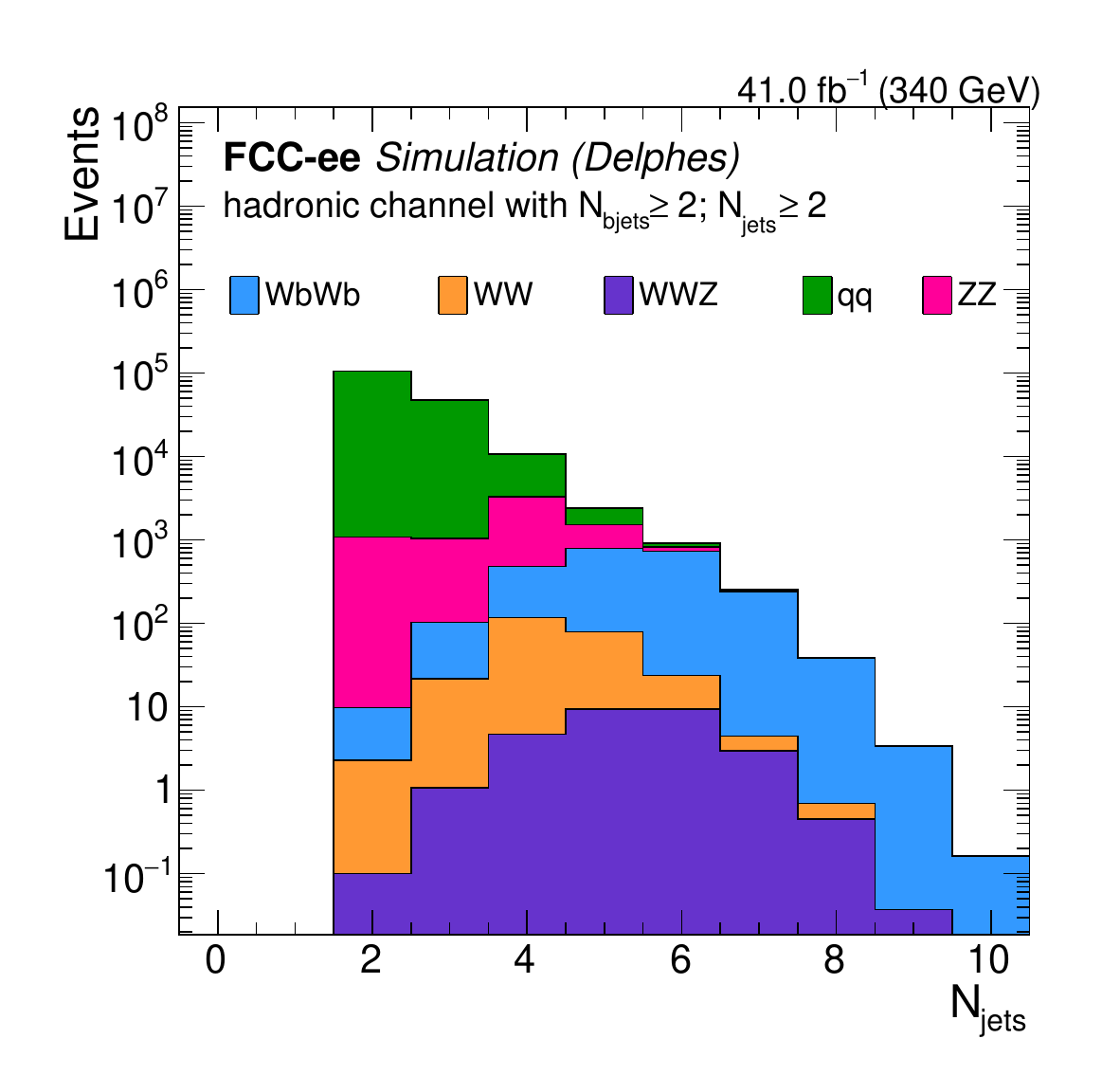}
    \caption{Same as Figure~\ref{fig:fit-distrib-345}, but for $\sqrts = 340\GeV$.}
    \label{fig:fit-distrib-340}
    \end{figure}

\begin{figure}[htbp]
    \centering
    \includegraphics[width=0.495\linewidth]{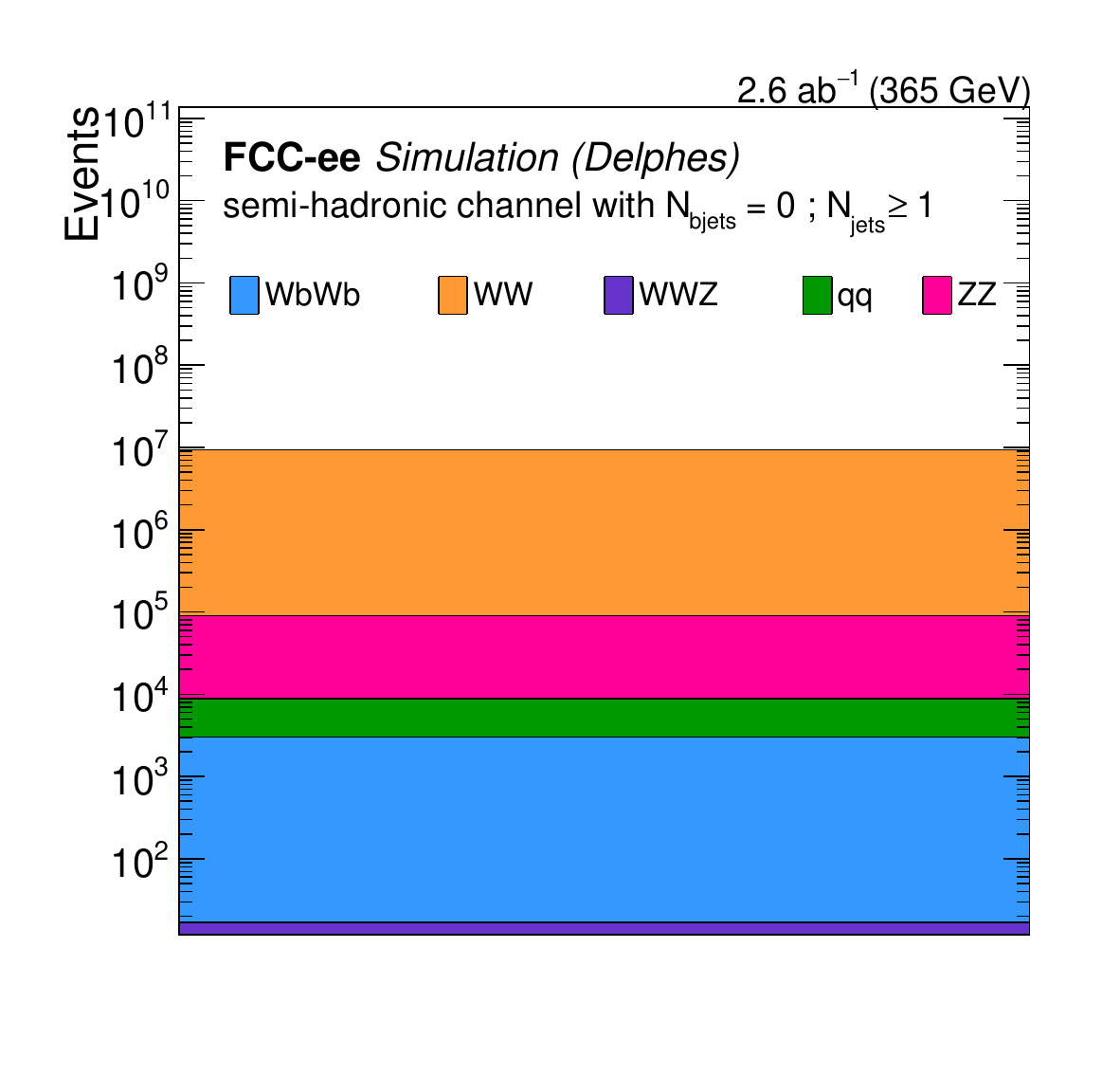}
    \includegraphics[width=0.495\linewidth]{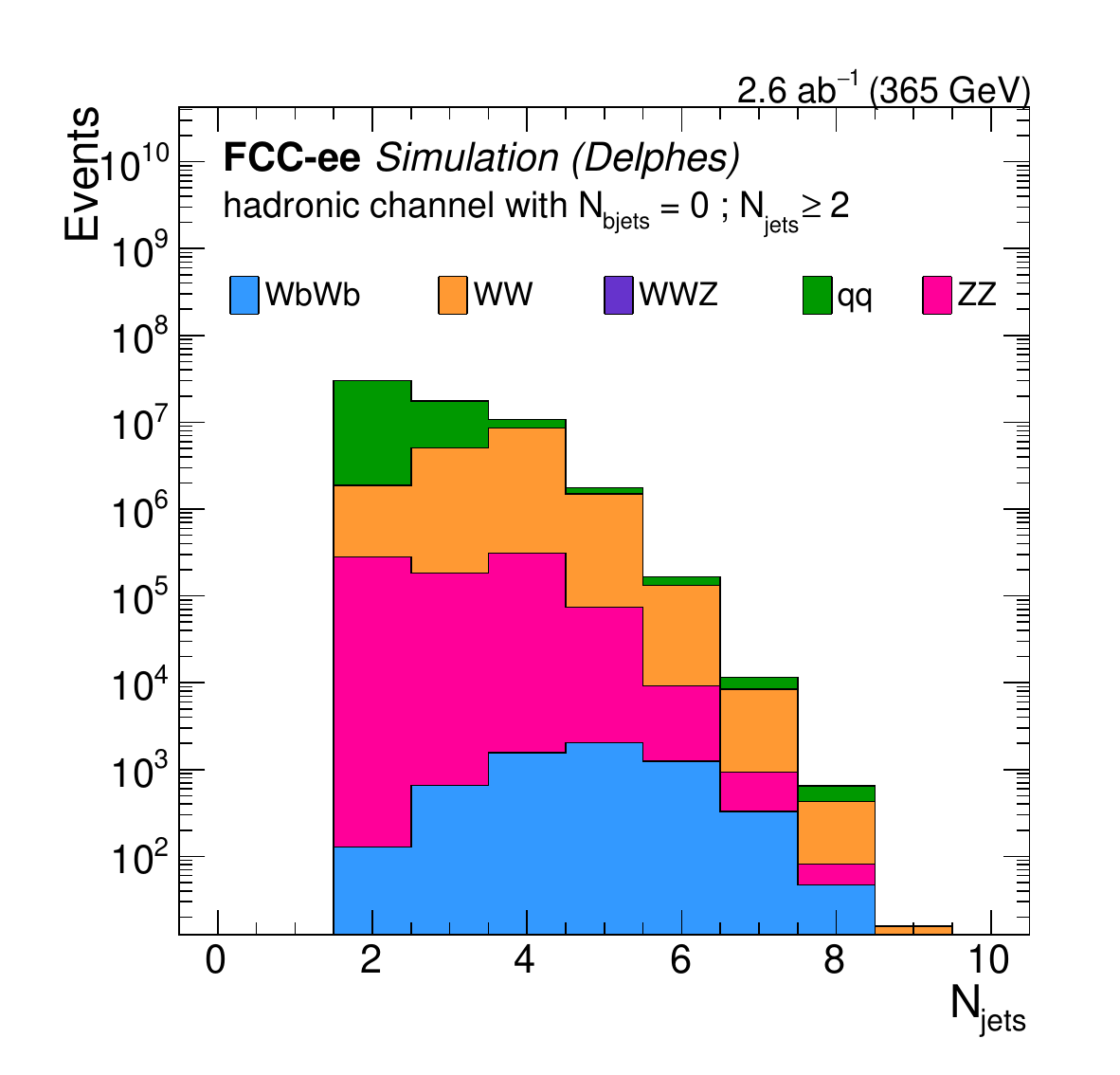}\\

    \includegraphics[width=0.495\linewidth]{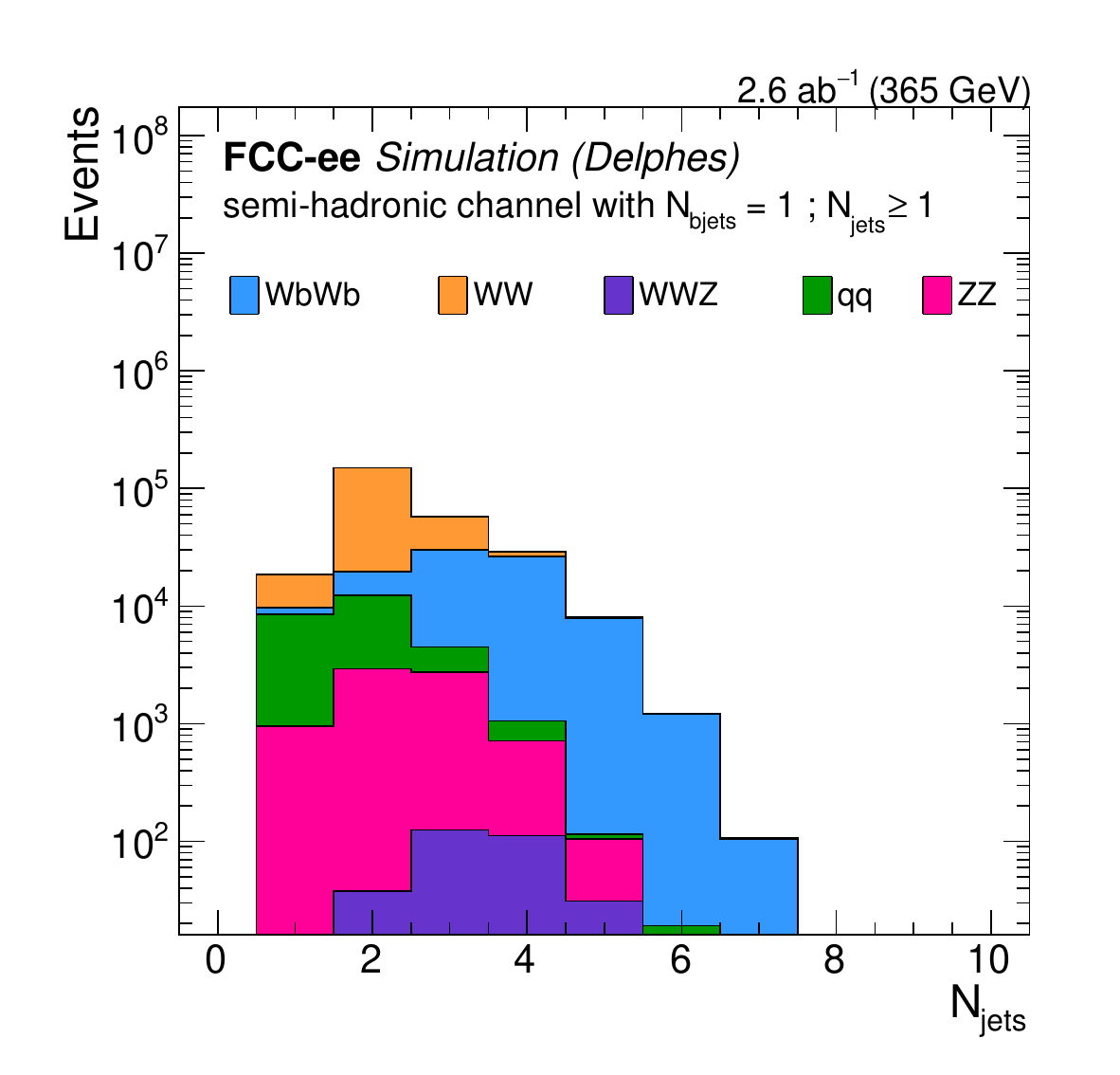}
    \includegraphics[width=0.495\linewidth]{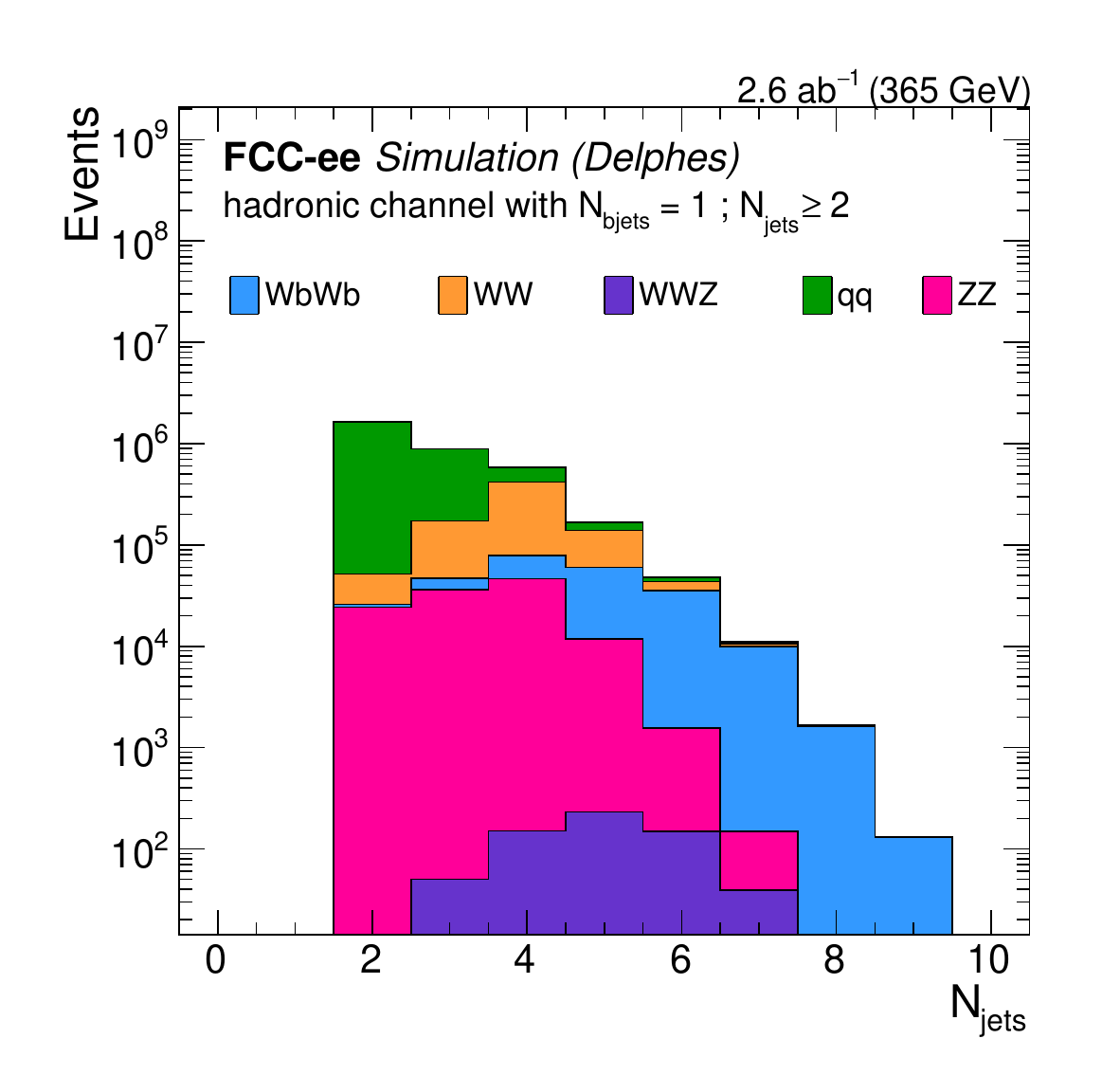}\\

    \includegraphics[width=0.495\linewidth]{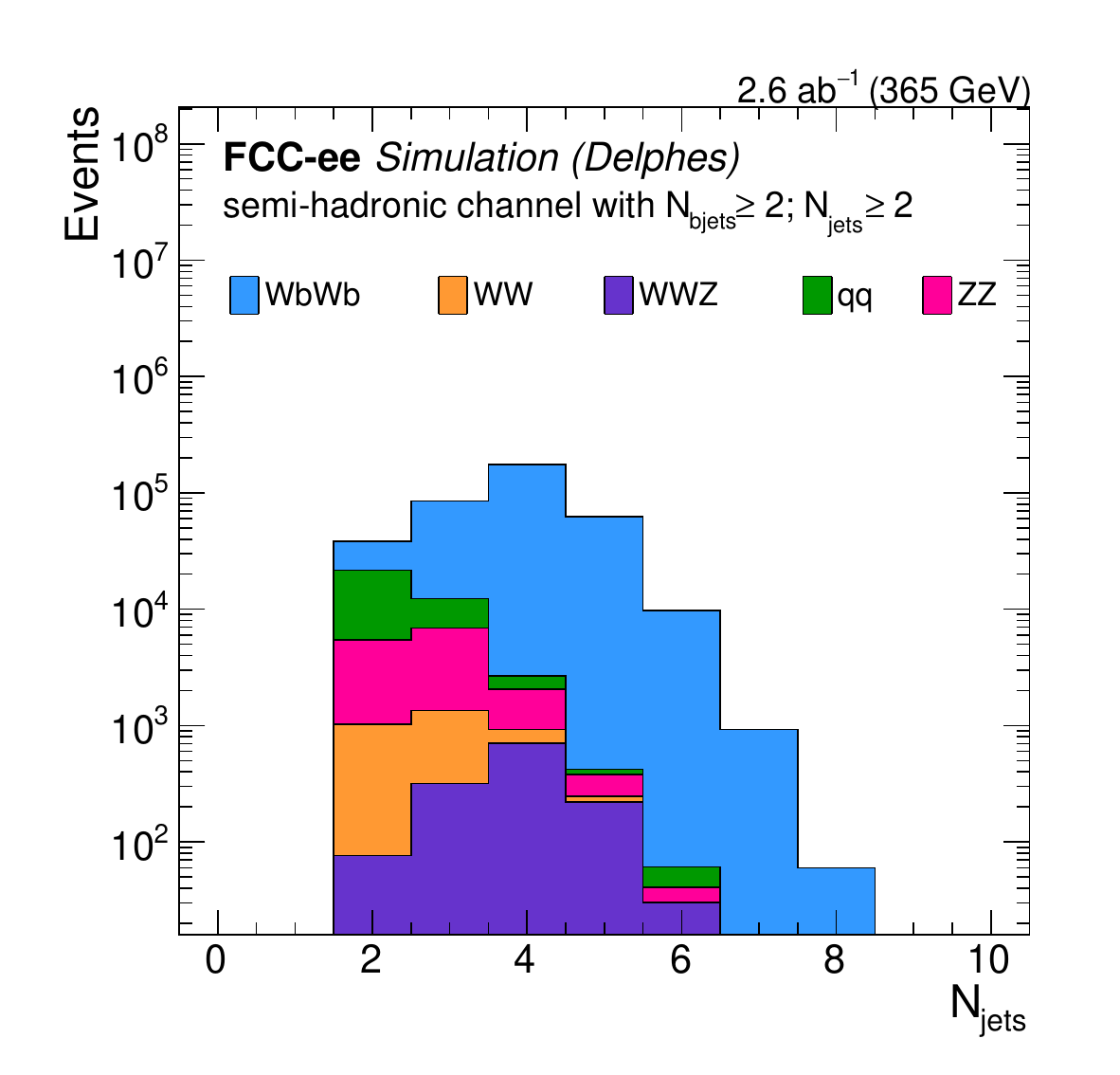}
    \includegraphics[width=0.495\linewidth]{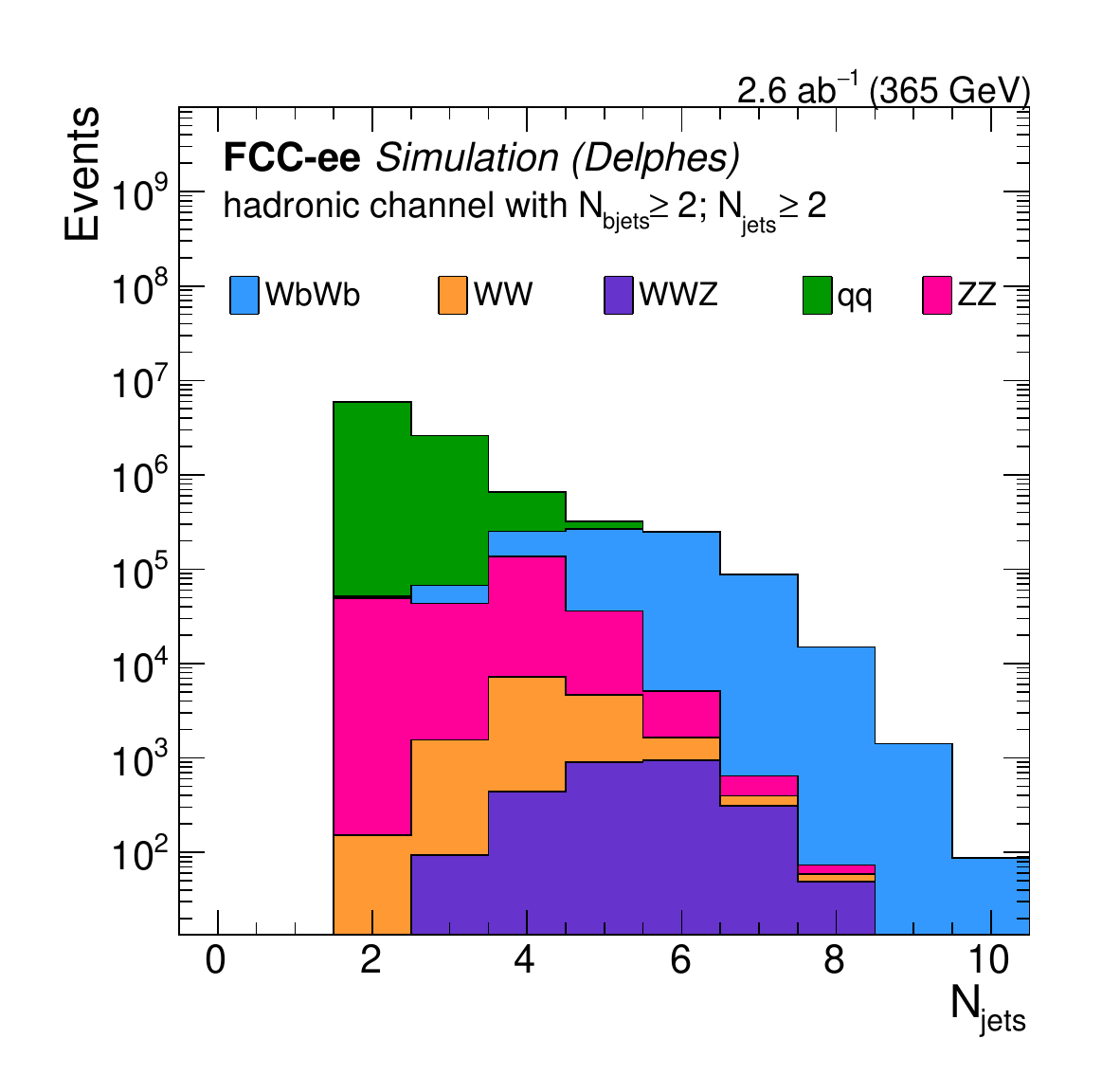}
    \caption{Same as Figure~\ref{fig:fit-distrib-345}, but for $\sqrts = 365\GeV$.}
    \label{fig:fit-distrib-365}
    \end{figure}

\begin{figure}[htbp]
    \centering
    \includegraphics[width=0.8\linewidth]{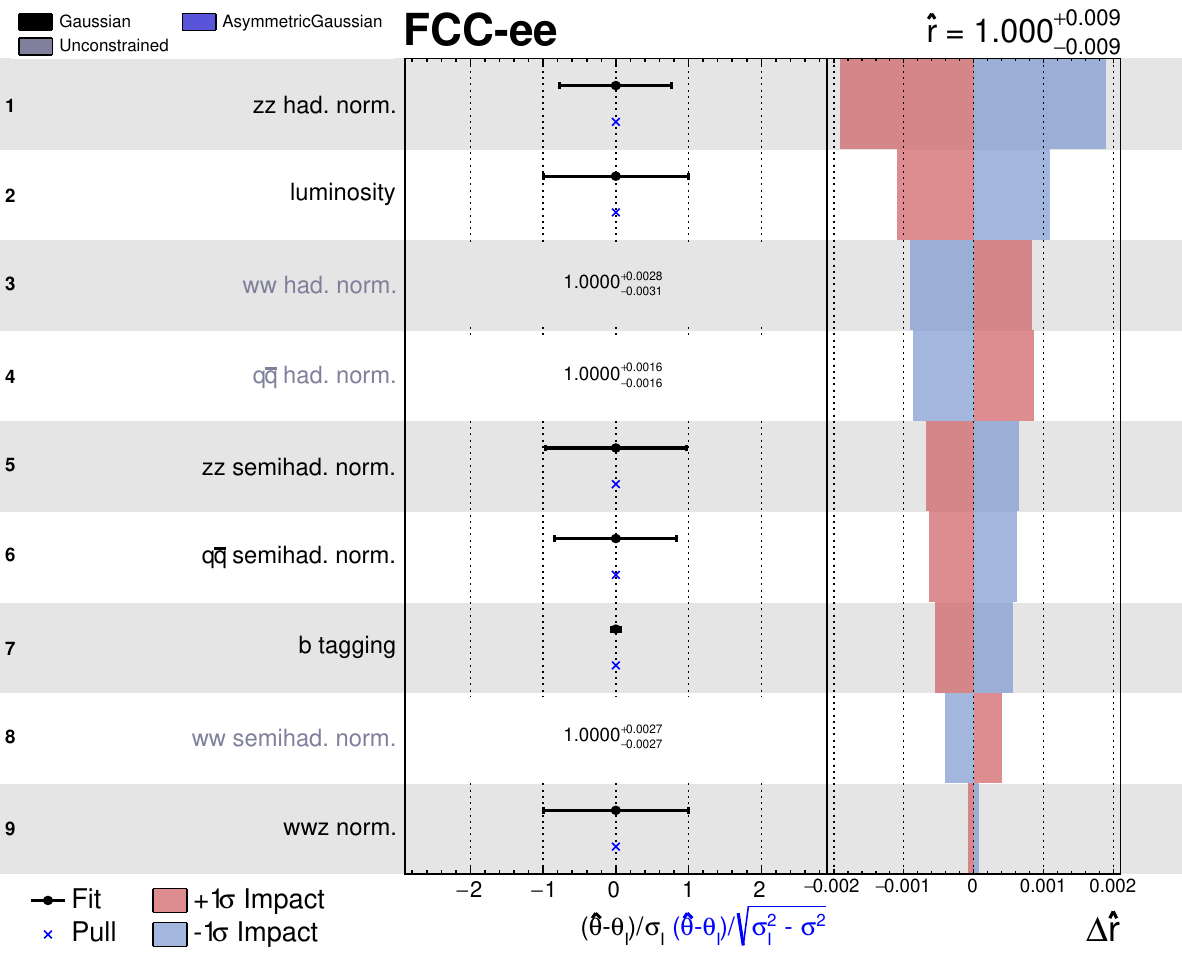}
    \caption{Impacts and constraints on the background normalisation and systematic uncertainties in the fit of the \WbWb production cross section at $\sqrts = 345\GeV$.}
    \label{fig:impacts_345}
\end{figure}

\begin{figure}[htbp]
    \centering
    \includegraphics[width=0.8\linewidth]{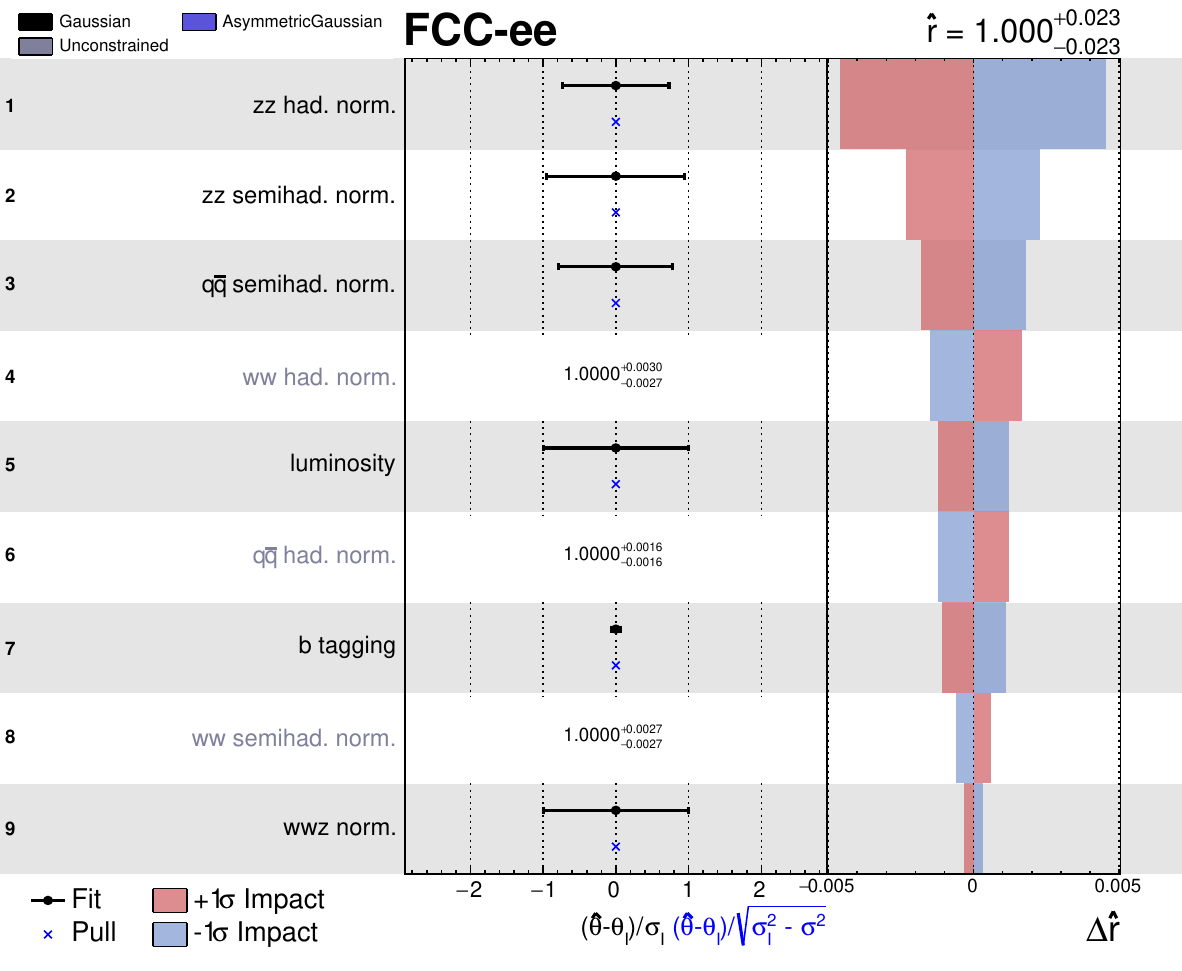}
    \caption{Same as Figure~\ref{fig:impacts_345}, but for $\sqrts = 340\GeV$.}
    \label{fig:impacts_340}
\end{figure}

\begin{figure}[htbp]
    \centering
    \includegraphics[width=0.8\linewidth]{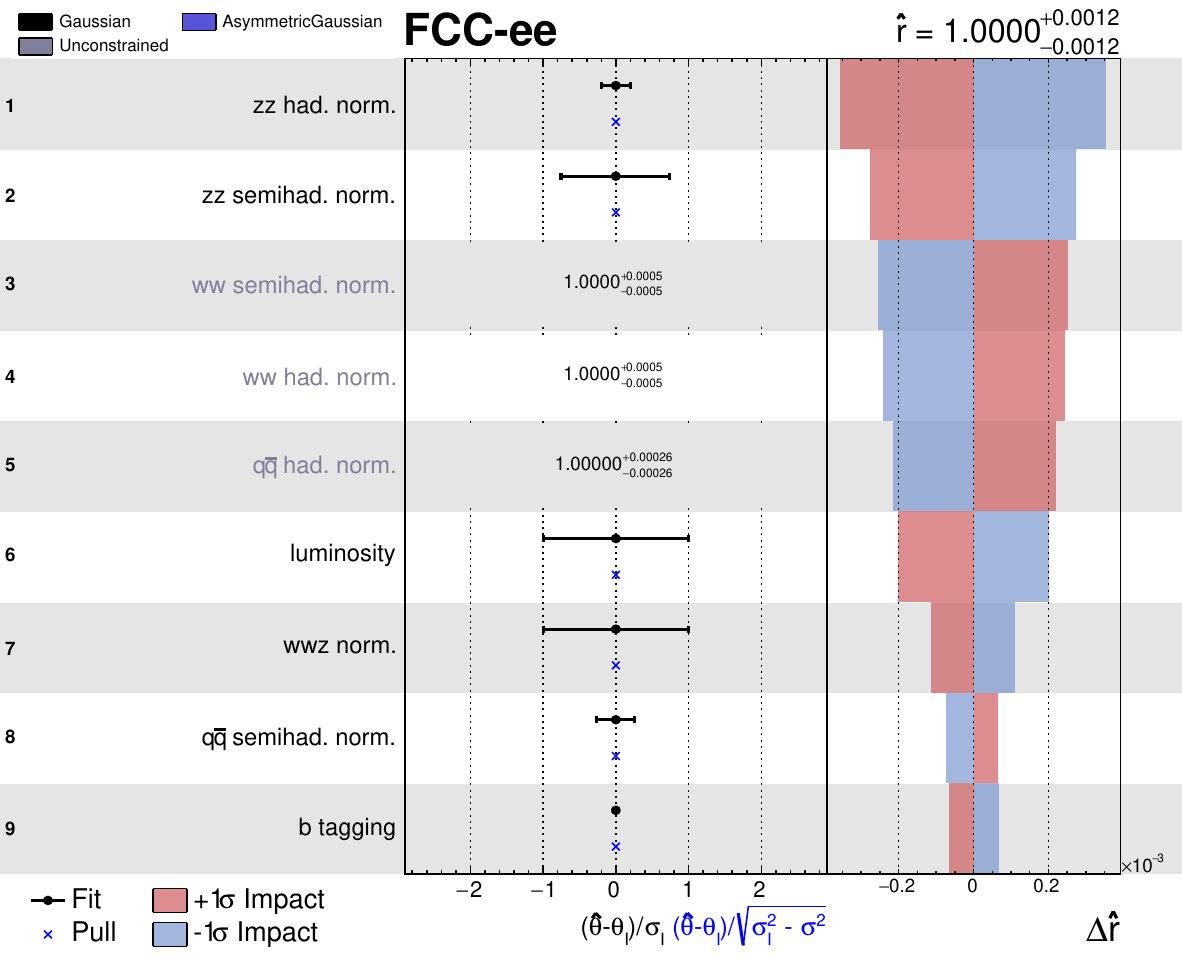}
    \caption{Same as Figure~\ref{fig:impacts_345}, but for $\sqrts = 365\GeV$.}
    \label{fig:impacts_365}
\end{figure}

Figures~\ref{fig:fit-distrib-340} and~\ref{fig:fit-distrib-365} show the input distributions to the fit for $\sqrts = 340$ and 365\GeV. The choice of input distributions is the same as for 345\GeV (Figure~\ref{fig:fit-distrib-345}). Figures~\ref{fig:impacts_345}--\ref{fig:impacts_365} show the impact of the various sources of systematic uncertainties on the measured cross sections, as well as the constraints obtained from the fit. The plots have been obtained using the Combine~\cite{CMS:2024onh} software.

\end{document}